\documentclass[numberedappendix,apj,twocolumn]{emulateapj}
\usepackage{color,url}

\newcommand{\isedfit}{{\tt iSEDfit}}
\newcommand{\kcorrect}{{\tt kcorrect}}

\newcommand{\bootes}{Bo\"{o}tes}
\newcommand{\zbootes}{zBo\"{o}tes}
\newcommand{\xbootes}{XBo\"{o}tes}

\newcommand{\lz}{\mbox{\ensuremath{L-Z}}}
\newcommand{\mz}{\mbox{\ensuremath{\mathcal{M}-Z}}}
\newcommand{\hii}{\textrm{H}~\textsc{ii}}
\newcommand{\mb}{\ensuremath{M_{B}}}
\newcommand{\kms}{\ensuremath{\textrm{km~s}^{-1}}}
\newcommand{\cms}{\ensuremath{\textrm{cm~s}^{-1}}}
\newcommand{\mass}{\ensuremath{\mathcal{M}}}
\newcommand{\mstar}{\ensuremath{\mathcal{M}^{\ast}}}
\newcommand{\msun}{\ensuremath{\mathcal{M}_{\sun}}}

\newcommand{\flunits}{\textrm{erg~s\ensuremath{^{-1}}~cm\ensuremath{^{-2}}~\AA\ensuremath{^{-1}}}}
\newcommand{\fluxunits}{\textrm{erg~s\ensuremath{^{-1}}~cm\ensuremath{^{-2}}}}
\newcommand{\lsun}{\ensuremath{L_{\sun}}}
\newcommand{\logmmsun}{\ensuremath{\log\,(\mathcal{M}/\mathcal{M_{\odot}})}}

\newcommand{\ha}{\textrm{H}\ensuremath{\alpha}}
\newcommand{\hb}{\textrm{H}\ensuremath{\beta}}
\newcommand{\hg}{\textrm{H}\ensuremath{\gamma}}

\newcommand{\oii}{[\textrm{O}~\textsc{ii}]}
\newcommand{\oiii}{[\textrm{O}~\textsc{iii}]}

\newcommand{\sii}{[\textrm{S}~\textsc{ii}]}
\newcommand{\siii}{[\textrm{S}~\textsc{iii}]}
\newcommand{\nii}{[\textrm{N}~\textsc{ii}]}

\newcommand{\halam}{\textrm{H}\ensuremath{\alpha~\lambda6563}}
\newcommand{\hblam}{\textrm{H}\ensuremath{\beta~\lambda4861}}
\newcommand{\hglam}{\textrm{H}\ensuremath{\gamma~\lambda4340}}
\newcommand{\hdlam}{\textrm{H}\ensuremath{\delta~\lambda4101}}

\newcommand{\hfivelam}{\textrm{H}\ensuremath{5~\lambda3889}}
\newcommand{\oiilam}{\oii~\ensuremath{\lambda3727}}
\newcommand{\oiiilam}{[\textrm{O}~\textsc{iii}]~\ensuremath{\lambda5007}}
\newcommand{\niilam}{[\textrm{N}~\textsc{ii}]~\ensuremath{\lambda6584}} 

\newcommand{\mgiilam}{\textrm{Mg}~\textsc{ii}~\ensuremath{\lambda2800}}
\newcommand{\nevlam}{[\textrm{Ne}~\textsc{v}]~\ensuremath{\lambda3426}} 
\newcommand{\neonlam}{[\textrm{Ne}~\textsc{iii}]~\ensuremath{\lambda3869}} 

\newcommand{\oiidoublet}{[\textrm{O}~\textsc{ii}]~\ensuremath{\lambda\lambda3726,3729}}
\newcommand{\oiiidoublet}{[\textrm{O}~\textsc{iii}]~\ensuremath{\lambda\lambda4959,5007}}
\newcommand{\niidoublet}{[\textrm{N}~\textsc{ii}]~\ensuremath{\lambda\lambda6548,6584}}
\newcommand{\siidoublet}{[\textrm{S}~\textsc{ii}]~\ensuremath{\lambda\lambda6716,6731}}

\newcommand{\ewoii}{\textrm{EW}(\oii)}
\newcommand{\ewoiii}{\textrm{EW}(\oiii)}

\newcommand{\ewhb}{\textrm{EW}(\textrm{H}\ensuremath{\beta})}

\newcommand{\niiha}{\nii/\ha}
\newcommand{\hahb}{\ha/\hb}

\newcommand{\ioniz}{\ensuremath{O_{32}}}
\newcommand{\pagel}{\ensuremath{R_{23}}}
\newcommand{\logu}{\ensuremath{\log\,(U)}}
\newcommand{\logq}{\ensuremath{\log\,(q)}}

\newcommand{\ewioniz}{\textrm{EW}(\ensuremath{O_{32}})}
\newcommand{\ewpagel}{\textrm{EW}(\ensuremath{R_{23}})}

\newcommand{\ionizcor}{(\ioniz)\ensuremath{_{\rm cor}}}

\newcommand{\logioniz}{\ensuremath{log(O_{32})}}

\newcommand{\logoh}{\ensuremath{12+\log\,(\textrm{O}/\textrm{H})}}

\newcommand{\ohstar}{\ensuremath{12+\log\,(\textrm{O}/\textrm{H})^{\ast}}}

\newcommand{\sfr}{\ensuremath{\psi}}

\newcommand{\sersic}{S\'{e}rsic}

\shorttitle{Evolution of the \mz{} Relation} 
\shortauthors{Moustakas et~al.}
\slugcomment{{\sc Submitted to ApJ}}

\begin{document}

\title{Evolution of the Stellar Mass-Metallicity Relation
  Since $\lowercase{z}=0.75$}

\author{John Moustakas\altaffilmark{1}, Dennis
  Zaritsky\altaffilmark{2}, Michael Brown\altaffilmark{3}, Richard
  J.~Cool\altaffilmark{4,5}, Arjun Dey\altaffilmark{6}, Daniel
  J.~Eisenstein\altaffilmark{7},  \\ Anthony
  H. Gonzalez\altaffilmark{8}, Buell Jannuzi\altaffilmark{6}, 
  Christine Jones\altaffilmark{7}, Chris~S. Kochanek\altaffilmark{9},
  \\ Stephen S. Murray\altaffilmark{7,10}, {\sc and} Vivienne
  Wild\altaffilmark{11}}  

\altaffiltext{1}{Center for Astrophysics and Space Sciences,
  University of California, San Diego, 9500 Gilman Drive, La Jolla,
  California, 92093, USA; jmoustakas@ucsd.com}
\altaffiltext{2}{Steward Observatory, University of Arizona, 933
  North Cherry Avenue, Tucson, AZ 85721, USA} 
\altaffiltext{3}{School of Physics, Monash University, Clayton,
  Victoria 3800, Australia} 
\altaffiltext{4}{Carnegie-Princeton Fellow; The Observatories of the
  Carnegie Institution of Washington, 813 Santa Barbara Street,
  Pasadena, CA 91101, USA}
\altaffiltext{5}{Department of Astrophysical Sciences, Princeton
  University, Peyton Hall, Princeton, NJ 08544}
\altaffiltext{6}{National Optical Astronomy Observatory, 950 North
  Cherry Avenue, Tucson, AZ 85719, USA}
\altaffiltext{7}{Harvard-Smithsonian Center for Astrophysics, 60
  Garden Street, Cambridge MA 02138, USA}
\altaffiltext{8}{Department of Astronomy, Bryant Space Science Center,
University of Florida, Gainesville, FL 32611, USA}
\altaffiltext{9}{Department of Astronomy, The Ohio State University,
  140 West 18th Avenue, Columbus, OH 43210, USA}
\altaffiltext{10}{Department of Physics and Astronomy, Johns Hopkins
  University, Baltimore, MD 21218, USA}
\altaffiltext{11}{Institute for Astronomy, University of Edinburgh,
  Royal Observatory, Blackford Hill, Edinburgh EH9 3HJ, UK} 
\setcounter{footnote}{11}

\begin{abstract} 
We measure the gas-phase oxygen abundances of $\sim3000$ star-forming
galaxies at $z=0.05-0.75$ using optical spectrophotometry from the AGN
and Galaxy Evolution Survey (AGES), a spectroscopic survey of $I_{\rm
  AB}<20.45$ galaxies over $7.9$~deg$^{2}$ in the NOAO Deep Wide Field
Survey (NDWFS) \bootes{} field.  We use state-of-the-art techniques to
measure the nebular emission lines and stellar masses, and explore and
quantify several potential sources of systematic error, including the
choice of metallicity diagnostic, aperture bias, and contamination
from unidentified active galactic nuclei (AGN).  Combining
volume-limited AGES samples in six independent redshift bins and
$\sim75,000$ star-forming galaxies with $r_{\rm AB}<17.6$ at
$z=0.05-0.2$ selected from the Sloan Digital Sky Survey (SDSS) that we
analyze in the identical manner, we measure the evolution of the
stellar mass-metallicity (\mz) between $z=0.05$ and $z=0.75$.  We find
that at fixed stellar mass galaxies at $z\sim0.7$ have just
$30\%-60\%$ the metal content of galaxies at the present epoch, where
the uncertainty is dominated by the strong-line method used to measure
the metallicity.  Moreover, we find no statistically significant
evidence that the \mz{} relation evolves in a mass-dependent way for
$\mass\simeq10^{9.8}-10^{11}$~\msun{} star-forming galaxies.  Thus,
for this range of redshifts and stellar masses the \mz{} relation
simply shifts toward lower metallicity with increasing redshift
without changing its shape.
\end{abstract}


\section{Introduction}\label{sec:intro}

Like stellar mass, the gas-phase metallicity of a galaxy is a
sensitive observational diagnostic of its past star formation history
and present-day evolutionary state for the simple reason that in a
closed system metallicity increases monotonically with each successive
generation of massive stars.\footnote{Throughout this paper we use the
  terms {\em metallicity} and {\em abundance} interchangeably to mean
  the heavy-element content of the warm ($T\approx10^{4}$~K)
  interstellar medium of galaxies.  In particular, we make the
  reasonable assumption that the nebular oxygen abundance,
  historically written as \logoh, traces the total gas-phase
  metallicity, $Z_{\rm gas}$.}  In detail, however, galaxies are not
closed systems: infall of cold, metal-poor gas from the intergalactic
medium, rapid gas accretion via minor and major mergers, and
supernova-driven winds of gas and metals can modulate the metallicity
of individual galaxies according to their large-scale environment, gas
supply, and assembly history.  Therefore, accurate abundance
measurements provide valuable insight into the interplay between many
fundamental processes in galaxy evolution, including star formation,
gas accretion, and supernova-driven feedback across cosmic time
\citep{tinsley80a, pagel97a, pettini04b}.

Nebular emission lines---ubiquitous in the rest-frame optical spectra
of star-forming galaxies---provide a particularly powerful way to
study the chemical abundances of both nearby and distant galaxies.
Among the most commonly observed lines are the Balmer \halam, \hblam,
and \hglam{} hydrogen recombination lines, and the collisionally
excited \oiidoublet, \neonlam, \oiiidoublet, \niidoublet, and
\siidoublet{} forbidden lines.  Because these lines originate
principally in star-forming (\hii) regions, they trace the physical
conditions in the gas from which the current generation of massive
stars is forming.  The metal lines in particular are the principal
coolants in \hii{} regions, making them sensitive to the total
abundance of oxygen and other heavy elements in the interstellar
medium.  Moreover, the relative line-strengths can be used to infer
the interstellar pressure, density, temperature, ionizing radiation
field strength, dust reddening, and the presence of an active galactic
nucleus (AGN).  Finally, from an observational standpoint, many of the
nebular lines are intrinsically strong, making them measurable in even
relatively low signal-to-noise spectra over a broad range of
redshifts.

Following the discovery of the correlation between dynamical mass and
oxygen abundance in dwarf irregular galaxies (\citealt{lequeux79a};
see also \citealt{garnett87a, oey93a}), numerous subsequent studies
demonstrated that star-forming galaxies ranging from the
lowest-luminosity dwarfs to massive disk galaxies obey a well-defined
luminosity-metallicity (\lz) correlation spanning several orders of
magnitude in $B$-band luminosity \citep{garnett87a, skillman89a,
  zaritsky94a, pilyugin04a, lama04a, salzer05a, lee06b, moustakas10a}.
Metallicity was also found to correlate with morphological type
\citep{edmunds84a}, surface mass density \citep{mccall82a, ryder95a,
  garnett97a}, and maximum rotational velocity \citep{zaritsky94a,
  garnett02a, dalcanton07a}.  However, because these global properties
all correlate with one another, the underlying physical driver of the
\lz{} relation remained elusive.  

\citet{tremonti04a} was the first to leverage the tremendous
statistical power of the Sloan Digital Sky Survey
\citep[SDSS;][]{york00a} spectroscopic database to show that the
gas-phase metallicity of a galaxy correlates best with stellar mass,
\mass, now known as the stellar mass-metallicity (\mz) relation.  The
\mz{} relation reveals that the gas-phase metallicity of star-forming
galaxies increases monotonically with stellar mass and then remains
relatively constant above $\mass\approx3\times10^{10}~\msun$.
Although subsequent studies have found weak residual correlations from
the \mz{} relation with stellar mass density \citep{tremonti04a,
  liang10a}, size \citep{ellison08a}, large- and small-scale
environment \citep{mouhcine07a, cooper08a, ellison08b, ellison09a,
  peeples09a}, star formation rate \citep[SFR;][]{mannucci10a,
  lara-lopez10a, yates11a, cresci11a}, and the existence of bars
\citep{ellison11a}, the intrinsic dispersion in the \mz{} relation is
$\lesssim0.1$~dex, making it among the tightest empirical correlations
known.

The physical origin of the \mz{} relation remains under debate.  One
possibility is that low-mass galaxies started forming stars later than
massive galaxies \citep[i.e., they are ``younger'';][]{noeske07b,
  leitner11b} and have been less efficient at synthesizing metals via
star formation \citep{brooks07a, mouhcine08a, calura09a}.  This
interpretation is qualitatively consistent with the measured low gas
fractions in massive galaxies relative to lower-mass galaxies
\citep{mcgaugh97a, geha06a, garcia-appadoo09a}, and with the observed
correlation between stellar mass and SFR \citep[the {\em star
    formation sequence};][]{brinchmann04a, salim07a, elbaz07a,
  noeske07a}.  Another popular interpretation is that supernova-driven
galactic winds preferentially expel metals from low-mass galaxies
\citep{larson74a, garnett02a, tremonti04a, dalcanton07a}.  Indeed,
state-of-the-art hydrodynamic and semianalytic theoretical models
\emph{require} metal-enriched outflows to match many observed galaxy
properties, including the \mz{} relation \citep{kobayashi07a,
  finlator08a, dutton09a, peeples11a, dave11b}.  A related outstanding
question is whether gas and metals ejected in previous star formation
episodes are re-accreted onto galaxies \citep{delucia04a,
  oppenheimer10a}, or whether galaxies are predominantly fed by infall
of cold, metal-poor gas from the intergalactic medium
\citep{koppen99a, faucher11a}.  Alternatively, \citet{koppen07a} show
that a SFR-dependent, and therefore mass-dependent, stellar initial
mass function (IMF) naturally explains the observed \mz{} relation
without needing to invoke metal-enriched outflows from galaxies.

One way of gaining insight into the complex and multifaceted
interrelationship between chemical abundance measurements, star
formation, gas accretion, and the role of supernova feedback during
galaxy growth is to measure the evolution of the \mz{} relation.  The
time since $z=1$, spanning $\sim60\%$ of the age of the Universe, is
important for many reasons.  Measurements of the star formation
sequence \citep{noeske07a, wuyts11a} suggest that star-forming
galaxies over this redshift range evolve smoothly as a population,
driven by continuous, secular processes like gas consumption and {\em
  in situ} star formation \citep{bell05a, noeske09a, leitner11b}.
Bolstered by measurements of their morphological distribution
\citep{bell05a, konishi11a} and clustering properties \citep{coil08a,
  zehavi11a}, these results suggest that we can draw direct
evolutionary connections linking star-forming galaxies across this
redshift range.  In addition, the metallicity-sensitive \pagel{}
parameter (defined in \S\ref{sec:oh}) can be measured from
ground-based optical spectroscopy of galaxies at least to $z\sim0.8$,
ensuring that metallicities can be derived using a single consistent
abundance calibration.

Unfortunately, previous studies have reported widely varying results
on the evolution of both the shape and normalization of the \mz{}
relation since $z=1$.  \citet{savaglio05a} and \citet{zahid11a} find
that the mean metallicity of galaxies at $z\sim0.8$ with
$\mass\sim10^{10.3}$~\msun{} differs by just $\sim0.05$~dex (factor of
$\sim1.12$) relative to similarly massive star-forming galaxies at
$z\sim0.1$, whereas they find that galaxies with
$\mass\sim10^{9.5}$~\msun{} undergo a factor of $2-3$ more chemical
evolution over the same redshift interval.  On the other hand,
\citet{liang06a} and \citet{cowie08a} report $0.2-0.3$~dex (factor of
$1.6-2$) of chemical evolution for galaxies with
$\mass\sim10^{10.3}$~\msun{} since $z\sim0.7$ and no statistically
significant evidence for evolution in the shape of the \mz{} relation
\citep[see also][]{rodrigues08a}.  \citet{lama09a} and
\citet{perez-montero09a} find an even greater amount of metallicity
evolution, $\sim0.35$~dex (factor of $2.2$) since $z=0.6-1$, and a
\emph{flattening} of the \mz{} relation at intermediate redshift.
Other studies based on the evolution of the $B$-band \lz{} relation
report that at fixed luminosity star-forming galaxies at $z=0.5-1$ are
$0.1-0.7$~dex (factor of $1.25-5$) more metal-poor than local
star-forming galaxies \citep{kobulnicky99b, carollo01a, lilly03a,
  kobulnicky03a, kobulnicky04a, maier04a, maier05a, maier06a,
  lama09a}.  However, the significant amount of luminosity evolution
experienced by blue, star-forming galaxies since $z=1$
\citep{blanton06a, faber07a, cool12a} renders the interpretation of
the \lz{} relation less straightforward.

The origin of these widely varying results on the evolution of the
\mz{} relation can be attributed to various issues, including cosmic
variance, small sample size, heterogenous selection criteria,
systematic differences in the methods used to infer nebular
abundances, an inconsistent analysis of the local \mz{} relation,
spectroscopy with insufficient spectral coverage, signal-to-noise
(S/N) ratio, or instrumental resolution, and uncertain stellar mass
estimates due to limited broadband photometric coverage.
Consequently, the efficacy of existing observational metallicity
constraints on state-of-the-art theoretical models of galaxy formation
\citep{delucia04a, brooks07a, dutton09a, dave11b} has been fairly
limited.  From an observational standpoint, there exists a clear need
for a large, homogeneously selected sample of galaxies with
high-quality spectroscopy and reliable stellar mass estimates to
better constrain the evolution of the \mz{} relation at intermediate
redshift.

To address this need, we measure the evolution of the \mz{} and
$B$-band \lz{} relations at intermediate redshift using oxygen
abundances of $\sim3000$ star-forming galaxies at $z=0.05-0.75$
observed as part of the AGN and Galaxy Evolution Survey
\citep[AGES;][]{kochanek11a}, and $\sim75,000$ galaxies at
$z=0.05-0.2$ selected from the SDSS.  The AGES main galaxy survey
consists of optical spectrophotometry for $\sim12,000$ galaxies in the
$\sim9$~deg$^{2}$ NOAO Deep Wide Field Survey
\citep[NDWFS;][]{jannuzi99a, brown03a, brown07a, brown08a} \bootes{}
field at a median redshift of $z\sim0.3$.  This sample is
statistically complete over $7.9$~deg$^{2}$ for galaxies with $I_{\rm
  AB}<20.45$ \citep{kochanek11a}, reaching $\sim2$~mag deeper than the
SDSS main galaxy sample \citep{strauss02a} over a considerably larger
area than other recent or ongoing surveys of intermediate-redshift
galaxies such as DEEP2 \citep{davis03a}, AEGIS \citep{davis07a}, VVDS
\citep{lefevre04a, lefevre05a, garilli08a}, and zCOSMOS
\citep{lilly09a}.\footnote{For galaxies at $z\lesssim0.5$, the Galaxy
  and Mass Assembly (GAMA) survey will ultimately supplant all these
  surveys by obtaining optical spectroscopy for several hundred
  thousand galaxies brighter than $r_{\rm AB}\approx19.8$ over
  $360$~deg$^{2}$ \citep{driver11a}.}  The statistical completeness,
large sample size, and availability of high-quality optical
spectroscopy and deep optical and near-infrared photometry enables us
to investigate the evolution of both the \mz{} and optical \lz{}
relations of star-forming galaxies at $z=0.05-0.75$ using a single,
self-consistent abundance diagnostic over the entire redshift range.

The plan of the paper is as follows.  In \S\ref{sec:data} we present
the ground-based optical and near-infrared imaging we use, summarize
the AGES observations, and describe how we measure the nebular
emission lines for the galaxies in our sample.  In \S\ref{sec:sed} we
describe the methodology used to derive rest-frame luminosities,
colors, and stellar masses, and in \S\ref{sec:sample} we select a
subset of the AGES galaxies for chemical abundance analysis.  We
present the methods we use to derive oxygen abundances in
\S\ref{sec:oh}, and our principal results in \S\ref{sec:evol}, where
we quantify the mass-dependent evolution of the \mz{} relation for
star-forming galaxies since $z=0.75$.  In \S\ref{sec:syseffects} we
address the effect of various potential sources of systematic
uncertainty on our results, and in \S\ref{sec:discussion} we discuss
recent theoretical work on the origin and evolution of the \mz{}
relation.  Finally, we summarize our principal conclusions in
\S\ref{sec:summary}.

We adopt a concordance cosmology with $\Omega_{\rm m}=0.3$,
$\Omega_{\Lambda}=0.7$, and $h_{70}\equiv H_{0}/100=0.7$, the AB
magnitude system \citep{oke83a}, and the \citet{chabrier03a} initial
mass function (IMF) from $0.1-100~\mathcal{M}_{\sun}$ unless otherwise
indicated.  For reference, the conversion from Vega to AB for the
$I$-band filter used to select AGES targets is $+0.45$~mag, that is
$I_{\rm AB}=I_{\rm Vega}+0.45$. 

\section{Observations}\label{sec:data}

In the following sections we present the multiwavelength imaging of
the \bootes{} field that we use (\S\ref{sec:phot}), describe the AGES
optical spectroscopy and emission-line measurements
(\S\ref{sec:ages}), and construct a comparison sample of local
galaxies from the SDSS (\S\ref{sec:sdss}).

\subsection{Multiwavelength Photometry}\label{sec:phot} 

\subsubsection{Optical and Near-infrared Imaging}\label{sec:ndwfs}  

Our baseline optical observations consist of deep $B_{W}RI$ imaging
available as part of the NDWFS third public data release.\footnote{The
  $B_{W}$ filter is a ``wide'' filter with a bluer effective
  wavelength ($\lambda_{\rm eff}\approx4200$~\AA) than the standard
  Johnson-Morgan $B$-band filter \citep{bessell90a}.}  The imaging was
carried out using the MOSAIC-I camera at the KPNO/Mayall 4~m
telescope, reaching a $5\sigma$ depth of $\sim26.5$, $\sim25.5$, and
$\sim25.3$~mag in a $2\arcsec$ diameter circular aperture in $B_{W}$,
$R$, and $I$, respectively.  Note that although $K$-band imaging from
the NDWFS is also available for $\sim60\%$ of the \bootes{} survey
region, we do not use these older data in favor of the more recent
near-infrared (near-IR) observations described below.

We supplement the NDWFS photometry with $U$- and $z$-band observations
obtained as part of two other imaging campaigns.  F.~Bian
et~al. (2012, in prep.) have obtained deep $U$-band imaging of the
\bootes{} field using the Large Binocular Camera
\citep[LBC;][]{giallongo08a} mounted at the primary focus of the
$8.4$~m Large Binocular Telescope (LBT); these observations reach a
$5\sigma$ depth of $\sim25.2$~mag for point sources.  Second, the
\zbootes\footnote{\url{http://archive.noao.edu/nsa/zbootes.html}}
survey \citep{cool07a} imaged $7.62$~deg$^{2}$ of the \bootes{} field
to a $3\sigma$ depth of $22.4$~mag in a $3\arcsec$ diameter aperture
using the {\sc 90prime} prime-focus wide-field imager mounted at the
Bok $2.3$~m telescope \citep{williams04a}.

Finally, we utilize deep $JHK_{s}$ near-IR imaging of the entire
\bootes{} field obtained using the NOAO Extremely Wide-Field Infrared
Mosaic \citep[NEWFIRM;][]{autry03a} instrument at the KPNO/Mayall 4~m
telescope.  The NEWFIRM observations achieved a $5\sigma$ depth of
$\sim22.9$, $\sim22.1$, and $\sim21.4$~mag in a $3\arcsec$ diameter
aperture in $J$, $H$, and $K_{s}$, respectively.  More details
regarding the NEWFIRM observations and reductions will be described in
an upcoming paper (A.~Gonzalez, 2011, private communication). 

\subsubsection{Ancillary X-ray, Mid-Infrared, and Radio
  Imaging}\label{sec:ancillary}

We use observations of the \bootes{} field in the X-ray from the {\em
  Chandra X-ray Observatory}, at $3.6-8$~\micron{} from the {\em
  Spitzer Space Telescope} \citep{werner04a}, and at $1.4$~GHz from
the Westerbork Synthesis Radio telescope to help identify and remove
AGN from our sample (see \S\ref{sec:agn}).  Here, we briefly describe
these ancillary data.  

The $5$~ks X-ray observations were obtained with the \emph{Chandra}
Advanced CCD Imaging Spectrometer (ACIS) instrument as part of the
\xbootes\footnote{\url{http://www.noao.edu/noao/noaodeep/XBootesPublic}}
survey \citep{murray05a, kenter05a}.  \xbootes{} covered
$\sim8.5$~deg$^{2}$ of the \bootes{} field to an on-axis limiting flux
of $\sim7.8\times10^{-15}$~\flunits{} in the $0.5-7$~keV band.  We use
the publically available catalog of $3213$ X-ray point-sources with
four or more X-ray counts matched to the NDWFS optical catalog by
\citet{brand06a} using a Bayesian matching algorithm.

In the infrared, the \emph{Spitzer} Deep Wide-Field Survey
\citep[SDWFS;][]{ashby09a} obtained deep, multi-epoch mid-IR imaging
of the \bootes{} field with the \emph{Spitzer} Infrared Array Camera
\citep[IRAC;][]{fazio04a}.  These observations reach
aperture-corrected $5\sigma$ depths of $\sim22.6$, $\sim22.1$,
$\sim20.2$, and $\sim20.2$~mag in a $4\arcsec$ diameter aperture at
$3.6$, $4.5$, $5.8$, and $8$~\micron, respectively.  In our analysis
we use the $3.6$~\micron-detected, aperture-corrected $4\arcsec$
diameter $3.6-8$~\micron{} aperture magnitudes available at the
NASA/IPAC Infrared Science
Archive\footnote{\url{http://irsa.ipac.caltech.edu/data/SPITZER/SDWFS}},
as described by \citet{ashby09a}.  

Finally, radio observations from the WSRT $1.4$~GHz radio survey
covered $\sim7$~deg$^{2}$ of the \bootes{} field to a $5\sigma$
limiting flux of $140~\mu$Jy \citep{devries02a}.  When matching to the
AGES catalog we use a $3\arcsec$ diameter search radius
\citep{hickox09a}. 

\subsection{AGES Optical Spectroscopy}\label{sec:ages}

The AGES main galaxy survey targeted galaxies brighter than $I_{\rm
  AB}=20.45$ and covered $7.9$~deg$^{2}$ of the NDWFS \bootes{} field,
making it among the largest wide-area spectroscopic surveys of
intermediate-redshift galaxies conducted to date.  We briefly
summarize the AGES observations in \S\ref{sec:redux}, and in
\S\ref{sec:ispec} we describe the procedure used to measure the
nebular emission-line strengths from these data.  A more thorough
description of the AGES experimental design, observations, and
redshift measurements can be found in \citet{kochanek11a} and
\citet{cool12a}.

\subsubsection{Observations, Reductions, and Spectroscopic
  Completeness}\label{sec:redux}  

The MMT/Hectospec multi-fiber optical spectrograph is fed by $300$
robotically controlled $1\farcs5$ diameter fibers, enabling efficient
data acquisition over a $1^{\circ}$ diameter field-of-view
\citep{roll98a, fabricant98a, fabricant05a}.  The $270$~line~mm$^{-1}$
grating used by AGES provides $3700-9200$~\AA{} spectra with
$1.2$~\AA{} pixels at $\sim6$~\AA{} FWHM resolution.  An atmospheric
dispersion corrector (ADC) built into the MMT f/5 wide-field lens
eliminates wavelength-dependent light-loss due to atmospheric
refraction, while relative spectrophotometric calibration is
facilitated by observing F sub-dwarf standard stars selected from the
SDSS \citep{abazajian04a, fabricant08a}.

We reduce the AGES spectra using standard procedures implemented in
{\sc
  hsred}\footnote{\url{http://www.astro.princeton.edu/$\sim$rcool/hsred}},
a customized Hectospec data reduction package based on the SDSS
spectroscopic data reduction
pipeline.\footnote{\url{http://spectro.princeton.edu}} Briefly, we
bias- and overscan-subtract the data, and then extract,
wavelength-calibrate, flat-field, and sky-subtract each galaxy
spectrum.  We combine multiple exposures using inverse variance
weights, and robustly reject cosmic rays \citep{dokkum01a}.  We then
flux-calibrate each spectrum, divide by a suitably scaled telluric
absorption spectrum, and correct for foreground Galactic reddening
\citep[$R_{V}\equiv A_{V}/E(B-V)=3.1$;][]{odonnell94a, schlegel98a}.
Note that five of the Hectospec configurations\footnote{A
  configuration refers to a given geometrical layout of the $300$
  Hectospec fibers.} obtained in 2004 could not be flux-calibrated,
and have been excluded from the present analysis.  Finally, we
determine redshifts using two independent codes and cross-validate the
results by visual inspection (see \citealt{kochanek11a} for details).

Because we are interested in faint emission lines at intermediate
redshift, accurate sky subtraction is crucial.  Therefore we
implemented an automated principal component analysis technique
originally developed for the SDSS to suppress the amplitude of the
sky-subtraction residuals redward of $\sim6700$~\AA, which are
dominated by a ``forest'' of atmospheric OH sky lines \citep{wild05a}.
Briefly, the method reconstructs a model of the night-sky spectrum for
each configuration of the Hectospec fibers by exploiting the fact that
the OH sky lines are highly correlated both in wavelength and between
all the fibers.  On average, we obtain a factor of two improvement in
the S/N ratio of affected pixels.

The AGES spectra suffer from two additional data-reduction issues that
do not affect the redshift measurements, but which are relevant to our
emission-line analysis.  First, the ADC was operated incorrectly
during some of the observations that took place in 2004, leading to
systematic errors in the spectrophotometry blueward of
$\sim5000$~\AA{} for these early observations \citep{fabricant08a}.
And second, the spectra are occasionally contaminated by the
featureless continuum of one or both of the light-emitting diodes
(LEDs) located on the Hectospec fiber-positioners.  In most cases this
extra light appears as a rapidly rising continuum redward of
$8500$~\AA, although it can also bias the average sky spectrum,
resulting in an oversubtraction of the sky flux in the
$8500-9200$~\AA{} wavelength range.  We address these issues in
\S\ref{sec:ispec}.

We conclude this section with a brief discussion of the AGES
spectroscopic completeness (for details see \citealt{kochanek11a} and
\citealt{cool12a}).  AGES covered $7.9$~deg$^{2}$ of the \bootes{}
field with $15$ slightly overlapping pointings, each with three
configurations.  Within each field, AGES targeted $100\%$ of galaxies
with $I_{\rm AB}<18.95$ and employed a variety of sparse-sampling
criteria to observe a subset of galaxies with $18.95<I_{\rm AB}<20.45$
that were also bright at other wavelengths.  Because these sampling
fractions are known exactly, it is easy to recover a sample that is
statistically complete to $I_{\rm AB}<20.45$ by simply weighting by
the inverse of the sampling rate.  The other sources of incompleteness
in AGES are relatively small.  Approximately $5\%$ of objects in the
parent sample failed to be assigned a spectroscopic fiber ({\em fiber
  incompleteness}).  Among the objects that were observed, AGES failed
to measure a redshift for $\sim2\%$ ({\em redshift incompleteness}),
which is a weak function of surface brightness.  Finally, we estimate
that roughly $4\%$ of objects are missing from the targeting catalog
({\em photometric incompleteness}) due to proximity to a bright star
or problems with the photometry itself.  To correct for these effects,
we weight every object in AGES by the product of these terms,
resulting in a sample that is statistically complete to $I_{\rm
  AB}=20.45$.

\subsubsection{Emission-line Measurements}\label{sec:ispec}

High-resolution population synthesis models have become an
indispensible tool for investigating the integrated optical spectra of
galaxies \citep{bruzual03a, vazquez05a, grillo09a, conroy09a, percival09a,
  vazdekis10a}.  By modeling and subtracting the stellar continuum
from the observed spectrum, these models have made it possible to
study the optical emission lines free from the systematic effects of
Balmer and metal-line absorption \citep[e.g.,][]{panter03a,
  tremonti04a, cid05a, moustakas06a, sarzi06a, ocvirk06a, tojeiro07a,
  oh11a}.  Neglecting the effects of stellar absorption, particularly
under the Balmer emission lines, can lead to severe biases in the
nebular abundances, SFRs, and dust reddenings inferred from the
optical emission lines \citep[e.g.,][]{kenn92b, kobulnicky99a,
  rosa-gonzalez02a, moustakas06b, moustakas10a, asari07a}.

Our basic strategy is to model the observed spectrum of each galaxy as
a non-negative linear combination of simple (i.e.,
instantaneous-burst) population synthesis models of varying ages,
making some simplifying assumptions regarding the stellar metallicity
of the galaxy and the effects of dust attenuation.  We then subtract
the best-fitting continuum model from the data, and fit the residual
emission-line spectrum assuming Gaussian line-profiles.  We emphasize
that our principal goal is to obtain an emission-line spectrum that
has been corrected self-consistently for stellar absorption, not to
constrain the star formation and chemical evolution history of the
galaxy from its integrated stellar spectrum.  The multi-dimensional
parameter space of star formation history, age, metallicity, and dust
attenuation is highly degenerate (i.e., there are many local minima),
and our simple approach is ill-suited for finding the global solution
\citep[but see][and references therein, for more advanced techniques
  that tackle this issue]{walcher10a}.  Fortunately, the emission-line
strengths we measure do not depend sensitively on the simplifying
assumptions underlying our technique.

We construct our template set using the \citet[hereafter
  BC03]{bruzual03a} population synthesis models, based on the Padova
1994 stellar evolutionary tracks \citep[and references
  therein]{girardi96a} and the empirical {\sc stelib} stellar library
\citep{leborgne03a}.  We choose $10$ instantaneous-burst,
solar-metallicity ($Z=0.02$) models with ages spaced
quasi-logarithmically in time between $5$~Myr and $13$~Gyr, assuming
the \citet{chabrier03a} initial mass function (IMF) from
$0.1-100$~\msun.  We treat dust reddening as a free parameter, and
adopt the \citet{calzetti00a} dust attenuation law.  Finally, we find
the best-fitting, non-negative linear combination of the templates
using a modified version of the {\sc
  pPXF}\footnote{\url{http://www-astro.physics.ox.ac.uk/$\sim$mxc/idl}}
continuum-fitting code \citep{cappellari04a, moustakas10a}.

We verified that adopting different model parameters does not have a
significant effect on the derived emission-line strengths.
Specifically, we experimented with a wider range of stellar
metallicities, varied the number of templates (i.e.,
instantaneous-burst ages), tried several different extinction curves,
and assumed a different initial mass function.  These results indicate
that it is not necessary to find the \emph{global} solution to the
continuum fitting problem in order to be able to study the
emission-line properties of galaxies free from the systematic effects
of stellar absorption.

We fit the observed stellar continuum of each galaxy iteratively.
First, we isolate the rest-frame $\sim3600-4400$~\AA{} spectral range
to derive a more precise measurement of the absorption-line redshift,
$z_{\rm abs}$, and to obtain an estimate of the stellar velocity
dispersion, $\sigma_{\rm disp}$, accounting for the instrumental
resolution of our spectra ($\sim6$~\AA{} FWHM) and the resolution of
the BC03 models ($\sim3$~\AA{} FWHM; BC03).  Next, we fix $z_{\rm
  abs}$ and $\sigma_{\rm disp}$ at the derived values and model the
full wavelength range, aggressively masking pixels that might be
affected by emission lines, sky-subtraction residuals, telluric
absorption, or the red leak (see \S\ref{sec:redux}).  For
approximately one-quarter of the observed Hectospec configurations
(largely from 2004; see \S\ref{sec:redux}) there are non-negligible
errors in the spectrophotometry shortward of $\lambda_{\rm
  obs}\sim5000$~\AA.  Therefore, for these configurations, we refit
the spectra, masking out the affected blue observed-frame wavelengths,
and then divide the observed spectrum by its best-fitting model, in
effect constructing a sensitivity function correction for each galaxy.
We then compute the median sensitivity function correction using all
the spectra on a given configuration, and divide all the spectra by
the median correction.  Finally, we refit those galaxy spectra one
last time, this time using the full wavelength range.  We assess the
effect of this correction on our abundance analysis below.

The last step before fitting the emission lines is to subtract the red
leak and any remaining low-level residuals due to imperfect
sky-subtraction, template mismatch, or imperfect flux-calibration.
Given the best-fitting continuum model for each galaxy, we subtract it
from the data and remove these residual differences (typically of
order a few percent) using a sliding $151$-pixel median filter.
Finally, we subtract this median residual spectrum from the original
data, and then refit and subtract the stellar continuum one last time,
leaving a pure emission-line spectrum.

Following \citet{moustakas10a}, we fit the emission-line spectrum of
each galaxy iteratively using a modified version of the {\sc
  gandalf}\footnote{\url{http://www.strw.leidenuniv.nl/sauron}}
emission-line fitting code, assuming Gaussian line-profiles
\citep{sarzi06a, schawinski07a, oh11a}.  Specifically, we fit the
\nevlam, \oiilam, \neonlam, \oiiidoublet, \niidoublet, and
\siidoublet{} forbidden lines, and the first five lines in the Balmer
series: \halam, \hblam, \hglam, \hdlam, and \hfivelam.  On the first
iteration we tie the redshifts and intrinsic velocity widths
(accounting for the instrumental resolution) of all the lines together
to aid in the detection and deblending of weak lines.  In addition, to
reduce the number of free parameters, we constrain the
$\oiii~\lambda5007/\oiii~\lambda4959$ and
$\nii~\lambda6584/\nii~\lambda6548$ doublet ratios to be 3:1
\citep{osterbrock06a}.  On the second iteration we relax most of these
constraints and use the best-fitting parameters from the first
iteration as initial guesses.  This second step is necessary because
of our imperfect knowledge of the wavelength-dependent instrumental
resolution, and to account for the fact that \oiilam{} is a doublet
which is better represented at the spectral resolution of our data as
a single, slightly broader Gaussian line than two closely-spaced
Gaussian line-profiles.  Note that even on the second iteration,
however, we (separately) constrain the redshifts and velocity widths
of the \oiii, \nii, and \sii{} doublets to have the same values.

\begin{figure}
\centering
\includegraphics[scale=0.4]{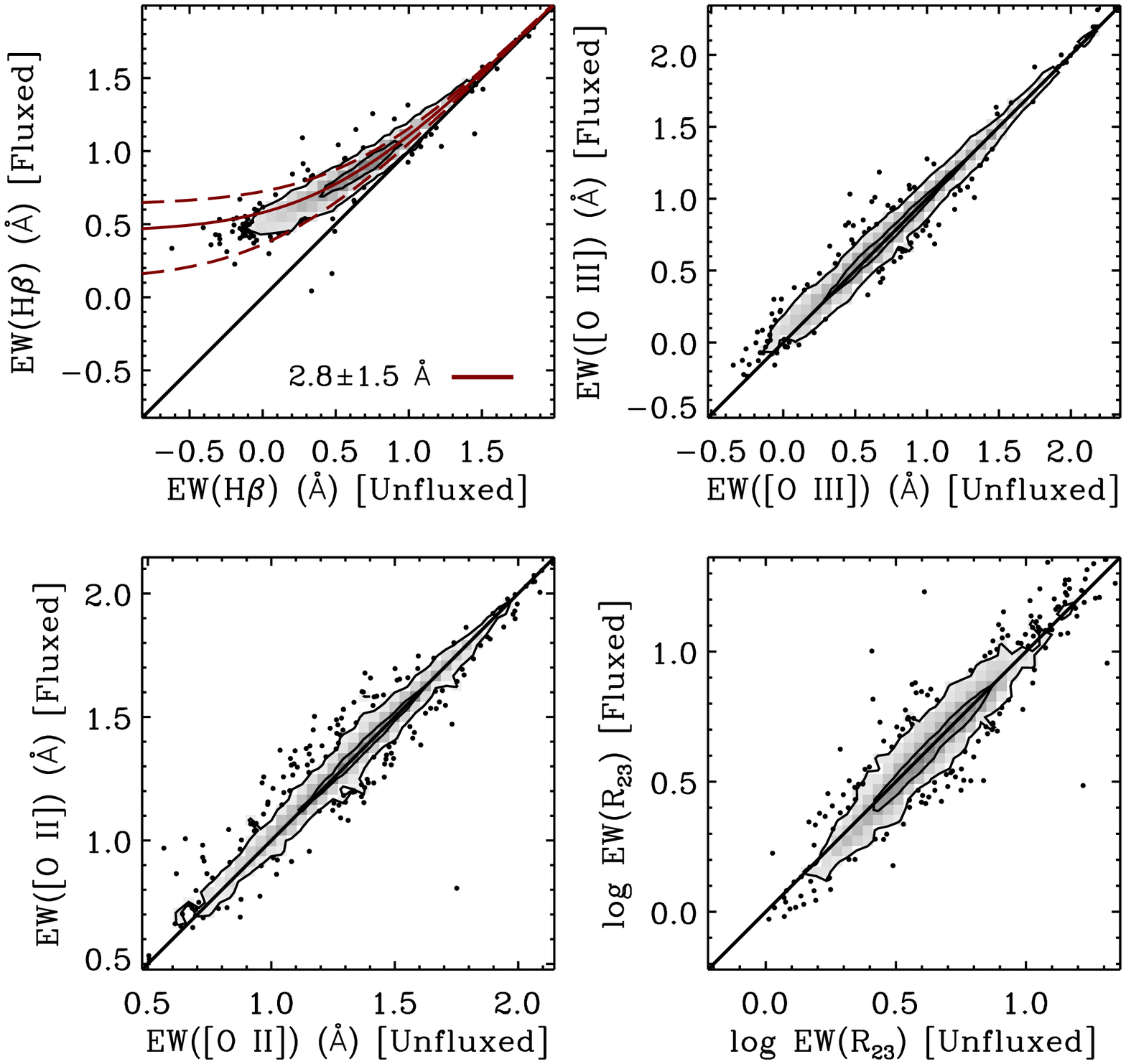}
\caption{Comparison of EWs measured from the fluxed and unfluxed AGES
  spectra for \hb, \oiiilam, and \oiilam.  In each panel the solid
  line shows the one-to-one relation, and the contours enclose $50\%$
  and $95\%$ of the points.  We find excellent statistical agreement
  among the two emission-line measurements, assuming that \ewhb{}
  measured from our unfluxed spectra is subject to $2.8\pm1.5$~\AA{}
  of stellar absorption (solid and dashed red lines in the upper-left
  panel).  In the lower-right panel we compare the
  metallicity-sensitive $\ewpagel\equiv(\ewoii+\ewoiii)/\ewhb$
  parameters measured from the fluxed and unfluxed spectra, after
  statistically correcting \ewhb{} measured from the unfluxed spectra
  for an average $2.8$~\AA{} of stellar absorption.  These comparisons
  indicate that in spite of the spectrophotometric errors in AGES, our
  oxygen abundance estimates are not adversely
  affected. \label{fig:ew_tests}}
\end{figure}

We evaluate the significance of each emission line by comparing the
amplitude of the best-fitting Gaussian model, $A_{l}$, against the
standard deviation of the residual (continuum-subtracted) spectrum,
$\sigma_{c}$, around the line; we define a significant line as having
$A_{l}>3\sigma_{c}$.  For each undetected emission line we estimate
the $1\sigma$ upper limit on the total line-flux assuming a Gaussian
line-profile using

\begin{equation}
F = \sqrt{2\pi}\,\sigma_{c}\,\delta_{v},
\label{eq:limit}
\end{equation}

\noindent where $\delta_{v}$ is the mean velocity width of all the
detected emission lines.  Finally, we estimate the equivalent width
(EW) of each emission line by dividing the integrated flux of the
Gaussian model by the mean flux of the stellar continuum immediately
blueward and redward of the given line.

We conclude this section by evaluating the effect of
spectrophotometric errors on our line-measurements by measuring the
\oii, \hb, and \oiiilam{} emission-line EWs from the \emph{unfluxed}
AGES spectra.  For these spectra we model the stellar continuum as a
low-order B-spline with iterative outlier rejection, subtract it from
the data, and measure the EW of each emission line as described above.
In Figure~\ref{fig:ew_tests} we compare the \oii, \hb, and \oiii{}
emission-line EWs measured from the fluxed and unfluxed spectra, as
well as the metallicity-sensitive
$\ewpagel\equiv(\ewoii+\ewoiii)/\ewhb$ parameter (see
\S\ref{sec:ohpagel}).  We render the data in greyscale with the
contours enclosing $50\%$ and $95\%$ of the points, and plot the
one-to-one relation as a solid line.  Overall we find good statistical
agreement among the EWs measured from the two sets of spectra,
although \hb{} is clearly affected by stellar absorption.  However,
applying a mean $2.8\pm1.5$~\AA{} correction to the \ewhb{} values
measured from the unfluxed spectra, shown as solid and dashed red
lines in the upper-left panel of Figure~\ref{fig:ew_tests}, would
bring the two EW measurements into good agreement.  Meanwhile, the
statistical agreement for \ewoii{} and \ewoiii{} is excellent.  In the
lower-right panel we compare the two \ewpagel{} estimates after
correcting \ewhb{} measured from the unfluxed spectra for an average
$2.8$~\AA{} of stellar absorption.  These comparisons indicate that
our oxygen abundance estimates are not adversely affected by the
spectrophotometric errors in AGES.

\subsection{SDSS Comparison Sample}\label{sec:sdss}

At low redshift ($z\sim0.1$) AGES is significantly affected by
variations in large-scale structure (i.e., cosmic variance) because of
the relatively small cosmological volume probed.  Therefore, we use
observations obtained as part of the SDSS as a local benchmark for
studying the chemical abundance properties of the more distant
galaxies observed by AGES.

The SDSS has obtained $ugriz$ broadband photometry and high-precision
optical ($3800-9200$~\AA) spectrophotometry for roughly two million
objects over nearly one-quarter of the sky \citep{york00a,
  stoughton02a}.\footnote{\url{http://www.sdss.org}} We draw our SDSS
comparison sample from the NYU Value-Added Galaxy
Catalog\footnote{\url{http://sdss.physics.nyu.edu/vagc}}
\citep[NYU-VAGC;][]{blanton05b} corresponding to the SDSS Data
Release~7 \citep[DR7;][]{adelman08a}.  From this database we select
$427,504$ galaxies over $6955$~deg$^{2}$ that satisfy the main sample
criteria defined by \citet{strauss02a}, and have Galactic extinction
corrected \citep{schlegel98a} Petrosian magnitudes $14.5<r<17.6$, and
redshifts $0.05<z<0.2$.  We choose the lower redshift limit to ensure
that \oiilam{} lies within the SDSS spectral range
\citep{stoughton02a}, while the upper redshift limit rejects $<2\%$ of
galaxies in the SDSS main sample.  From the VAGC we retrieve for each
object the spectroscopic redshift, the Galactic extinction-corrected
$ugriz$ Petrosian and {\tt cmodel} \citep[i.e., total;
  see][]{abazajian04a} magnitudes, integrated $JHK_{s}$ photometry
(available for $\sim55\%$ of the sample) from the Two Micron All Sky
Survey Extended Source Catalog \citep[2MASS/XSC;][]{jarrett00a,
  strutskie06}, and the statistical weight for each galaxy that
corrects for the $\sim8\%$ spectroscopic incompleteness of the survey
\citep{blanton05b}.  Corrected for incompleteness the effective number
of galaxies in our SDSS sample is $464,145$.

We cross-match this sample against the public database of
spectroscopic measurements for SDSS/DR7 galaxies available at the
MPA-JHU team
website.\footnote{\url{http://www.mpa-garching.mpg.de/SDSS/DR7}} The MPA-JHU
database provides fluxes and EWs for all the optical emission lines of
interest after carefully modeling and subtracting the stellar
continuum of each galaxy using an updated, but as-yet unpublished
version of the BC03 population synthesis models \citep{tremonti04a,
  brinchmann04a, bruzual07a}.  A negligible number of objects, roughly
$0.3\%$ of galaxies in the parent VAGC sample, do not appear in the
MPA-JHU database, and have been excluded.  We post-process this
database by adding a flag indicating the presence of a significant
emission line based on the amplitude of the line relative to the
variance of the continuum as described in \S\ref{sec:ispec}, and by
computing upper limits for each line using equation~(\ref{eq:limit}).

As a consistency check, we retrieved the SDSS spectra of a randomly
selected sample of $500$ galaxies from the SDSS data
archive\footnote{\url{http://www.sdss.org/dr7}} and modeled them using the
continuum and emission-line fitting code used to fit the AGES spectra
in \S\ref{sec:ispec}.  In general the measured line-fluxes and EWs are
consistent within the statistical uncertainties.

\section{Spectral Energy Distribution Modeling}\label{sec:sed}  

Accurate rest-frame colors, luminosities, and stellar masses require
photometry that has been measured over the same physical aperture for
each galaxy.  In \S\ref{sec:apphot} we describe the procedure we use
to build aperture-matched optical to near-IR SEDs of the galaxies in
our sample, and in \S\ref{sec:kcorr} and \S\ref{sec:mass} we use these
data in conjunction with the spectroscopic redshifts from AGES to
derive $K$-corrections and stellar masses for the full sample,
respectively.

\subsection{Aperture-Matched Photometry}\label{sec:apphot}  

Our general strategy is to determine the apparent colors of each
galaxy using aperture photometry measured from PSF-matched images, and
to use an estimate of the total magnitude in the $I$-band to set the
overall normalization of the SED.  This procedure is appealing because
aperture colors in general have a higher S/N ratio than the total
magnitude in each band, and it minimize contamination from neighboring
sources.

To detect sources and to obtain an estimate of the total $I$-band
magnitude of each galaxy we run {\sc SExtractor} \citep{bertin96a} in
single-image mode on the unsmoothed $I$-band mosaics.  We use the {\sc
  mag\_auto} (Kron-like) magnitude, $I_{\rm AUTO}$, as an estimate of
the integrated flux of each galaxy \citep{kron80a}.  Note that
although {\sc mag\_auto} may miss a significant amount of flux at
faint apparent magnitudes, and for galaxies with very extended
surface-brightness profiles \citep[e.g.,][]{graham05a}, by inserting
artifical galaxies into the NDWFS $I$-band mosaics and re-running {\sc
  SExtractor}, \citet{brown07a} find that $I_{\rm AUTO}$ recovers the
total (input) galaxy magnitude to within $\sim5\%$ for galaxies
brighter than $I_{\rm AB}\approx21$, which is sufficient for our
purposes.

Unfortunately, $I_{\rm AUTO}$ is occasionally biased toward bright
magnitudes by the low surface-brightness tails of bright stars.
Although AGES intentionally excluded regions around bright stars
\citep{kochanek11a}, $I_{\rm AUTO}$ appears to be affected for roughly
$10\%$ of the sample.  Because these low surface-brightness halos are
not present in the $R$-band mosaics, we re-run {\sc SExtractor} on the
$R$-band images and use the $R$-band {\sc mag\_auto} magnitude to
obtain a second estimate of the total $I$-band magnitude, $I_{R}$,
using the $R-I$ color measured from the PSF-matched images in a
$6\arcsec$ diameter aperture (see below).  These aperture colors are
considerably less affected by this extended, low surface-brightness
light.  Following \citet{cool12a}, we obtain a new estimate of the
total $I$-band magnitude, $I^{\prime}$, using a statistic that chooses
the fainter of $I_{R}$ and $I_{\rm AUTO}$ if they differ
significantly, and averages them otherwise.  For $\sim7\%$ ($\sim3\%$)
of the sample the magnitude correction computed this way exceeds
$0.1$~mag ($0.5$~mag).  However, for the remainder of the sample there
is a very tight correlation between $I^{\prime}$ and $I_{\rm AUTO}$,
with a median difference of $<0.01$~mag and a $1\sigma$ scatter of
$\sim2\%$.  To account for this small amount of scatter, in
\S\ref{sec:sample} we will restrict our statistical (flux-limited)
galaxy sample to $I_{\rm AB}^{\prime}<20.4$, which excludes $\sim2\%$
of the parent sample \citep{cool12a}.  Hereafter, we adopt
$I^{\prime}$ as the total $I$-band magnitude of each galaxy and for
clarity drop the prime superscript.

Next, we smooth the optical and near-IR mosaics to a common spatial
resolution to ensure that the apparent colors are measured from the
same physical area on each galaxy \citep{brown07a}.  The mosaics with
the best spatial resolution (seeing) are the $B_{W}$-, $R$-, $I$-,
$H$- and $K_{s}$-band mosaics, so we smooth these such that the
resulting stellar PSF is a \citet{moffat69a} profile with $1\farcs35$
FWHM and $\beta=2.5$.  The seeing of the $U$-, $z$-, and $J$-band
mosaics is slightly worse, so we convolve these mosaics to an
equivalent $1\farcs6$ FWHM Moffat PSF.  We then measure aperture
photometry in fixed apertures ranging in diameter from $1\arcsec$ to
$20\arcsec$ centered on the $I$-band position of each source using
customized software.  We use the {\sc SExtractor} segmentation maps to
account for pixels contaminated by neighboring objects and missing
pixels, and estimate the photometric error in each aperture by
computing the interval that encompasses $68\%$ of the flux measured in
$100$ blank-sky apertures placed randomly within $2\arcmin$ of each
source.  Finally, we construct the SED of each galaxy using the
$4\arcsec$ diameter aperture magnitudes, and then scale everything by
the difference between the $I$-band aperture magnitude and the total
$I$-band magnitude.  We use $4\arcsec$ aperture magnitudes as a
compromise between choosing the largest possible aperture to maximize
the S/N, while simultaneously minimizing contamination from
neighboring objects.  However, we did verify that using the $6\arcsec$
diameter aperture magnitudes yields the same colors to within
$\lesssim\pm4\%$ and with no statistically significant systematic
differences.

\subsection{$K$-corrections}\label{sec:kcorr}

To estimate the rest-frame colors and luminosities of the galaxies in
our sample, we compute $K$-corrections using the publically available
code
\kcorrect\footnote{\url{http://howdy.physics.nyu.edu/index.php/Kcorrect}}
\citep[v4.1.4;][]{blanton07a}.  For each galaxy \kcorrect{} takes as
input the redshift, the observed photometry, and the corresponding
filter response curves, and fits the data with a non-negative linear
combination of five basis templates representing a diversity of galaxy
star formation histories.  We have verified that the more detailed SED
modeling performed in \S\ref{sec:mass} yields the same rest-frame
quantities, but adopt \kcorrect{} here because of its ease-of-use and
speed.

For our AGES sample we fit the aperture-matched $UB_{W}RIzJHK_{s}$
photometry assembled in \S\ref{sec:apphot}, and for our SDSS sample we
fit the observed $ugrizJHK_{s}$ photometry (see \S\ref{sec:sdss}), all
corrected for foreground Galactic extinction
\citep[$R_{V}=3.1$;][]{odonnell94a, schlegel98a}.  We compute
$K$-corrections for each galaxy on a case-by-case basis by selecting
the bandpass that minimizes the transformation from the observed frame
to the rest frame \citep{hogg02a, blanton03a}.  For example, for our
AGES sample we compute the absolute $B$-band magnitude
\citep{bessell90a} from the observed $B_{W}$ magnitude at
$z\lesssim0.2$, from the $R$-band magnitude at $0.2\lesssim
z\lesssim0.65$, and from the $I$-band magnitude at $z\gtrsim0.65$.  We
use the best-available filter curve for each bandpass by taking into
account the detector quantum efficiency, telescope throughput, and
atmospheric transmission.  Before fitting, we add a minimum
photometric uncertainty of $0.02$~mag in $B_{W}RIHK_{s}$ in quadrature
to the statistical error in each bandpass.  To account for the
slightly larger PSF of the $U$-, $z$-, and $J$-band mosaics (see
\S\ref{sec:apphot}), we assume a minimum error of $0.05$~mag in these
bands.  For the SDSS we adopt minimum errors of $0.02$~mag in $gri$,
$0.05$~mag in $u$, and $0.03$~mag in $z$ \citep{ivezic04a}.  Finally,
we also adjust the SDSS $ugriz$ magnitudes by $-0.036$, $0.012$,
$0.010$, $0.028$, and $0.040$~mag, respectively, to place them on an
absolute AB magnitude system \citep{blanton07a}.

After fitting our AGES sample, we identified small residual,
systematic offsets between the observed photometry and the magnitudes
synthesized from the best-fitting models \citep[e.g.,][]{ilbert05a}.
These relative zeropoint differences can arise from errors in the
photometric calibration, slightly mismatched physical apertures (e.g.,
due to imperfect PSF matching), imperfect filter response curves, and
template mismatch.  Therefore, we fit the sample iteratively, each
time adjusting the observed photometry in each bandpass by the median
zeropoint difference between the measured and synthesized magnitudes.
The zeropoints converged to better than $0.5\%$ in five iterations.
The resulting zeropoint corrections we apply to the data, all relative
to the $I$-band, are: $-0.073$, $-0.032$, $-0.016$, $-0.037$,
$-0.108$, $+0.042$, and $-0.053$~mag in $U$, $B_{W}$, $R$, $z$, $J$,
$H$, and $K_{s}$, respectively.

Finally, we tested the absolute photometric calibration of our AGES
sample relative to the SDSS.  We used the best-fitting \kcorrect{}
model to synthesize observed photometry in the SDSS filters for each
galaxy in our AGES sample.  We then queried the SDSS/DR7 database for
$ugriz$ Petrosian magnitudes of the sample, and compared the
synthesized and observed magnitudes.  We found a negligible
($\lesssim0.01$~mag) systematic difference to $i\approx18.5$, and a
median difference of $0.01-0.02$~mag to $i\approx21$, with a random
scatter of $\sim0.05$~mag.  Therefore, we conclude that the absolute
photometric calibration of our AGES photometry is consistent with the
SDSS to better than $\sim5\%$.

\subsection{Stellar Masses}\label{sec:mass}

We estimate stellar masses for the galaxies in our sample using
\isedfit, a Bayesian SED-fitting code that uses population synthesis
models to infer the physical properties of a galaxy given its observed
broadband SED (see J. Moustakas et~al. 2012, in prep., for additional
details).  As in \S\ref{sec:ispec}, we adopt the BC03 population
synthesis models based on the Padova isochrones, the {\sc stelib}
stellar library, and the \citet{chabrier03a} IMF.  Given the broadband
fluxes $F_{i}$ of a galaxy at redshift $z$ in $i=1,N$ filters,
\isedfit{} uses Bayes' theorem to compute the posterior probability
distribution function (PDF)

\begin{equation}
p({\mathbf Q}|F_{i},z) = p({\mathbf Q})\times p(F_{i},z|{\mathbf Q})
\label{eq:posterior}
\end{equation}

\noindent for a given set of model parameters ${\mathbf Q}$ (stellar
mass, age, metallicity, etc.).  Here, $p({\mathbf Q})$ is the prior
probability of the model parameters, $p(F_{i},z|{\mathbf Q})$ is the
likelihood $\mathcal{L}\propto \exp[-\chi^2(F_{i}, z,{\mathbf Q})/2]$
of the data given the model, and $\chi^{2}$ is the usual
goodness-of-fit statistic appropriate for normally distributed
photometric uncertainties.  We draw each model parameter from a
specified prior probability distribution using a Monte Carlo
technique, which is equivalent to multiplying the likelihood by the
prior probability \citep{walcher10a}.  Once $\chi^{2}$ has been
computed for every model, the marginalized posterior PDF of the
parameter of interest, for example $p(\mass)$ for the stellar mass,
follows from equation~(\ref{eq:posterior}) after integrating over the
other ``nuisance'' parameters \citep{kauffmann03a, salim07a, auger09a,
  taylor11a}.  The advantage of this approach over traditional
best-fitting (maximum likelihood) techniques is that it accounts for
both photometric uncertainties and physical degeneracies among
different models.  We adopt the median of the posterior PDF as the
best estimate of the stellar mass, and the uncertainty as $1/4$ of the
$2.3-97.7$ percentile range of the $p(\mass)$ distribution, which
would be equivalent to $1\sigma$ for a Gaussian distribution.

We assume exponentially declining star formation histories of the form
$\psi(t)\propto \exp(-t/\tau)$, but allow each model galaxy to
experience one or more stochastic bursts of star formation.  We draw
$1/\tau$ from a uniform distribution in the range
$0.01-10$~Gyr$^{-1}$, corresponding to a characteristic timescale for
star formation ranging from $\tau=0.1$~Gyr (similar to an
instantaneous burst), to $\tau=100$~Gyr (continuous star formation).
We allow the age $t$ (time for the onset of star formation) of each
model to range with equal probability between $0.1-13$~Gyr, although
we disallow ages older than the age of the Universe at the redshift of
each galaxy.  We assume a uniform prior on stellar metallicity in the
range $0.004<Z<0.04$, and adopt the time-dependent attenuation curve
of \citet{charlot00a}, in which stellar populations older than
$10$~Myr are attenuated by a factor $\mu$ times less than younger
stellar populations.  We draw $\mu$ from an order four Gamma
distribution that ranges from zero to unity centered on a typical
value $\langle\mu\rangle=0.3$ \citep{charlot01a, wild11a}, and the
$V$-band optical depth from an order two Gamma distribution that peaks
around $A_{V}\approx1.2$~mag, with a tail to $A_{V}\approx6$~mag.
Finally, we allow each model galaxy to experience one or more random
bursts of star formation at time $t_{b}>t$ in each $2$~Gyr time
interval over the lifetime of the galaxy with a probability of $50\%$.
We characterize each burst by a duration $\Delta t_{b}$, drawn from a
logarithmic distribution in the range $30-300$~Myr, and a fractional
stellar mass formed $F_{b}$, which we draw from a logarithmic
distribution from $0.01-4$ \citep{salim07a, wild09a}.  For simplicity,
we assume that the stellar population that forms in the burst has the
same stellar metallicity and is attenuated by the same amount of dust
as the pre-burst stellar population.

The largest systematic uncertainty affecting our stellar mass
estimates is likely due to our choice of IMF, although if the IMF is
spatially and temporarily invariant (i.e., universal) as is commonly
assumed, then adopting a different IMF should simply shift all our
stellar masses by a fixed amount.  For example, the
\citet{salpeter55a} and \citet{kroupa01a} IMFs are $0.25$~dex and
$0.03$~dex heavier than the \citet{chabrier03a} IMF, respectively,
while the so-called ``diet Salpeter'' IMF of \citet{bell01b} is
$0.1$~dex lighter.  Although there are recent tantalizing suggestions
of a Salpeter (or heavier) IMF in early-type galaxies
\citep{dokkum10a, treu10a}, current observations heavily favor a
universal Chabrier- or Kroupa-like IMF in nearby star-forming disk
galaxies, including the Milky Way \citep[and references
  therein]{bastian10a}.  

Varying our prior distributions---for example, adopting different
metallicity or attenuation priors, or assuming smooth, burst-free
star-formation histories---changes our stellar masses by
$\lesssim0.1$~dex in the mean, and by less than a factor of two for
individual objects.  We also test the effect of adopting different
population synthesis models \citep{maraston05a, conroy10b} and find
comparable systematic differences in the derived stellar masses for
both our SDSS and AGES samples, i.e. typically $\lesssim0.1$~dex.  We
refer the interested reader to \citet{kannappan07a},
\citet{marchesini09a}, and \citet{muzzin09a} for recent detailed
discussions of the uncertainties associated with deriving stellar
masses from broadband photometry.

\section{Sample Selection}\label{sec:sample}    

The AGES main galaxy sample contains $12,473$ galaxies brighter than
$I_{\rm AB}=20.45$ with well-measured redshifts over $7.9$~deg$^{2}$
of the \bootes{} field \citep{kochanek11a}.  We exclude from this
sample five Hectospec configurations that could not be flux-calibrated
(see \S\ref{sec:redux}), and two additional configurations with very
poor fluxing.  However, we preserve the statistical completeness of
the sample by upweighting the remaining objects on a field-by-field
basis to account for the missing configurations.  Finally, we restrict
our analysis to galaxies with $15.45<I_{\rm AB}<20.4$ and
$0.05<z<0.75$, leaving $10,838$ galaxies (Table~\ref{table:samples}),
or an effective sample of $24,525$ galaxies after correcting for
sparse-sampling and other sources of incompleteness (see
\S\ref{sec:redux}).  In \S\ref{sec:selection} we select a subset of
these objects with the requisite nebular emission lines to estimate
their oxygen abundances, and in \S\ref{sec:agn} we apply a variety of
multiwavelength criteria to identify and remove AGN from our sample.
Finally, in \S\ref{sec:properties} we present some of the basic
properties of our final sample of star-forming emission-line galaxies,
and define volume-limited subsamples.

\subsection{Selecting Emission-Line Galaxies}\label{sec:selection} 

\begin{deluxetable}{lcc}
\tablecaption{AGES and SDSS Galaxy Samples\label{table:samples}}
\tablewidth{0pt}
\tablehead{
\colhead{Sample} & 
\colhead{$N_{\rm gal}$\tablenotemark{a}} & 
\colhead{Section\tablenotemark{b}} 
}
\startdata
AGES &   &   \\ 
\hspace{0.2cm}Parent &  $10838$ &  \ref{sec:sample} \\ 
\hspace{0.2cm}Emission-Line &  $4033$ &  \ref{sec:selection} \\ 
\hspace{0.5cm}SF\tablenotemark{c} &  $3205$ &  \ref{sec:agn} \\ 
\hspace{0.5cm}AGN &  $828$ &  \ref{sec:agn} \\ 
\hspace{0.2cm}Abundance\tablenotemark{d} &   &   \\ 
\hspace{0.5cm}KK04 &  $2975$ &  \ref{sec:ohsummary} \\ 
\hspace{0.5cm}T04 &  $3191$ &  \ref{sec:ohsummary} \\ 
\hspace{0.5cm}M91 &  $2969$ &  \ref{sec:ohsummary} \\ 
\cline{1-3} \\
SDSS &   &   \\ 
\hspace{0.2cm}Parent &  $427504$ &  \ref{sec:sdss} \\ 
\hspace{0.2cm}Emission-Line &  $108040$ &  \ref{sec:selection} \\ 
\hspace{0.5cm}SF\tablenotemark{c} &  $76411$ &  \ref{sec:agn} \\ 
\hspace{0.5cm}AGN &  $31629$ &  \ref{sec:agn} \\ 
\hspace{0.2cm}Abundance\tablenotemark{d} &   &   \\ 
\hspace{0.5cm}KK04 &  $75378$ &  \ref{sec:ohsummary} \\ 
\hspace{0.5cm}T04 &  $75976$ &  \ref{sec:ohsummary} \\ 
\hspace{0.5cm}M91 &  $75313$ &  \ref{sec:ohsummary}
\enddata
\tablenotetext{a}{Number of galaxies in each sample.}
\tablenotetext{b}{Section containing details regarding how we select each sample.}
\tablenotetext{c}{Star-forming galaxies.}
\tablenotetext{d}{Number of galaxies with well-measured oxygen abundances based on the \citet[M91]{mcgaugh91a}, \citet[T04]{tremonti04a}, and \citet[KK04]{kobulnicky04a} abundance calibrations.}
\end{deluxetable}

We select a sample of emission-line galaxies based on the strength of
\hb.  The principal advantages of using \hb{} are that its strength is
proportional to the instantaneous SFR, and it is less sensitive to
variations in dust attenuation, excitation, and metallicity than other
strong optical emission lines \citep{kenn92b, moustakas06b,
  gilbank10a}.  From a practical perspective, \hb{} also has the
advantage that it is observable in both AGES and the SDSS over the
full redshift range of interest.

\begin{figure}
\centering
\includegraphics[scale=0.4]{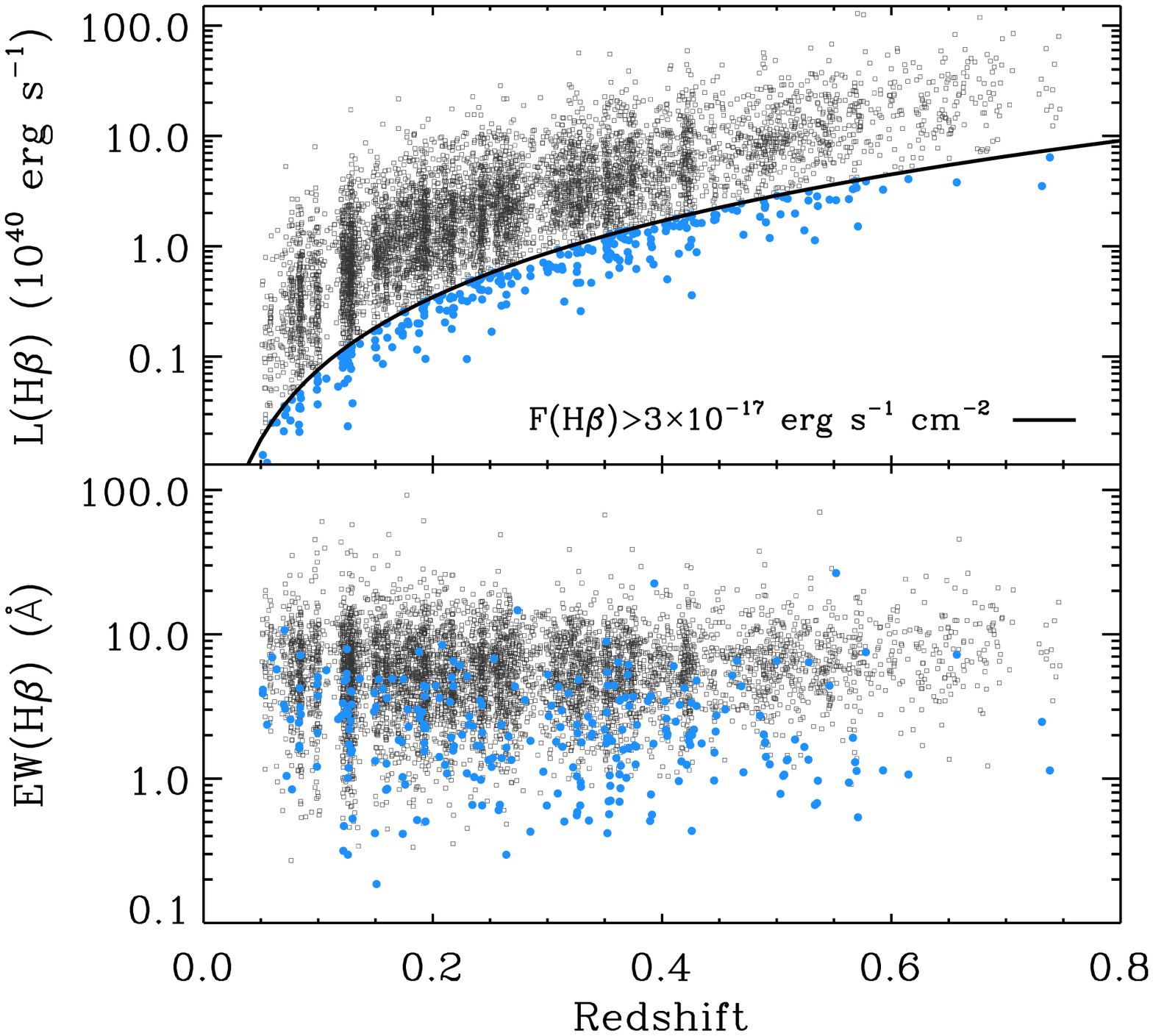}
\caption{Selection of emission-line galaxies in AGES based on the
  \hb{} line-strength.  In the top panel we plot \hb{} luminosity,
  $L(\hb)$, vs.~redshift, and select all galaxies ({\em small black
    points}) with an observed \hb{} line-flux
  $F(\hb)>3\times10^{-17}$~\fluxunits{} ({\em solid black curve}).
  The bottom panel plots rest-frame \ewhb{} vs.~redshift for the
  resulting emission-line sample, where for comparison we also show
  the objects that fail our fiducial flux cut ({\em light blue
    points}).  Our selection on \hb{} luminosity is physically
  motivated because $L(\hb)$ is proportional to the instantaneous
  SFR. \label{fig:zvhb}}
\end{figure}

In the top panel of Figure~\ref{fig:zvhb} we plot the \hb{}
luminosity, $L(\hb)$, versus redshift for all the galaxies in AGES
with a significant \hb{} emission line.  For simplicity we ignore both
relative and absolute aperture effects in our selection (see also
\S\ref{sec:apbias}).  We select all objects ({\em small black points})
above the black curve, which corresponds to a flux limit of
$F(\hb)>3\times10^{-17}$~\fluxunits.  In the bottom panel we plot
rest-frame \ewhb{} against redshift for the resulting emission-line
sample, where for comparison we also show the objects that fail our
fiducial flux cut ({\em light blue points}).  The \ewhb-redshift plot
shows that our sample spans a wide range of \ewhb{} at all redshifts,
although the distribution narrows at higher redshift due to the
increasing limiting $L(\hb)$ luminosity.

\begin{figure}[b]
\centering
\includegraphics[scale=0.4]{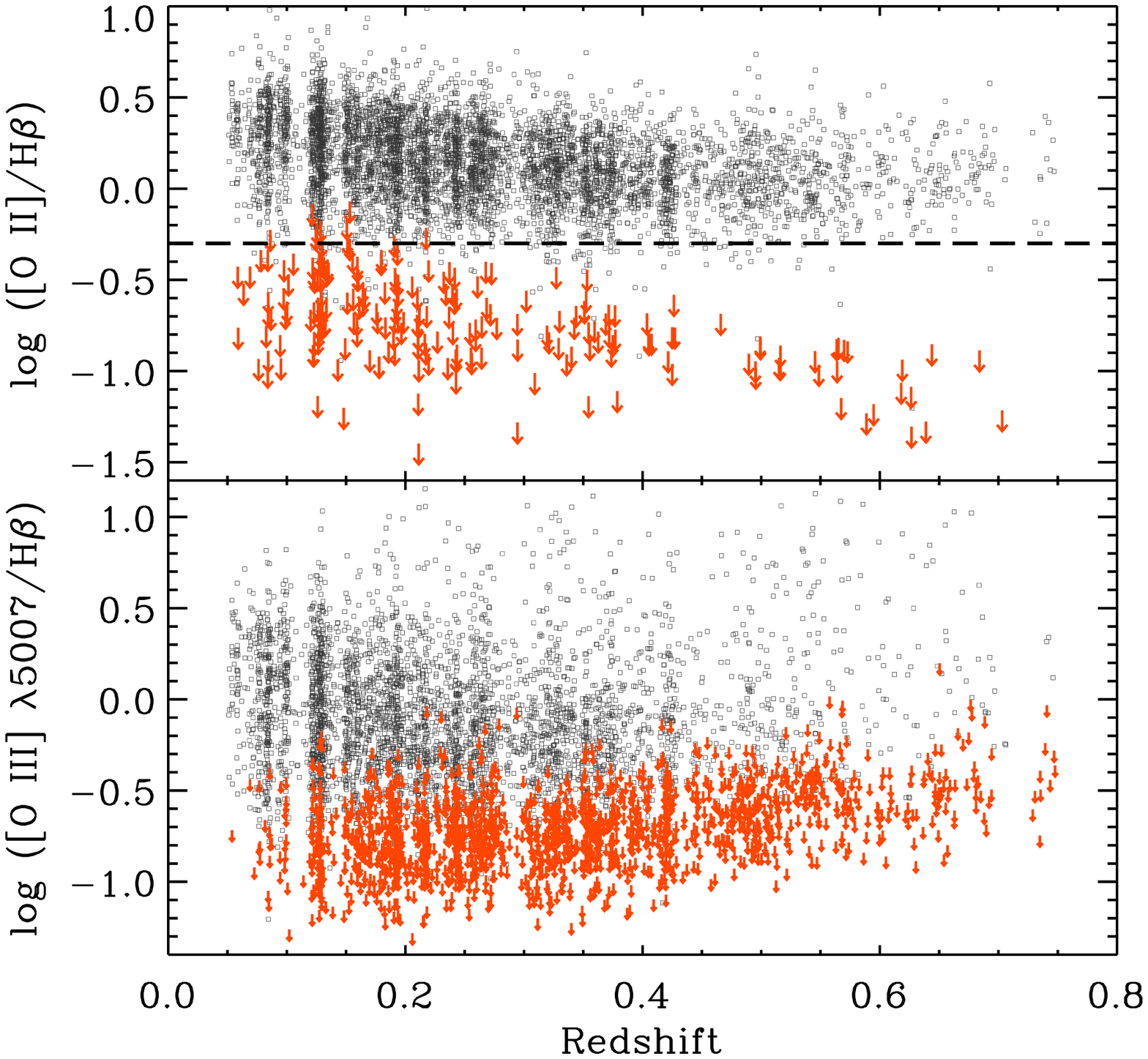}
\caption{Metallicity-sensitive \oii/\hb{} and \oiiilam/\hb{}
  line-ratios vs.~redshift for our \hb-limited sample of galaxies (see
  Fig.~\ref{fig:zvhb}).  We represent galaxies with well-measured
  line-ratios using small black points and upper limits using orange
  arrows.  In the upper panel the dashed line at
  $\log\,(\oii/\hb)=-0.3$ represents the criterion we use to select
  galaxies with significant \oii{} emission, which eliminates just
  $\sim6\%$ of the \hb-selected sample.  By contrast, the bottom panel
  shows that the fraction of galaxies with only upper limits on
  \oiiilam{} is not negligible due the greater sensitivity of \oiii{}
  to variations in metallicity and excitation.
\label{fig:zvoiioiii}}
\end{figure}

In order to estimate the gas-phase oxygen abundances of the galaxies
in our sample we require a measurement of the \oii{} and \oiiilam{}
emission lines (see \S\ref{sec:ohpagel}).  In
Figure~\ref{fig:zvoiioiii} we plot \oii/\hb{} and \oiii/\hb{} versus
redshift for our \hb-selected AGES sample.  The small black points
correspond to well-measured line-ratios, while the orange arrows
represent upper limits.  Focusing on \oii{} first
(Fig.~\ref{fig:zvoiioiii}, \emph{top}), we find that the data exhibit
a distinct lower envelope around $\log\,(\oii/\hb)=-0.3$ ({\em dashed
  line}), below which we find only $\sim6\%$ of our sample, including
nearly all the upper limits.  Visual inspection of the spectra with
low \oii/\hb{} ratios reveals that many exhibit a highly reddened
stellar continuum, i.e., they are dusty.  However, because these
objects constitute such a small fraction of the parent sample, and
because their redshift distribution is consistent with the redshift
distribution of the parent \hb{} sample, we do not expect that
excluding them will bias our results.  Therefore, in addition to our
\hb{} flux cut, we further restrict our sample to have a well-measured
\oii{} emission line with $\log\,(\oii/\hb)>-0.3$.

In contrast to \oii/\hb, the \oiii/\hb{} ratios for approximately
one-quarter of the \hb-selected AGES galaxies are upper limits
(Fig.~\ref{fig:zvoiioiii}, \emph{bottom}).  This result is not
terribly surprising because \oiii{} is especially sensitive to
variations in metallicity and excitation; in particular, the
\oiii/\hb{} ratio decreases precipitously with increasing metallicity
and decreasing excitation \citep{kewley02a}.  However, we show in
\S\ref{sec:effects} that the potential bias against metal-rich
galaxies in an \oiii-limited sample does not significantly affect our
measurement of the \mz{} relation.  Therefore, to significantly
simplify the subsequent analysis, we proceed by requiring the galaxies
in our emission-line sample to have a well-measured \oiii{} emission
line (the black points in Fig.~\ref{fig:zvoiioiii}, \emph{bottom}).

The final number of emission-line galaxies in AGES satisfying our \hb,
\oii, and \oiii{} selection criteria is $4033$, or $37\%$ of our
parent sample.  For our SDSS sample we apply the same selection
criteria described above with the exception of a correspondingly
brighter \hb{} flux limit, $F(\hb)>1\times10^{-16}$~\fluxunits,
leaving $108,040$ galaxies, or $25\%$ of the parent sample.  For
reference, not requiring our SDSS sample to have a well-measured
\oiii{} line would increase the sample by approximately one-third.  In
Table~\ref{table:samples} we summarize the total number of objects in
these and all subsequent subsamples.

\subsection{Identifying and Removing AGN}\label{sec:agn}

The Balmer and collisionally excited forbidden lines present in the
integrated optical spectra of galaxies can arise from photoionization
by young, massive stars, or by the non-thermal photoionizing continuum
of an AGN.  Therefore, when deriving abundances from optical emission
lines it is crucial that AGN be identified and removed from the
sample.  We leverage the tremendous multiwavelength coverage of the
\bootes{} field to identify AGN using a variety of criteria based on
optical, X-ray, mid-IR, and radio observations.

\begin{figure}
\centering
\includegraphics[scale=0.38]{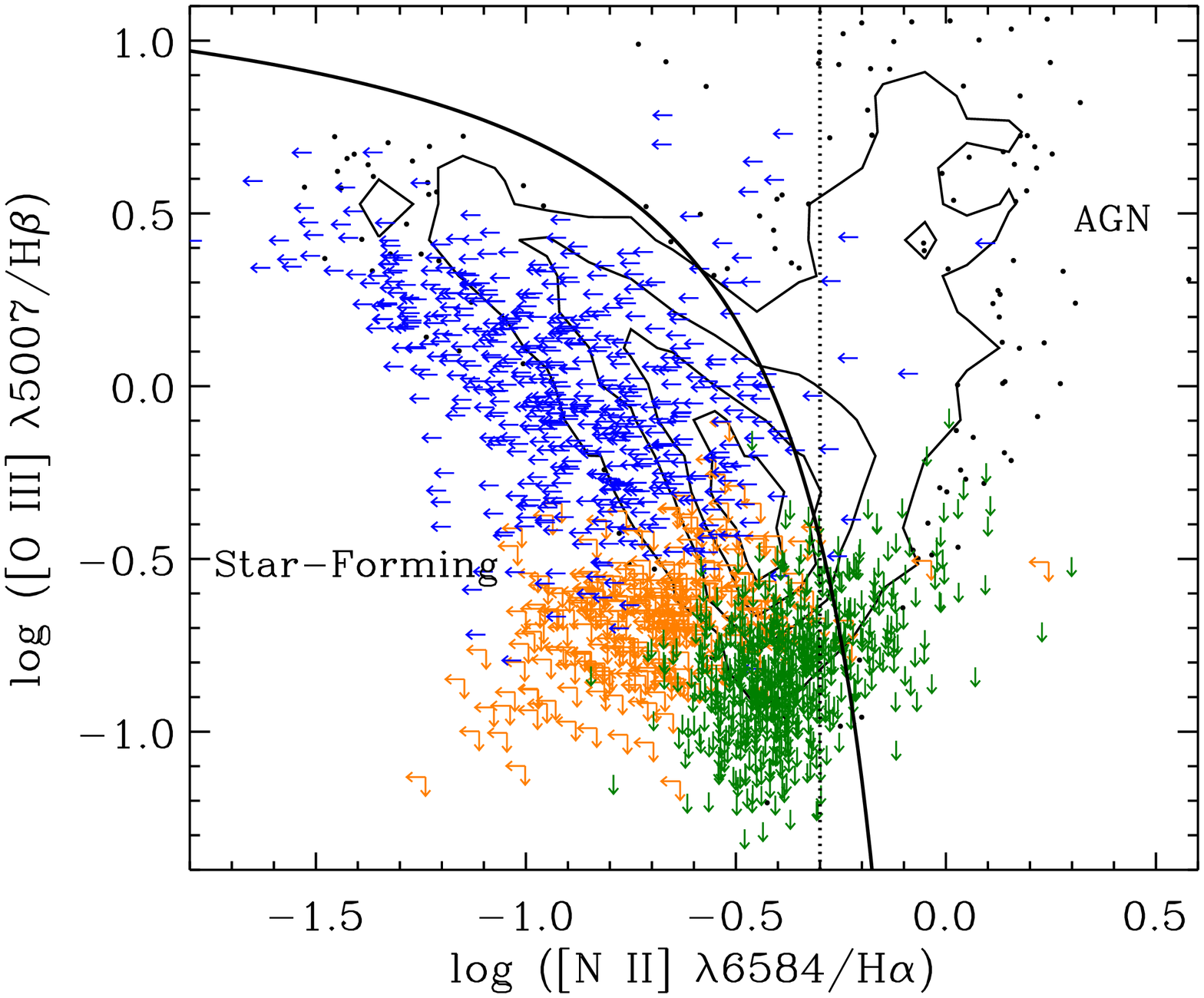}
\caption{AGES emission-line diagnostic (BPT) diagram used to classify
  our sample into star-forming galaxies or AGN based on their position
  below or above the solid curve, respectively \citep{kauffmann03c}.
  The contours, enclosing $25\%$, $50\%$, $75\%$, and $95\%$ of the
  sample, represent galaxies with well-measured \nii/\ha{} and
  \oiii/\hb{} ratios; the small black points are individual galaxies
  outside the $95\%$ contour level.  The arrows represent objects with
  $1\sigma$ upper limits on either: (1) \nii/\ha{} \emph{and}
  \oiii/\hb{} ({\em orange left- and down-pointing arrows}); (2) just
  \nii/\ha{} ({\em blue left-pointing arrows}); or (3) just
  \oiii/\hb{} ({\em green down-pointing arrows}).  Objects with upper
  limits on any of these line-ratios can be classified as star-forming
  if they lie in the star-forming region of the BPT diagram;
  otherwise, their classification is ambiguous.  However, objects with
  upper limits on \oiii/\hb{} can be classified as AGN nearly
  unambiguously if they have $\log(\nii/\ha)>-0.3$ ({\em vertical
    dotted line}).  
\label{fig:bpt}}
\end{figure}

Beginning with our parent sample of emission-line galaxies (see
\S\ref{sec:selection}), we identify type~1 (broad-line) AGN as objects
with broad (typically, $\sigma>800$~\kms{} FWHM) \ha, \hb, or
\mgiilam{} emission, depending on which lines are observable given the
AGES spectral range.  In our SDSS sample broad-line AGN have been
removed already.  In addition, we remove any object with significant
\nevlam{} emission, which is a hallmark of AGN activity, or an
enhanced \neonlam/\hb{} ratio, which is also indicative of a
non-thermal photoionizing radiation field \citep{rola97a,
  dessauges00a, perez-montero07a}.\footnote{Strong \neonlam{} emission
  is also observed in metal-poor dwarf galaxies
  \citep[e.g.,][]{izotov98a}; however, these types of objects are only
  present in our sample at low redshift and easily identified through
  visual inspection of their optical spectrum.}

Next, we use the standard \oiiilam/\hb{} versus \niilam/\ha{}
emission-line diagnostic diagram (Fig.~\ref{fig:bpt}) to identify
type~2 (narrow-line) AGN (\citealt{baldwin81a, veilleux87a,
  kewley01b}; hereafter, the BPT diagram).  We do not show the SDSS
BPT diagram because it exhibits the same basic features \citep[see,
  e.g.,][]{kauffmann03c}.  We identify objects located below and to
the left of the empirical demarcation ({\em solid curve}) proposed by
\citet{kauffmann03c} as star-forming galaxies, and the remaining
objects as AGN.  \citet{stasinska06a} estimate that the maximum
contribution to the integrated \hb{} line-flux from an AGN for an
object classified as star-forming using the \citet{kauffmann03c}
criterion is $\sim3\%$.  For comparison, the ``maximum starburst''
curve proposed by \citet{kewley01b} allows up to a $20\%-30\%$
contribution to \hb{} from an AGN.  We adopt the stricter
\citet{kauffmann03c} demarcation to minimize the metallicity bias
incurred from contamination to the nebular emission lines from AGN
(see \S\ref{sec:agncontam}).

\begin{figure*}
\centering
\includegraphics[scale=0.55,angle=90]{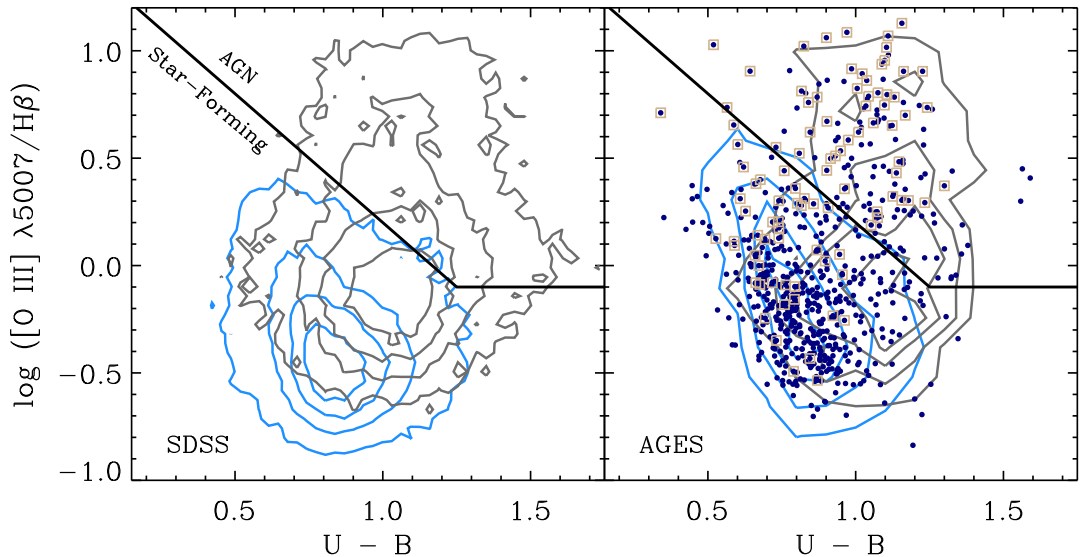}
\caption{BPT-like diagnostic diagram for our (\emph{left}) SDSS and
  (\emph{right}) AGES galaxy samples.  This diagram has been proposed
  by \citet{yan11a} as an empirical means of segregating star-forming
  galaxies ({\em below the solid line}) from AGN ({\em above the solid
    line}) when the ionization-sensitive \nii/\ha{} ratio cannot be
  measured.  The blue contours enclose $25\%$, $50\%$, $75\%$, and
  $95\%$ of the galaxies classified as star-forming using the BPT
  diagram (see Fig.~\ref{fig:bpt} and \S\ref{sec:agn}), while the grey
  contours show the distribution of BPT-classified AGN.  The dark blue
  points represent galaxies that cannot be classified using the BPT
  diagram.  For clarity we do not plot the small number of
  unclassified SDSS galaxies, although nearly all of them lie in the
  AGN portion of this diagnostic diagram.  Within the AGES sample, the
  tan squares represent galaxies classified as AGN using other
  multiwavelength (X-ray, mid-IR, and radio) criteria.  With respect
  to objects classified as star-forming galaxies using the BPT
  diagram, we find that this technique has high \emph{completeness},
  but low \emph{purity}.  In other words, while $>98\%$ of
  BPT-classified star-forming galaxies are successfully classified as
  star-forming using this diagnostic diagram, more than $50\%$ of AGN
  are \emph{also} identified as star-forming.  As noted by
  \citet{yan11a}, however, the bulk of these objects are
  \emph{composite} systems, with an admixture of ongong star-formation
  and AGN activity.  \label{fig:yan}}
\end{figure*}

In Figure~\ref{fig:bpt} we differentiate between four different
scenarios: (1) galaxies with well-measured \nii/\ha{} and \oiii/\hb{}
line-ratios ({\em contours and greyscale}); (2) objects with
\nii/\ha{} upper limits ({\em blue arrows}); (3) objects with
\oiii/\hb{} upper limits ({\em green arrows}); and (4) galaxies with
both \nii/\ha{} \emph{and} \oiii/\hb{} upper limits ({\em orange
  arrows}).  An object with an upper limit on any of these line-ratios
can be classified as star-forming if it lies in the star-forming
region of the BPT diagram; otherwise, its classification is ambiguous,
with one exception: a galaxy can be classified as an AGN if it has an
upper limit on \oiii/\hb{} but a well-measured \nii/\ha{} ratio with
$\log(\nii/\ha)>-0.3$ ({\em vertical dotted line}).

In order to classify the rest of the AGES sample, including all the
objects at $z>0.4$ (which lack \nii/\ha{} measurements), we must rely
on other diagnostics of AGN activity.  Recognizing the difficulty of
measuring the \nii/\ha{} ratio of intermediate-redshift galaxies,
\citet{yan11a} have proposed a new BPT-like diagnostic diagram that
replaces the \nii/\ha{} ratio with the rest-frame $U-B$ color (see
\citealt{lama10a}, \citealt{lara-lopez10b}, and \citealt{juneau11a}
for alternative empirical diagnostic diagrams).  The physical
motivation behind this classification scheme is that AGN are more
likely to be found in galaxies dominated by evolved stellar
populations \citep[e.g.,][]{kauffmann03c}; therefore, at fixed
\oiii/\hb{} ratio, galaxies whose integrated emission-line spectrum is
dominated by star-formation are more likely to be blue in their $U-B$
color.

In Figure~\ref{fig:yan} we plot $U-B$ versus \oiii/\hb{} for our
(\emph{left}) SDSS and (\emph{right}) AGES galaxy samples.  The light
blue contours represent objects classified as star-forming galaxies
using the BPT diagram, the grey contours represent objects classified
as AGN, and the dark blue points indicate objects that cannot be
classified using the BPT diagram.  For clarity we do not include in
this figure the small number of SDSS galaxies that could not be
classified using the BPT diagram, although nearly all of them would be
classified as AGN using the \citet{yan11a} diagnostic diagram.  Among
our AGES galaxies, the tan squares represent objects identified as AGN
using various other optical, X-ray, mid-IR, and radio criteria, as
described below.  Finally, the solid line is the empirical boundary
proposed by \citet{yan11a} to separate star-forming galaxies from AGN,
which are located below and above the solid line, respectively.

Examining Figure~\ref{fig:yan}, we find that the \citet{yan11a}
diagnostic diagram successfully classifies as star-forming more than
$98\%$ of the BPT-classified star-forming galaxies; however, it
\emph{also} identifies as star-forming roughly $50\%$ of the objects
classified as AGN using the BPT diagram.  In other words, the diagram
has high completeness for star-forming galaxies, but at the expense of
a relatively high contamination rate (i.e., low purity).  However, as
emphasized by \citet{yan11a}, nearly all of these ``misclassified''
objects are, in fact, \emph{composite} systems with an admixture of
star-formation and AGN activity.  Moreover, the contribution from the
AGN to the integrated emission-line spectrum in composite systems is
always sub-dominant \citep[$5\%-30\%$;][]{stasinska06a}.

To mitigate the tendency for the \citet{yan11a} diagnostic diagram to
classify composite systems as pure star-forming galaxies, we utilize a
variety of complementary multiwavelength criteria to identify AGN in
our AGES sample \citep[see also][]{hickox09a, rujopakarn10a,
  assef11a}.  First, we classify as an AGN any object with more than a
$50\%$ probability of being an X-ray point source \citep{kenter05a,
  brand06a}.  Although the X-ray emission in some of these objects may
arise from processes associated with star formation
\citep[e.g.,][]{watson09a}, roughly $90\%$ of them have total
$0.5-7$~keV X-ray luminosities $L_{X}\gtrsim10^{42}$~erg~s$^{-1}$,
consistent with hosting an AGN \citep{georgakakis06a}.  We also
isolate objects lying in the ``Stern wedge'' of the IRAC $[3.6]-[4.5]$
versus $[5.8]-[8.0]$ color-color diagram \citep{stern05a}, which is
effective at identifying AGN that are both obscured and unobscured in
the X-ray \citep{hickox07a, gorjian08a, ashby09a}.  Finally, we
compute the radio power at $1.4$~GHz, $P_{1.4}$, of the radio sources
in our sample (see \S\ref{sec:ancillary}), $K$-corrected assuming an
intrinsic power-law spectrum $S_{\nu}\propto\nu^{-\alpha}$ with
$\alpha=0.5$ \citep{prandoni06a}, and identify AGN as sources having
$\log\,P_{1.4}>23.8$~W~Hz$^{-1}$ \citep{kauffmann08a, hickox09a}.

One potential objection to the preceding analysis is that AGN
identified at X-ray, mid-IR, and radio wavelengths should not
necessarily be removed from an \emph{optical} emission-line abundance
study.  For example, there exists an interesting (albeit rare) class
of objects with significant X-ray emission but with no discernable
evidence of AGN activity in the optical, so-called X-ray Bright
Optically Normal Galaxies \citep[XBONGs;][]{comastri02a, civano07a,
  yan11a}.  Similarly, AGN identified based on their mid-IR colors may
be so obscured at optical wavelengths as to not affect the optical
emission lines in any important way.  Nevertheless, we conservatively
reject any object (not otherwise classified as star-forming using the
BPT diagram) whose optical emission lines \emph{might} be contaminated
by an AGN.  The reason for adopting this position is that the oxygen
emission-line ratios of unidentified AGN are similar to those of
metal-poor star-forming galaxies, and therefore could bias the amount
of metallicity evolution that we infer, a point that we revisit in
\S\ref{sec:agncontam}.

To summarize, our final SDSS and AGES star-forming galaxy samples
contain $76,411$ and $3205$ galaxies, respectively (see
Table~\ref{table:samples}).  However, $432$ ($14\%$) of the AGES
emission-line galaxies, $85\%$ of which are at $z>0.4$, were
classified as star-forming using the \citet{yan11a} diagnostic
diagram, which carries a potential $\sim50\%$ contamination rate from
composite (i.e., low-level) AGN.  On the other hand, by using
additional multiwavelength AGN diagnostics it is likely that the
contamination rate in our sample is much lower.  Nevertheless, we
conservatively estimate that $\lesssim10\%$ of the galaxies in our
star-forming galaxy sample, or up to $\sim50\%$ of the objects at
$z>0.4$, may host weak (composite) AGN.  We show in
\S\ref{sec:agncontam}, however, that these unidentified optical AGN,
if present, have a minimal effect on our conclusions.

\subsection{Sample Properties}\label{sec:properties}  

\begin{figure}[b]
\centering
\includegraphics[scale=0.4]{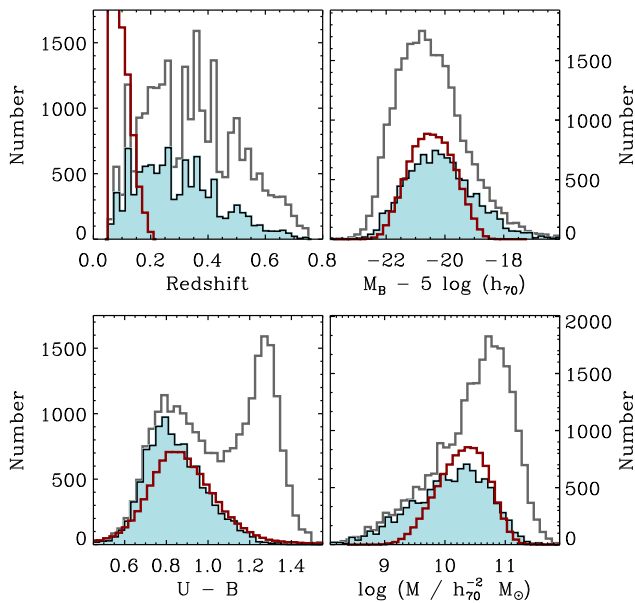}
\caption{Distribution of redshift, absolute $B$-band magnitude,
  rest-frame $U-B$ color, and stellar mass for the AGES parent sample
  ({\em unfilled light grey histograms}), and our sample of
  star-forming emission-line galaxies ({\em filled light blue
    histograms}), weighted to correct for the statistical
  incompleteness of the survey (see \S\ref{sec:redux}).  The unfilled
  dark red histograms show for comparison the distributions of the
  same properties for our SDSS star-forming galaxy sample, where we
  have scaled the SDSS distributions down by a factor of $15$ for
  comparison purposes.  As expected, our emission-line selection
  criteria preferentially remove massive, luminous, optically red
  galaxies.
  \label{fig:histograms}}
\end{figure}

In Figure~\ref{fig:histograms} we compare the distributions of
redshift, absolute $B$-band magnitude, $U-B$ color, and stellar mass
for the star-forming galaxies in AGES ({\em filled light blue
  histograms}) against the full parent sample ({\em unfilled light
  grey histograms}), both corrected for the statistical incompleteness
of the survey (see \S\ref{sec:redux}).  We also show for comparison
the distribution of properties for our SDSS star-forming galaxy sample
({\em unfilled dark red histograms}), where we have scaled the SDSS
distributions down by a factor of $15$ for visualization purposes.
Not unexpectedly, our emission-line cuts preferentially remove
massive, luminous galaxies with red optical colors, the majority of
which are presumably old, early-type galaxies with little or no
ongoing star formation (e.g., \citealt{kauffmann03b, zhu10a}, but see
the discussion in \S\ref{sec:effects}).  Corrected for incompleteness,
the median (mean) redshift of our SDSS sample is $z_{\rm med}=0.087$
($0.094$), compared to $z_{\rm med}=0.27$ ($0.30$) for our AGES
sample; otherwise, the distributions of luminosity, stellar mass, and
color for the two samples are very similar.

\begin{figure}
\centering \includegraphics[scale=0.45]{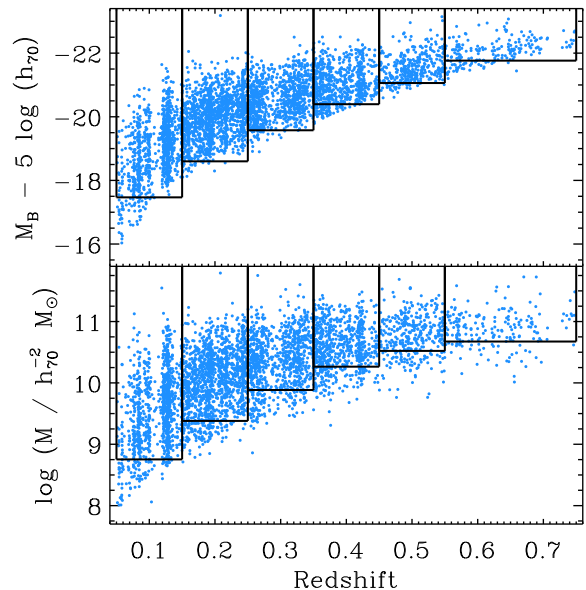}
\caption{Absolute $B$-band magnitude and stellar mass vs.~redshift for
  the star-forming galaxies in AGES.  The black boundaries demarcate
  volume-limited subsamples in six redshift bins centered on $z=0.1$,
  $0.2$, $0.3$, $0.4$, $0.5$, and $0.65$ (see \S\ref{sec:properties}
  and Table~\ref{table:limits}).
\label{fig:zvsmb}}
\end{figure}

In Figure~\ref{fig:zvsmb} we examine the properties of the
star-forming galaxies in AGES in more detail by plotting $B$-band
absolute magnitude and stellar mass versus redshift.  The distribution
of points follows the characteristic shape of a flux-limited survey,
in which intrinsically luminous, massive galaxies are observable over
the full range of redshifts, unlike low-luminosity galaxies (i.e.,
Malmquist bias).  This figure also reveals the relative dearth of
luminous, massive star-forming galaxies in AGES at low redshift due to
the limited cosmological volume probed.

We use Figure~\ref{fig:zvsmb} to calculate the minimum absolute
magnitude and stellar mass above which our sample is volume-limited as
a function of redshift.  The limiting \mb{} is directly proportional
to the survey flux limit, modulo small variations in $K$-corrections
among different galaxy types.  However, the limiting stellar mass has
an additional dependence on the mass-to-light ratio (i.e.,
star-formation history) of each galaxy.  At fixed stellar mass and
redshift a galaxy with a large (small) mass-to-light ratio will lie
preferentially below (above) the survey flux limit.  Since
observations suggest that galaxies of a given mass with high star
formation rates tend to be metal-poor \citep{mannucci10a, yates11a},
properly accounting for the stellar mass limit ensures that we do not
bias our measurement of the \mz{} relation toward lower metallicity.

To calculate the absolute magnitude and stellar mass limits we adopt
an empirical approach \citep[see, e.g.,][]{lama09a}.  First, we divide
AGES into six redshift bins centered on $z=0.1$, $0.2$, $0.3$, $0.4$,
$0.5$, and $0.65$, corresponding to approximately equal intervals of
cosmic time ($\sim1$~Gyr).  Next, we compute the absolute magnitude
and stellar mass each galaxy would have if its apparent magnitude,
$I$, equalled the magnitude limit of the survey, $I_{\rm
  AB,lim}=20.4$.  Specifically, we calculate

\begin{equation}
M_{B, {\rm min}} = M_{B} - (I-I_{\rm lim})
\end{equation}

\noindent and

\begin{equation}
\log\,(\mass_{\rm min}/\msun) = \log\,(\mass/\msun) + 0.4\,(I-I_{\rm
  lim})
\end{equation}

\noindent for each galaxy.  The distributions of $M_{B, {\rm min}}$
and $\mass_{\rm min}$ in each redshift slice reflect the variations in
$K$-corrections and mass-to-light ratio among different galaxies.  We
then define $M_{B, {\rm lim}}$ and $\mass_{\rm lim}$ as the luminosity
and stellar mass that includes more than $50\%$ of the galaxies.  In
Figure~\ref{fig:zvsmb} we show the boundaries that define our
volume-limited samples in each redshift interval, and list the $50\%$
completeness limits in Table~\ref{table:limits}.  For reference, the
$75\%$ and $95\%$ stellar mass completeness limits are roughly
$0.15$~dex and $0.4$~dex higher, while the corresponding limiting
\mb{} values are approximately $0.1$~mag and $0.35$~mag brighter,
respectively.

\begin{deluxetable}{ccc}
\tablecaption{AGES Limiting $B$-Band Magnitude \& Stellar Mass\tablenotemark{a}\label{table:limits}}
\tablewidth{0pt}
\tablehead{
\colhead{Redshift Range} & 
\colhead{$M_{B, {\rm lim}}$} & 
\colhead{$\log\,(\mass_{\rm lim}/\msun)$} 
}
\startdata
$0.05-0.15$ &  -17.47 &   8.75 \\ 
$0.15-0.25$ &  -18.60 &   9.38 \\ 
$0.25-0.35$ &  -19.58 &   9.88 \\ 
$0.35-0.45$ &  -20.40 &  10.26 \\ 
$0.45-0.55$ &  -21.06 &  10.52 \\ 
$0.55-0.75$ &  -21.76 &  10.67
\enddata
\tablenotetext{a}{Absolute $B$-band magnitude and stellar mass as a function of redshift above which the AGES star-forming galaxy sample is more than $50\%$ complete, accounting for variations in $K$-corrections and mass-to-light ratio.}
\end{deluxetable}

\section{Deriving Gas-Phase Oxygen Abundances}\label{sec:oh} 

\subsection{Background on Strong-Line Metallicity
  Calibrations}\label{sec:ohintro}   

The observational technique of using optical emission lines to derive
the physical conditions (density, temperature, chemical composition)
of the ionized gas in \hii{} regions and star-forming galaxies has a
long, rich history \citep{aller42a, shields90a, stasinska07a}.  In
particular, considerable effort has gone into using optical
spectroscopy to infer the abundance of oxygen in star-forming regions.
Oxygen is important because it is the most abundant element (by
number) in the Universe after hydrogen and helium \citep{asplund09a},
making it a crucial coolant in the interstellar medium; moreover, all
the principal oxygen transitions are observable in the rest-frame
optical.

The classical, so-called \emph{direct} method of inferring the
gas-phase oxygen abundance is to measure the electron temperature of
the ionized gas from the strength of an auroral line, usually
\oiii~$\lambda4363$, relative to a nebular line such as \oiiilam{}
\citep{dinerstein90a, garnett04b, osterbrock06a}.\footnote{The
  abundance of oxygen and other heavy elements can be inferred from
  the optical metal-recombination lines as well.  However, because
  these lines are extraordinarily weak, they have been measured in
  only a relatively small number of Galactic and extragalactic \hii{}
  regions and starburst dwarf galaxies \citep[and references
    therein]{esteban09a}.}  Unfortunately, \oiii~$\lambda4363$ and
other temperature-sensitive auroral lines (e.g., \nii~$\lambda5755$
\siii~$\lambda6312$, \oii~$\lambda7325$) become vanishingly weak with
increasing metallicity; therefore, they are rarely detected in the
integrated spectra of galaxies observed as part of flux-limited
surveys such as AGES and the SDSS, which typically include relatively
luminous, metal-rich galaxies.  For example, electron temperatures
have been measured for fewer than $\sim0.1\%$ of galaxies in the
entire SDSS database, all of which are relatively nearby, metal-poor
dwarfs \citep{kniazev04a, izotov06a}.

\begin{figure*}
\centering
\includegraphics[angle=90,scale=0.55]{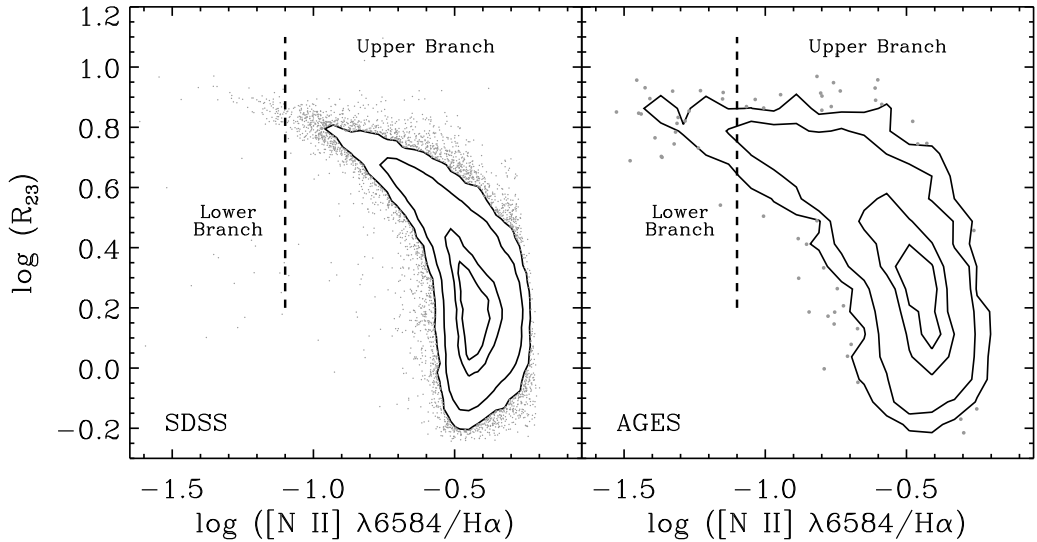}
\caption{Emission-line diagnostic diagram used to assign galaxies to
  the \emph{upper}, $\log\,(\niiha)>-1.1$, and \emph{lower}
  $\log\,(\niiha)<-1.1$, branches of the \pagel-O/H relation (see
  \S\ref{sec:ohpagel}) for our (\emph{left}) SDSS and (\emph{right})
  AGES star-forming galaxy samples with well-measured \nii/\ha{}
  ratios.  For reference, the contours enclose $25\%$, $50\%$, $90\%$,
  and $97.5\%$ of the sample, and the small grey points represent
  galaxies that lie outside the $97.5\%$ contour level.  This diagram
  demonstrates that the galaxies in both samples overwhelming belong
  to the upper \pagel{} branch. 
\label{fig:niiha_r23}}
\end{figure*}

In light of the observational challenges confronting direct,
electron-temperature abundance measurements, complementary methods
have been developed that rely exclusively on the strongest emission
lines in the rest-frame optical.  These so-called \emph{strong-line}
methods are statistical correlations between various line-ratios and
the gas-phase oxygen abundance, \logoh.  Unfortunately, different
calibrations yield widely different oxygen abundances {\em based on
  the same input emission-line spectrum}.  The magnitude of the
systematic differences in metallicity depends on which calibrations
are being compared, but they can be as large as $\sim0.7$~dex, or a
factor of $\sim5$ \citep{kenn03a, perez-montero05a, liang06b,
  nagao06a, bresolin04a, bresolin05a, bresolin07a, kewley08a,
  lopez10a, stasinska10a, moustakas10a}.  In particular, so-called
\emph{theoretical} strong-line methods, which are based on {\em ab
  initio} photoionization model calculations
\citep[e.g.,][]{mcgaugh91a, charlot01a, kewley02a}, yield oxygen
abundances that are, on average, several times higher than abundances
derived using either the direct method, or \emph{empirical}
strong-line methods, which are calibrated against \hii{} regions with
direct abundance estimates \citep[e.g.,][]{pilyugin00a, pilyugin01a,
  denicolo02a, pettini04a, perez-montero05a, pilyugin05a,
  stasinska06b, yin07a}.

\subsection{Deriving Metallicities Using the \pagel{}
  Parameter}\label{sec:ohpagel}

Because of the poorly understood systematic uncertainties in the
nebular abundance scale described in \S\ref{sec:ohintro}, it is
important that the requisite emission lines we use to derive
metallicities are observable over the full redshift range of interest,
$z=0.05-0.75$.  The combination of nebular lines that satisfies this
requirement for our SDSS \emph{and} AGES samples is the \pagel{}
parameter \citep{pagel79a},

\begin{equation}
\pagel \equiv \frac{\oiilam + \oiiidoublet}{\hblam}
\label{eq:r23}.
\end{equation}

\noindent The \pagel{} parameter is a popular and widely used
abundance diagnostic for several reasons.  First, its strength is
directly proportional to both principal ionization states of oxygen,
and therefore does not require large, uncertain ionization
corrections; the relative amount of neutral oxygen and O$^{3+}$ is
negligible in star-forming regions across a wide range of physical
conditions.  Second, \pagel{} is explicitly related to the oxygen
abundance, unlike other popular diagnostics that depend implicitly on
the relative abundance of additional elements like nitrogen (see
\S\ref{sec:calib}).  Finally, \pagel{} can be measured from the ground
using a combination of optical and near-infrared spectroscopy from
$z=0-4$, which allows the chemical enrichment histories of galaxies to
be studied using a consistent metallicity diagnostic over $90\%$ of
the age of the Universe \citep{kenn98c, maiolino08a, richard11a}.

To explore the effect on our conclusions of choosing a particular
strong-line calibration, we carry out our analysis using three
independent theoretical strong-line calibrations of \pagel:
\citet{mcgaugh91a}, \citet{kobulnicky04a}, and \citet{tremonti04a},
hereafter referred to as the M91, KK04, and T04 abundance
calibrations, respectively.  At a fixed value of \pagel, both
observations and photoionization modeling show that there is a
second-order dependence of O/H on the ionization parameter, $U$, or
the hardness of the ionizing radiation field \citep{shields90a,
  mcgaugh91a, pilyugin01a, kewley02a}.  The M91 and KK04 calibrations
constrain the ionization parameter using the \ioniz{} ratio, where

\begin{equation}
\ioniz \equiv \frac{\oiiidoublet}{\oiilam}
\label{eq:o32},
\end{equation}

\noindent while the T04 calibration ignores this effect.  All three
calibrations are based on photoionization models, and therefore they
yield metallicities that are considerably higher than those implied by
the direct method or empirical strong-line calibrations (e.g.,
\citealt{pettini04a, pilyugin05a}; see \citealt{kewley08a}).  However,
as \citet{moustakas10a} argue, existing empirical strong-line
calibrations should be applied to integrated galaxy spectra with great
care because they are being extrapolated into a part of physical
parameter space that is not well-constrained by current observations
of \hii{} regions.  In Appendix~\ref{appendix:calib} we compare the
M91, KK04, and T04 calibrations in more detail.

Despite its popularity and wide-spread use, \pagel-based abundances
suffer from several potential systematic uncertainties.  First, the
relationship between \pagel{} and O/H is famously double-valued (see,
e.g., Fig.~\ref{fig:r23}): a given value of \pagel{} can correspond to
a solution on the metal-rich {\em upper branch}, or on the metal-poor
{\em lower branch}; the region where the two branches overlap is known
as the {\em turn-around region}.  Fortunately, the bulk of the
galaxies in our SDSS and AGES samples are relatively luminous and
massive, and therefore overwhelmingly belong on the upper branch.  We
illustrate this point in Figure~\ref{fig:niiha_r23}, where we plot
\niilam/\ha{} versus \pagel{} for our (\emph{left}) SDSS and
(\emph{right}) AGES star-forming galaxy samples.  Objects that belong
on the lower \pagel{} branch generally have $\log\,(\niiha)<-1.1$,
while upper-branch galaxies have $\log\,(\niiha)>-1.1$
\citep{mcgaugh91a, contini02a, kewley08a}.  We find that fewer than
$1\%$ of SDSS galaxies, and only $\sim3\%$ of AGES galaxies belong to
the lower \pagel{} branch.  Of course, we do not have a measurement of
\nii/\ha{} for galaxies in AGES at $z>0.4$ because of the red
wavelength cut-off of our spectra.  However, we assume that all these
galaxies lie on the upper \pagel{} branch, which is not an
unreasonable assumption given their luminosities and stellar masses
(see Fig.~\ref{fig:zvsmb}).  In particular, this assumption is
conservative because it minimizes the amount of chemical evolution
these objects would have to undergo to lie on the present-day \mz{}
relation (see \S\ref{sec:mzlocal}).

The second issue to consider is that both \pagel{} and \ioniz{} are
sensitive to dust attenuation.  Unfortunately, we cannot correct for
dust reddening because the Balmer decrement, \ha/\hb, can be measured
only for galaxies in AGES at $z<0.4$, while the \hb/\hg{} ratio is
generally too noisy to give a reliable estimate of the reddening.
Therefore, following \citet{kobulnicky03a}, we use EWs to estimate
oxygen abundances.  Besides being relatively insensitive to dust
extinction as we show below, EWs have the added advantage of being
impervious to flux-calibration issues (see \S\ref{sec:ispec}).

With some simple assumptions it is easy to show that the EW-based
\pagel{} and \ioniz{} parameters are

\begin{equation}
\ewpagel \equiv \frac{\alpha\ewoii+\ewoiii}{\ewhb}, 
\label{eq:ewr23}
\end{equation}

\noindent and 

\begin{equation}
\ewioniz \equiv \frac{\ewoiii}{\alpha\ewoii},  
\label{eq:ewo32}
\end{equation}

\noindent where 

\begin{equation}
\alpha = \frac{I_{c}(4934)}{I_{c}(3727)} 10^{0.4\Delta
  E(B-V) [k(4934)-k(3727)]}
\label{eq:alpha}
\end{equation}

\noindent is a constant for each galaxy that depends on its recent
star-formation history and level of dust attenuation.  In
equation~(\ref{eq:alpha}), $I_{c}(4934)$ and $I_{c}(3727)$ are the
intrinsic (unattenuated) flux of the stellar continuum at
$\lambda=4934$~\AA{} (the wavelength midway between \hb{} and
\oiiilam) and $3727$~\AA, respectively; $\Delta E(B-V)$ is the
reddening difference between the ionized gas and the continuum
\citep[in general, $\Delta E(B-V)\geq0$;][]{calzetti94a, calzetti01a};
and $k(4934)$ and $k(3727)$ is the value of the assumed attenuation
law at $4934$~\AA{} and $3727$~\AA, respectively
\citep{kobulnicky03a}.  Using population synthesis models and some
basic assumptions about $\Delta E(B-V)$ and $k(\lambda)$,
\citet{kobulnicky03a} estimate $\alpha=0.84\pm0.3$.

\begin{figure}
\centering
\includegraphics[angle=0,scale=0.4]{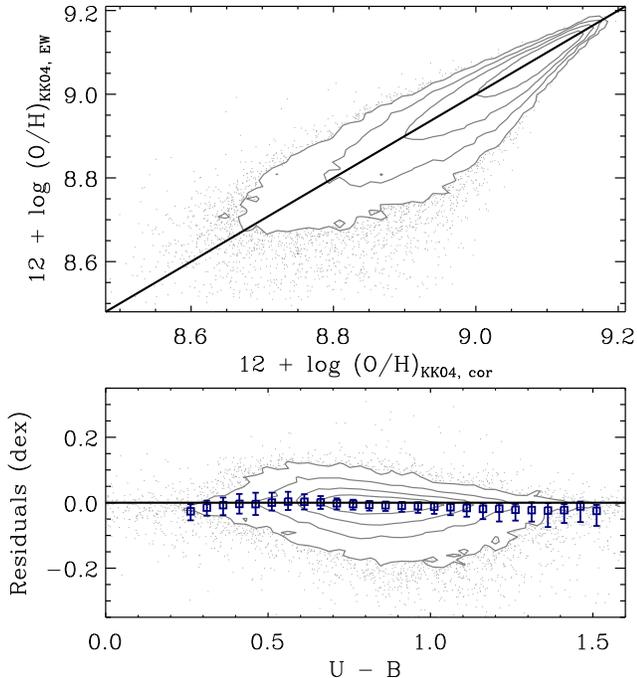}
\caption{({\em Top}) Correlation between oxygen abundances estimated
  using reddening-corrected fluxes and oxygen abundances derived using
  EWs and $\alpha=1.0$ (see \S\ref{sec:ohpagel}).  This comparison
  uses the KK04 abundance calibration, but the same conclusion holds
  if we use either the M91 or T04 calibrations.  (\emph{Bottom})
  Abundance residuals vs.~rest-frame $U-B$ color.  The open squares
  give the median residual metallicity in $0.05$~mag wide bins of
  $U-B$ color and the error bars indicate the interquartile range.
  The median residuals vary by $<0.05$~dex with a scatter of just
  $0.01-0.06$~dex; therefore, abundances derived using EWs provide a
  reliable way of estimating the metallicity when reddening-corrected
  fluxes are unavailable.  For reference, the contours in both panels
  enclose $50\%$, $75\%$, $90\%$, and $97.5\%$ of the sample, and the
  solid line shows the one-to-one relation; the small grey points are
  individual galaxies lying outside the $97.5\%$ contour level.
\label{fig:oh_compare}}
\end{figure}

\begin{deluxetable*}{lccccccccc}[!h]
\tablecaption{SDSS \mz{} and \lz{} Relations\label{table:mzlzlocal}}
\tablewidth{0pt}
\tablehead{
\colhead{} & 
\colhead{} & 
\multicolumn{4}{c}{\mz\tablenotemark{b}} & 
\colhead{} & 
\multicolumn{3}{c}{\lz\tablenotemark{c}} \\
\cline{3-6}\cline{8-10}
\colhead{Calibration\tablenotemark{a}} & 
\colhead{} & 
\colhead{$\ohstar$} & 
\colhead{$\log\,(\mstar/10^{9}\, \msun$)} & 
\colhead{$\gamma$} & 
\colhead{$\sigma$\tablenotemark{d}} & 
\colhead{} & 
\colhead{$c_{0}$} & 
\colhead{$c_{1}$} & 
\colhead{$\sigma$\tablenotemark{d}} 
}
\startdata
\multicolumn{2}{c}{} & \multicolumn{8}{c}{Reddening-Corrected Fluxes\tablenotemark{a}} \\
\cline{1-10}
KK04 & 
  & 
$9.115$ & 
$2.043$ & 
$1.41$ & 
$0.07$ & 
  & 
$9.066$ & 
$-0.231$ & 
$0.16$ \\
M91 & 
  & 
$9.004$ & 
$3.117$ & 
$1.27$ & 
$0.09$ & 
  & 
$8.919$ & 
$-0.256$ & 
$0.18$ \\
T04 & 
  & 
$9.098$ & 
$3.475$ & 
$1.22$ & 
$0.10$ & 
  & 
$9.014$ & 
$-0.291$ & 
$0.20$ \\
MPA-JHU & 
  & 
$9.137$ & 
$3.436$ & 
$0.89$ & 
$0.09$ & 
  & 
$9.026$ & 
$-0.248$ & 
$0.16$
\\
\cline{1-10}
\multicolumn{2}{c}{} & \multicolumn{8}{c}{Equivalent Widths\tablenotemark{a}} \\
\cline{1-10}
KK04 & 
  & 
$9.108$ & 
$2.001$ & 
$1.25$ & 
$0.07$ & 
  & 
$9.054$ & 
$-0.229$ & 
$0.16$ \\
M91 & 
  & 
$9.006$ & 
$3.151$ & 
$1.06$ & 
$0.09$ & 
  & 
$8.903$ & 
$-0.248$ & 
$0.17$ \\
T04 & 
  & 
$9.097$ & 
$3.575$ & 
$1.05$ & 
$0.10$ & 
  & 
$8.997$ & 
$-0.286$ & 
$0.20$
\enddata
\tablenotetext{a}{Results derived using the \citet[KK04]{kobulnicky04a}, \citet[M91]{mcgaugh91a}, and  \citet[T04]{tremonti04a} calibrations of the \pagel{} parameter, using both reddening-corrected line-fluxes and equivalent widths (see \S\ref{sec:ohpagel} for details). For comparison, we also derive the \mz{} and \lz{} relations using the oxygen abundances published by the MPA-JHU team (see \S\ref{sec:mzlocal}).}
\tablenotetext{b}{\mz{} relation given by $12+\log\,(\textrm{O}/\textrm{H}) = 12+\log\,(\textrm{O}/\textrm{H})^{\ast} -\log\, [1+(\mstar/10^{9}\msun)^{\gamma}]$.}
\tablenotetext{c}{$B$-band \lz{} relation given by $\logoh=c_{0}+c_{1}(\mb+20.5)$.}
\tablenotetext{d}{Residual $1\sigma$ scatter about the best-fitting relation in dex.}
\end{deluxetable*}

We obtain an independent, empirical estimate of $\alpha$ by
correlating the reddening-corrected \ioniz{} ratio, \ionizcor, and
\ewioniz{} using our SDSS sample.  First, we estimate the nebular
reddening for each galaxy using the observed Balmer decrement and
assume an intrinsic \hahb{} ratio of $2.86$.  Next, we derive
\ionizcor{} using the \citet{odonnell94a} Milky Way extinction curve
and $R_{V}=3.1$ \citep{moustakas06b, kenn09a}.  Equating \ewioniz{}
and \ionizcor, we find, on average, $\alpha=1.0\pm0.16$.  To a very
good approximation, in other words, equations~(\ref{eq:ewr23}) and
(\ref{eq:ewo32}) with $\alpha\approx1$ reduce to
equations~(\ref{eq:r23}) and (\ref{eq:o32}) but with EWs in place of
reddening-corrected fluxes.

In Figure~\ref{fig:oh_compare} we compare the EW-based abundances
against the abundances derived using reddening-corrected fluxes.  For
this comparison we use the KK04 calibration, although we reach
identical conclusions using either the T04 or M91 calibration.  In the
bottom panel of Figure~\ref{fig:oh_compare} we correlate the abundance
residuals against the rest-frame $U-B$ color.  We find an excellent
correlation between the two abundances, with a negligible systematic
difference ($<0.05$~dex) and a $1\sigma$ scatter that ranges from
$0.01-0.03$~dex at high metallicity, to $0.04-0.06$~dex for more
metal-poor galaxies, independent of the adopted abundance calibration.
Therefore, we conclude that EWs (assuming $\alpha=1$) provide an
effective means of estimating the gas-phase abundances of galaxies in
the absence of a measurement of the Balmer decrement (but for an
alternative conclusion see \citealt{liang07a}).  In
\S\ref{sec:mzlzews} we further show that using EWs in place of
reddening-corrected fluxes has a negligible effect on the inferred
shape and normalization of the \mz{} and \lz{} relations.

\subsection{From Equivalent Widths to
  Metallicity}\label{sec:ohsummary} 

Given the measured values of \ewpagel{} and \ewioniz, and the inferred
or assumed \pagel{} branch, we derive the gas-phase oxygen abundances
of all the star-forming galaxies in our SDSS and AGES samples using
the M91, KK04, and T04 strong-line calibrations (see
Appendix~\ref{appendix:calib}).  To derive the uncertainty in \logoh,
we apply the Monte Carlo-based method developed by
\citet{moustakas10a}, which uses the statistical uncertainties on
\ewoii, \ewoiii{} and \ewhb{} to obtain a realistic estimate of the
abundance error for each object.  This algorithm also relies on
quantitative criteria to identify galaxies that lie \emph{off} the
adopted abundance calibration, typically because they have
$\pagel\gtrsim10$ (see Fig~\ref{fig:r23}).  The final number of AGES
and SDSS galaxies with well-measured oxygen abundances is $\sim3000$
and $\sim75,000$, respectively, where the exact number depends on
which strong-line calibration is being used (see
Table~\ref{table:samples}).

\section{Analysis}\label{sec:evol}

Armed with stellar masses, optical luminosities, and gas-phase oxygen
abundances for both our SDSS and AGES samples, we now proceed to study
the change in the chemical abundances of star-forming galaxies since
$z=0.75$.  Although we focus our analysis on the evolution of the
\mz{} relation, we also examine the evolution of the $B$-band \lz{}
relation.  In \S\ref{sec:mzlzlocal} we establish the \mz{} and \lz{}
relations at $z\sim0.1$ using our SDSS sample, and in
\S\ref{sec:mzlzevol}, we use AGES to measure their evolution from
$z=0.05-0.75$.  We conclude in \S\ref{sec:compare} by comparing our
results with previous determinations of the \mz{} relation at
intermediate redshift.

\subsection{SDSS \mz{} and $B$-Band \lz{}
  Relations}\label{sec:mzlzlocal}

We begin by determining the \mz{} relation for star-forming galaxies
at $z\sim0.1$ in \S\ref{sec:mzlocal}, and the \lz{} relation in
\S\ref{sec:lzlocal}.  In particular, we examine the effect of adopting
a specific abundance calibration on the results.  In both sections we
derive metallicities using the reddening-corrected emission-line
fluxes (see \S\ref{sec:ohpagel}), although in \S\ref{sec:mzlzews} we
examine the effect of using EWs on our results.

\subsubsection{\mz{} Relation at $z\sim0.1$}\label{sec:mzlocal}   

\begin{figure}[b]
\centering
\includegraphics[angle=0,scale=0.4]{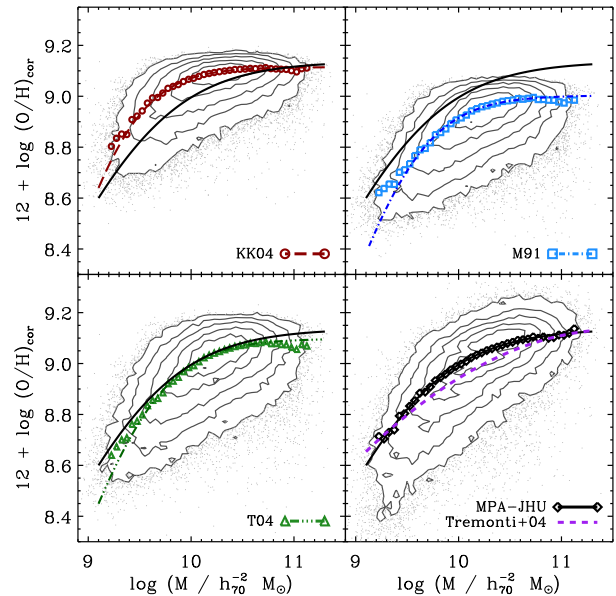}
\caption{SDSS \mz{} relation based on four different methods of
  deriving oxygen abundances from the reddening-corrected line-fluxes.
  The first three panels show the \mz{} relation based on the three
  \pagel-based strong-line calibrations adopted in this paper (KK04,
  M91, and T04), while the \mz{} relation in the lower-right panel
  uses the metallicities for SDSS/DR7 galaxies derived by the MPA-JHU
  team (see \S\ref{sec:mzlocal}).  The contours enclose $25\%$,
  $50\%$, $75\%$, $90\%$, and $97.5\%$ of the sample, and the small
  grey points are individual galaxies that lie outside the $97.5\%$
  contour level.  The symbols in each panel (as indicated in the
  legend) show the weighted mean metallicity in $0.05$-dex wide bins
  of stellar mass, and the corresponding curves show the best-fitting
  \mz{} relation given by equation~(\ref{eq:mzclosedbox}).  For
  reference, we reproduce in each panel the MPA-JHU \mz{} relation
  ({\em solid black curve}).  In the lower-right panel we also show
  for comparison the original SDSS \mz{} relation (based on the
  SDSS/DR4) published by \citet{tremonti04a} ({\em dashed purple
    curve}).  This figure demonstrates that both the normalization and
  the shape of the \mz{} relation depend on the method used to derive
  oxygen abundances.  \label{fig:mzlocal}}
\end{figure}

In Figure~\ref{fig:mzlocal} we plot the SDSS \mz{} relation using four
different abundance calibrations.  The first three panels show the
\mz{} relations based on the KK04, M91, and T04 abundance
calibrations, and the lower-right panel shows, for comparison, the
\mz{} relation using the oxygen abundances derived by the MPA-JHU
team.\footnote{\url{http://www.mpa-garching.mpg.de/SDSS/DR7}} The
MPA-JHU abundances are based on fitting the \citet{charlot01a}
photoionization models to \emph{all} the available nebular emission
lines, and therefore they serve as a useful cross-check of the
\pagel-based abundances.  We point out that the T04 calibration of
\pagel{} by \citet{tremonti04a} was derived by correlating \pagel{}
against the MPA-JHU metallicities; therefore, the statistical
agreement between the T04 and MPA-JHU \mz{} relations is not
unexpected.

We characterize the shape of the \mz{} relation by computing the
weighted mean metallicity in $0.05$~dex wide bins of stellar mass, and
plot the results in Figure~\ref{fig:mzlocal} using red circles, blue
squares, green triangles, and black diamonds for each of the four
calibrations, as indicated in the figure.  To fit the \mz{} relation
we adopt the following functional form:

\begin{equation}
12+\log\,(\textrm{O}/\textrm{H}) = 12+\log\,(\textrm{O}/\textrm{H})^{\ast} -
\log\, \left[1+\left(\frac{\mstar}{10^{9}\, \msun}\right)^{\gamma}\right],
\label{eq:mzclosedbox}
\end{equation}

\begin{figure*}
\centering
\includegraphics[angle=0,scale=0.55]{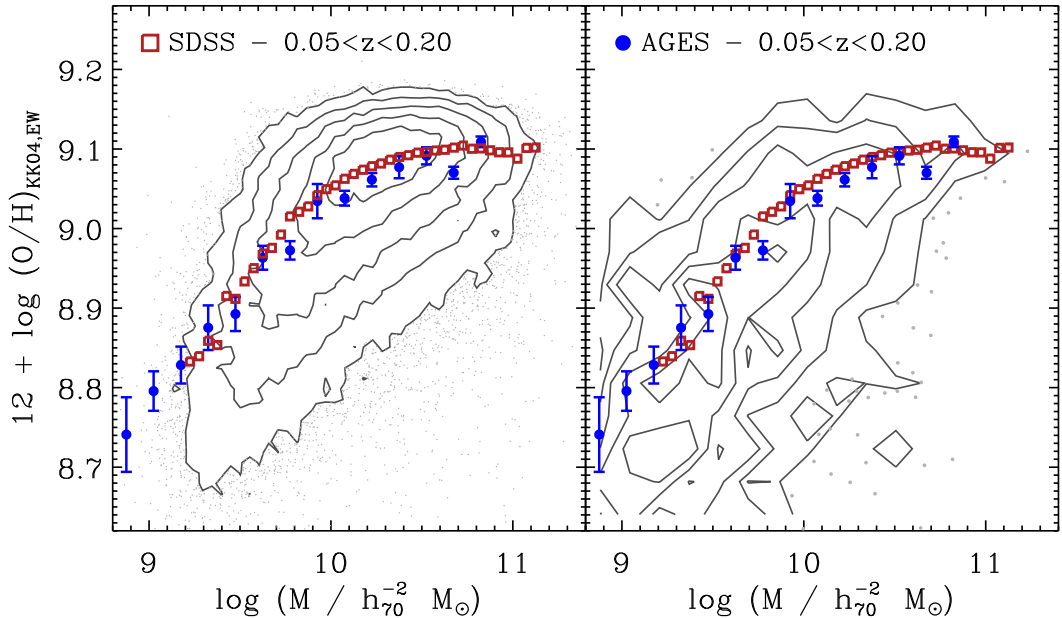}
\caption{\mz{} relation at $z\sim0.1$ based on our (\emph{left}) SDSS
  and (\emph{right}) low-redshift ($0.05<z<0.20$) AGES samples.  In
  both panels the contours enclose $25\%$, $50\%$, $75\%$, $90\%$, and
  $97.5\%$ of the sample, and the small grey points are individual
  galaxies that lie outside the $97.5\%$ contour level.  The open red
  squares and filled blue points represent the median metallicity in
  $0.05$ and $0.15$~dex wide bins of stellar mass for the SDSS and
  AGES, respectively, and the error bars on the AGES points are the
  standard error in the mean metallicity in each stellar mass bin.
  Where they overlap, the \mz{} relations of the two samples are
  statistically consistent with one another.  Note however, the
  relative paucity of massive, metal-rich galaxies in AGES over this
  redshift range, due to the relatively small amount of cosmological
  volume probed.  \label{fig:mzlocal_ages}}
\end{figure*}

\noindent where \ohstar{} and \mstar{} are the characteristic
metallicity and stellar mass, respectively.  We discuss and justify
this particular parameterization of the \mz{} relation compared to the
more commonly used polynomial model in Appendix~\ref{appendix:mzform}.
We show the best-fitting KK04, M91, T04, and MPA-JHU \mz{} relations
in Figure~\ref{fig:mzlocal} using long-dashed, dot-dashed,
triple-dot-dashed, and solid lines, respectively, and list the
best-fitting coefficients in Table~\ref{table:mzlzlocal}.  Note that
the formal statistical uncertainties on the coefficients are
negligible ($<0.01$~dex), and therefore have not been tabulated.  For
reference, we reproduce the MPA-JHU \mz{} relation in every panel of
Figure~\ref{fig:mzlocal}.  In addition, in the lower-right panel we
show the \mz{} relation derived by \citet{tremonti04a} from a sample
of $\sim53,000$ galaxies from the SDSS Data Release 4 (DR4) as a
short-dashed purple line.  The origin of the discrepancy between the
\mz{} relation published by \citet{tremonti04a} using DR4 and the
MPA-JHU \mz{} relation we derive using DR7 is not clear, given that
the metallicities were derived using the same maximum likelihood
technique and \citet{charlot01a} photoionization models.
Nevertheless, evolutionary studies and theoretical models that are
calibrated against the original \citet{tremonti04a} \mz{} relation
should bear these differences in mind.

Returning to Figure~\ref{fig:mzlocal}, we find that all four abundance
calibrations yield a tight, well-defined \mz{} relation---the
$1\sigma$ scatter ranges from $0.07-0.10$~dex---that increases
monotonically with stellar mass and then flattens by varying amounts
above a characteristic mass around $\sim10^{10}~\msun$.  However, both
the normalization and shape of the \mz{} relation vary significantly
among the different calibrations, a result that has been explored in
detail by \citet{kewley08a}.  For example, the KK04 \mz{} relation is
much flatter ($\gamma$ is larger), and the characteristic mass is a
factor of $\sim1.6$ smaller compared to the other three \mz{}
relations, although the characteristic metallicities of the four \mz{}
relations are within $\sim0.1$~dex ($25\%$) of one another.

Understanding the origin of the differences between the \mz{}
relations implied by these strong-line calibrations is beyond the
scope of this paper \citep[but see the discussion in][]{moustakas10a}.
The main points to take away from the preceding discussion and
Figure~\ref{fig:mzlocal} are that both the normalization \emph{and}
shape of the \mz{} relation depend on the method used to derive
gas-phase abundances.  This result has important implications for
theoretical semianalytic and hydrodynamic models of galaxy formation
that use the \mz{} relation to calibrate and tune so-called
``recipes'' for star formation efficiency, supernova feedback, and
galactic winds \citep[e.g.,][]{brooks07a, finlator08a, dutton09a,
  dave11b}.  From an observational standpoint, these results emphasize
the importance of focusing on the \emph{differential} metallicity
evolution, and of considering multiple abundance diagnostics to
ascertain whether the \emph{rate} of metallicity evolution depends on
the adopted abundance calibration (see especially \S\ref{sec:mzevol}),
an issue that heretofore has largely been ignored.

We conclude this section by verifying that the \mz{} relations
measured from our SDSS sample at $z\sim0.1$ and the subset of AGES
galaxies in our lowest redshift bin, $0.05<z<0.20$ (hereafter, the
low-redshift AGES sample), are consistent.  In
Figure~\ref{fig:mzlocal_ages} we plot the KK04-based \mz{} relations
for these two samples.  We emphasize that the M91 and T04 abundance
calibrations yield identical conclusions.  The red squares and blue
circles show the median metallicity in $0.05$ and $0.15$~dex wide bins
of stellar mass for our SDSS and low-redshift AGES sample,
respectively.  The error bars on the AGES points reflect the standard
error of the mean metallicity in each stellar mass bin (the SDSS error
bars are generally smaller than the symbols).

Examining Figure~\ref{fig:mzlocal_ages}, we find that the two \mz{}
relations are statistically consistent with one another.  This figure
also strikingly demonstrates that the SDSS sample is crucial for
constraining the massive end of the \mz{} relation; our low-redshift
AGES sample simply contains too few massive, metal-rich galaxies due
to the limited cosmological volume probed by the survey.  Moreover,
the consistency of the two \mz{} relations suggests that the relative
aperture bias between the two surveys is negligible, even though the
fibers used by AGES are a factor of two smaller (see also the detailed
discussion of aperture effects in \S\ref{sec:apbias}).  Therefore, we
conclude that our SDSS sample provides an unbiased sample for
measuring the evolution of the \mz{} relation at intermediate redshift
from AGES.

\subsubsection{$B$-band \lz{} Relation at $z\sim0.1$}\label{sec:lzlocal} 

\begin{figure}
\centering
\includegraphics[angle=0,scale=0.4]{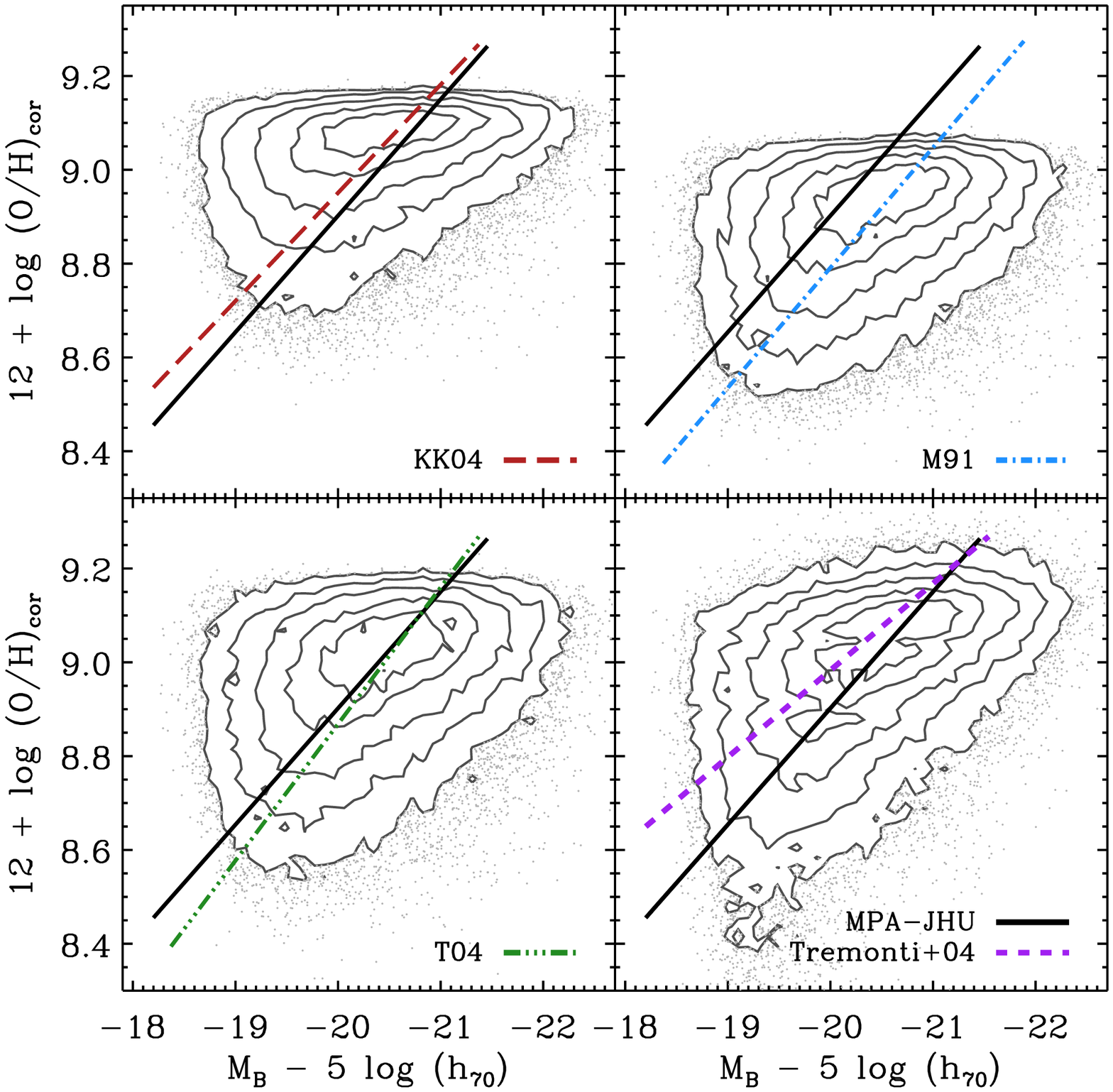}
\caption{$B$-band \lz{} relation from the SDSS based on four different
  abundance calibrations.  The contours enclose $25\%$, $50\%$,
  $75\%$, $90\%$, and $97.5\%$ of the sample, and the small grey
  points represent galaxies lying outside the $97.5\%$ contour level.
  The dashed red, dot-dashed blue, triple-dot-dashed green, and solid
  black lines represent the ordinary least-squares bisector fits to
  the KK04, M91, T04, and MPA-JHU \lz{} relations, respectively.  We
  reproduce the MPA-JHU \lz{} relation in each panel for reference,
  and in the lower-right panel we also show the \lz{} relation derived
  by \citet{tremonti04a} as a short-dashed purple line.  The
  best-fitting slopes range from $-0.231$~dex~mag$^{-1}$ to
  $-0.291$~dex~mag$^{-1}$ (see Table~\ref{table:mzlzlocal}),
  demonstrating that the choice of abundance calibration can have a
  significant effect on the inferred \lz{} relation.  The large
  difference in slope with respect to \citet{tremonti04a}, whose
  sample extended to lower absolute magnitudes, also demonstrates that
  the \lz{} relation is sensitive to the range of luminosities spanned
  by the sample. \label{fig:lzlocal}}
\end{figure}

We turn our attention next to measuring the $B$-band \lz{} relation at
$z\sim0.1$ using our SDSS sample.  Although the optical \lz{} relation
is considerably more sensitive to variations in star formation history
(age or mass-to-light ratio, recent bursts of star formation, etc.)
and dust attenuation than the \mz{} relation \citep[e.g.,][]{jlee04a,
  salzer05a}, many previous studies of the nebular abundances of
star-forming galaxies at both low and high redshift have investigated
the $B$-band \lz{} relation as a proxy for the \mz{} relation.
However, we will see in \S\ref{sec:lzevol} that the optical \lz{}
relation is an ineffective means of quantifying the chemical evolution
of star-forming galaxies due to its greater sensitivity to the effects
of \emph{luminosity} evolution.

In Figure~\ref{fig:lzlocal} we show the $B$-band \lz{} relations from
the SDSS based on the same four abundance calibrations presented in
Figure~\ref{fig:mzlocal}.  As has been known for nearly three decades
(see \S\ref{sec:intro}), star-forming galaxies exhibit a strong
correlation between optical luminosity and gas-phase abundance;
however, the \lz{} relation is clearly sensitive to the adopted
abundance calibration, among other factors.  To quantify the observed
differences we defer to historical precedent and model the \lz{}
relation using a simple linear model, given by

\begin{equation}
12+\log\,(\textrm{O}/\textrm{H}) = c_{0} + c_{1}(\mb+20.5),
\label{eq:lzfit}
\end{equation}

\noindent where $c_{0}$ is the metallicity at $\mb=-20.5$ and $c_{1}$
is the slope in dex~mag$^{-1}$.  We solve equation~(\ref{eq:lzfit})
using an ordinary least-squares bisector fit, which is appropriate
when the scatter in the observed correlation is larger than the
measurement error in either axis \citep{isobe90a}.  We show the
best-fitting \lz{} relations in Figure~\ref{fig:lzlocal} based on the
KK04, M91, T04, and MPA-JHU metallicities as long-dashed red,
dot-dashed blue, triple-dot-dashed green, and solid black lines,
respectively, and list the best-fitting coefficients in
Table~\ref{table:mzlzlocal}.  Again, the statistical uncertainties on
the coefficients are negligible given the sample size and have not
been tabulated.  For comparison, we also show in the lower-right panel
as a short-dashed purple line the $B$-band \lz{} relation determined
by \citet{tremonti04a} using the MPA-JHU abundances of $\sim53,000$
galaxies from the SDSS/DR4.

We find that the slope of the $B$-band \lz{} relation we derive ranges
from $-0.231$~dex~mag$^{-1}$ when using the KK04 calibration, to
$-0.291$~dex~mag$^{-1}$ based on the T04 abundance calibration.
\citet{tremonti04a}, on the other hand, inferred a \lz{} slope of
$-0.185$~dex~mag$^{-1}$ using a sample that extended to much fainter
luminosities.  Other studies studying both low-luminosity dwarfs and
massive disk galaxies have obtained slopes ranging from
$-0.130$~dex~mag$^{-1}$ to $-0.280$~dex~mag$^{-1}$
\citep[e.g.,][]{skillman89a, jlee04a, lama04a, salzer05a,
  moustakas10a}.  We conclude, therefore, that the slope of the \lz{}
relation is sensitive to both the adopted abundance calibration,
\emph{and} the range of luminosities spanned by the parent sample.
Finally, we note that for a given abundance calibration, the $1\sigma$
scatter in the $B$-band \lz{} relation is roughly a factor of two
larger than the scatter in the corresponding \mz{} relation.

\begin{figure*}
\centering
\includegraphics[angle=0,scale=0.65]{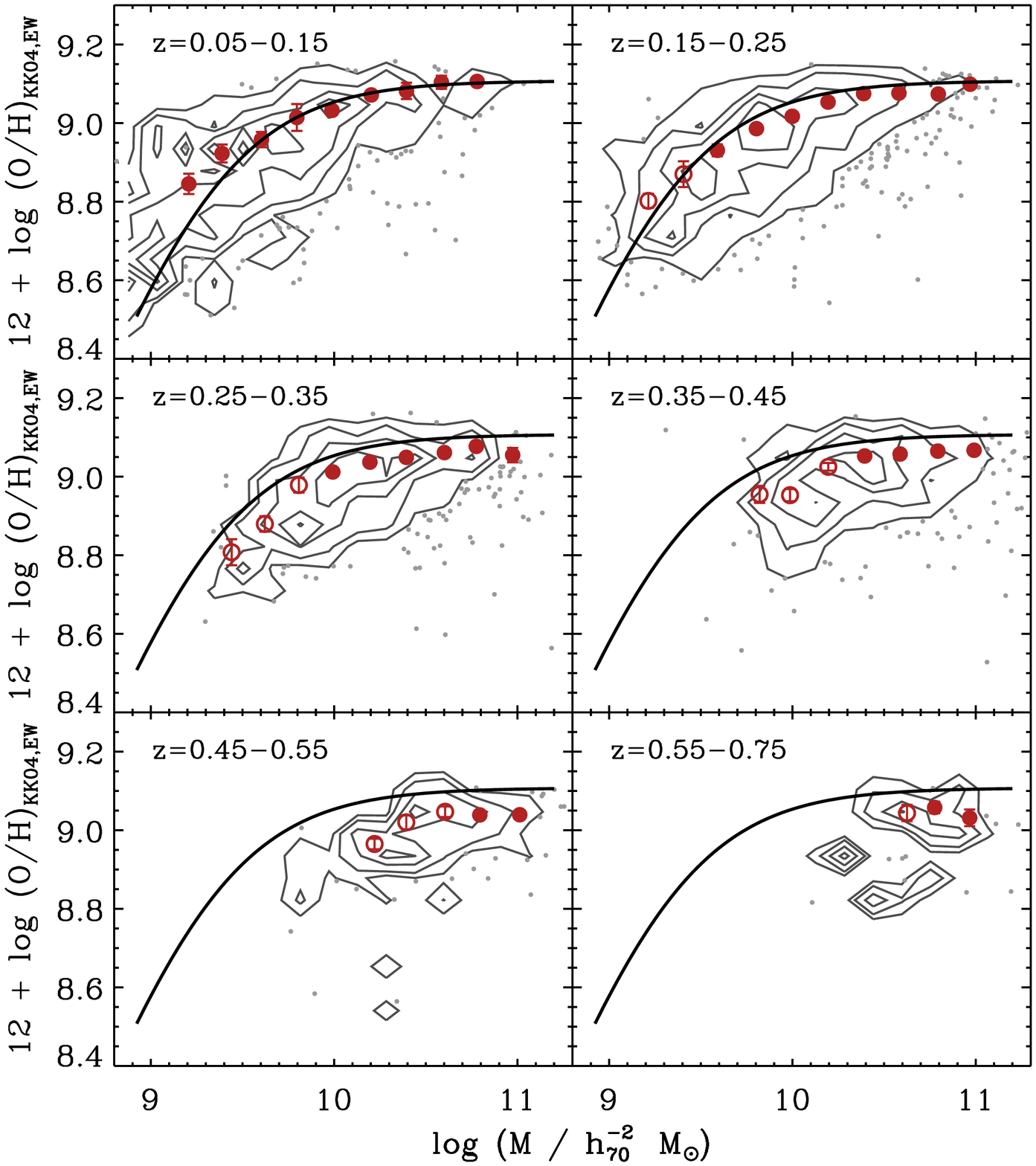}
\caption{AGES \mz{} relation at $z=0.05-0.75$ based on the KK04
  abundance calibration.  For reference, the contours enclose $25\%$,
  $50\%$, $75\%$, and $90\%$ of the galaxies in each redshift
  interval, and the small grey points represent individual galaxies
  that lie outside the $90\%$ contour level.  The filled (open) points
  show the weighted mean metallicity in $0.2$~dex wide bins of stellar
  mass for galaxies above (below) our completeness limit (see
  \S\ref{sec:properties}).  The solid black curve, reproduced in every
  panel for reference, is the local \mz{} relation from the SDSS (see
  \S\ref{sec:mzlocal}).  We find that galaxies experience relatively
  little chemical evolution over this redshift range.  Note that the
  T04 and M91 abundance calibrations yield similar
  results. \label{fig:mzevol}}
\end{figure*}

\subsubsection{Effect of Using \ewpagel{} on the Local \mz{}
  and \lz{} Relations}\label{sec:mzlzews}

The preceding analysis has all been based on abundances derived from
reddening-corrected fluxes; however, to measure the evolution of the
\mz{} relation from AGES (see \S\ref{sec:mzlzevol}) we must rely on
oxygen abundances derived from emission-line EWs (see
\S\ref{sec:ohpagel}).  Therefore, in Table~\ref{table:mzlzlocal} we
also list the coefficients of the SDSS \mz{} and \lz{} relations using
our EW-based abundances.  With respect to the \mz{} relations derived
using reddening-corrected fluxes, we find that the EW-based \mz{}
relations are slightly steeper ($\gamma$ is smaller by $15\%-20\%$),
but that the characteristic stellar masses and characteristic
metallicities agree to within $\lesssim0.01$~dex.  These differences
translate into maximum metallicity differences of $\sim0.02$~dex
around $\sim10^{9.8}$~\msun, and $\lesssim0.01$~dex around
$10^{11}$~\msun.  Similarly, the \lz{} relations derived using either
reddening-corrected fluxes or EWs have nearly identical slopes and
zeropoints.  Therefore, we conclude that using EWs to measure the
evolution of the \mz{} relation should not bias our conclusions.

\subsection{Evolution of the \mz{} and $B$-Band \lz{} Relations Since
  $z=0.75$}\label{sec:mzlzevol}

Building on our measurement of the local \mz{} and \lz{} relations in
the previous section, we now use AGES to measure their evolution since
$z=0.75$.

\subsubsection{Evolution of the \mz{} Relation}\label{sec:mzevol}  

In Figure~\ref{fig:mzevol} we plot the \mz{} relation from AGES in six
redshift bins centered on $z=0.1$, $0.2$, $0.3$, $0.4$, $0.5$, and
$0.65$ (see also Fig.~\ref{fig:zvsmb}).  Here we only show results
using the KK04 abundance calibration, although below we explicitly
explore how the choice of abundance calibration affects the amount of
inferred evolution.  For reference, we overplot the local \mz{}
relation from the SDSS in every panel as a solid black curve (see
Table~\ref{table:mzlzlocal}).  To quantify the observed evolution we
compute the weighted mean metallicity in $0.2$~dex wide bins of
stellar mass, requiring a minimum of $10$ galaxies with well-measured
abundances in each bin.  We weight each galaxy $i$ by $\sigma_{{\rm
    eff},i}^{-2}$, the effective inverse variance, given by

\begin{equation}
\sigma_{{\rm eff},i}^{-2}\equiv w_{i} \sigma_{\log\,(\rm{O}/\rm{H})_{i}}^{-2},
\label{eq:sigmaeff}
\end{equation}

\begin{deluxetable*}{ccccccc}[!h]
\tablecaption{Mean Oxygen Abundance in Bins of Stellar Mass and Redshift\tablenotemark{a}\label{table:oh_bymass}}
\tablewidth{0pt}
\tablehead{
\colhead{Redshift Range} & 
\colhead{$N$\tablenotemark{b}} & 
\colhead{$\langle z\rangle$\tablenotemark{c}} & 
\colhead{$\langle \log\,(\mass/\msun) \rangle$\tablenotemark{d}} & 
\colhead{$\langle12+\log\,(\textrm{O}/\textrm{H})_{\rm KK04}\rangle$} & 
\colhead{$\langle12+\log\,(\textrm{O}/\textrm{H})_{\rm M91}\rangle$} & 
\colhead{$\langle12+\log\,(\textrm{O}/\textrm{H})_{\rm T04}\rangle$} 
}
\startdata
\multicolumn{7}{c}{$10.9<\log\,(\mass/\msun)<11.1$} \\
\cline{1-7}
$0.05-0.20$ & 
$1217$ & 
$0.14$ & 
$10.98$ & 
$9.095\pm0.010$ & 
$8.975\pm0.010$ & 
$9.055\pm0.010$ \\
$0.15-0.25$ & 
$15$ & 
$0.20$ & 
$10.97$ & 
$9.099\pm0.010$ & 
$8.964\pm0.015$ & 
$9.057\pm0.020$ \\
$0.25-0.35$ & 
$27$ & 
$0.32$ & 
$10.97$ & 
$9.055\pm0.018$ & 
$8.896\pm0.027$ & 
$8.976\pm0.029$ \\
$0.35-0.45$ & 
$41$ & 
$0.39$ & 
$10.99$ & 
$9.067\pm0.010$ & 
$8.923\pm0.015$ & 
$8.997\pm0.019$ \\
$0.45-0.55$ & 
$18$ & 
$0.50$ & 
$11.01$ & 
$9.039\pm0.010$ & 
$8.891\pm0.013$ & 
$8.965\pm0.016$ \\
$0.55-0.75$ & 
$16$ & 
$0.62$ & 
$10.97$ & 
$9.032\pm0.021$ & 
$8.884\pm0.033$ & 
$8.952\pm0.031$
\\
\cline{1-7}
\multicolumn{7}{c}{$10.7<\log\,(\mass/\msun)<10.9$} \\
\cline{1-7}
$0.05-0.20$ & 
$4070$ & 
$0.12$ & 
$10.78$ & 
$9.102\pm0.010$ & 
$8.985\pm0.010$ & 
$9.073\pm0.010$ \\
$0.05-0.15$ & 
$16$ & 
$0.12$ & 
$10.78$ & 
$9.106\pm0.010$ & 
$8.977\pm0.016$ & 
$9.072\pm0.019$ \\
$0.15-0.25$ & 
$35$ & 
$0.22$ & 
$10.80$ & 
$9.074\pm0.010$ & 
$8.933\pm0.013$ & 
$9.015\pm0.015$ \\
$0.25-0.35$ & 
$65$ & 
$0.31$ & 
$10.78$ & 
$9.077\pm0.010$ & 
$8.945\pm0.016$ & 
$9.025\pm0.018$ \\
$0.35-0.45$ & 
$64$ & 
$0.39$ & 
$10.79$ & 
$9.065\pm0.010$ & 
$8.921\pm0.015$ & 
$9.004\pm0.015$ \\
$0.45-0.55$ & 
$32$ & 
$0.51$ & 
$10.79$ & 
$9.039\pm0.010$ & 
$8.880\pm0.012$ & 
$8.963\pm0.014$ \\
$0.55-0.75$ & 
$20$ & 
$0.61$ & 
$10.78$ & 
$9.058\pm0.015$ & 
$8.913\pm0.021$ & 
$8.968\pm0.040$
\\
\cline{1-7}
\multicolumn{7}{c}{$10.5<\log\,(\mass/\msun)<10.7$} \\
\cline{1-7}
$0.05-0.20$ & 
$8760$ & 
$0.11$ & 
$10.59$ & 
$9.098\pm0.010$ & 
$8.980\pm0.010$ & 
$9.067\pm0.010$ \\
$0.05-0.15$ & 
$19$ & 
$0.12$ & 
$10.58$ & 
$9.104\pm0.017$ & 
$8.982\pm0.022$ & 
$9.068\pm0.030$ \\
$0.15-0.25$ & 
$75$ & 
$0.20$ & 
$10.58$ & 
$9.076\pm0.010$ & 
$8.938\pm0.017$ & 
$9.019\pm0.018$ \\
$0.25-0.35$ & 
$84$ & 
$0.31$ & 
$10.60$ & 
$9.061\pm0.010$ & 
$8.910\pm0.013$ & 
$8.994\pm0.013$ \\
$0.35-0.45$ & 
$80$ & 
$0.39$ & 
$10.59$ & 
$9.058\pm0.010$ & 
$8.909\pm0.011$ & 
$8.990\pm0.012$ \\
$0.45-0.55$ & 
$31$ & 
$0.50$ & 
$10.61$ & 
$9.047\pm0.013$ & 
$8.898\pm0.020$ & 
$8.968\pm0.018$ \\
$0.55-0.75$ & 
$14$ & 
$0.60$ & 
$10.63$ & 
$9.044\pm0.021$ & 
$8.885\pm0.031$ & 
$8.963\pm0.033$
\\
\cline{1-7}
\multicolumn{7}{c}{$10.3<\log\,(\mass/\msun)<10.5$} \\
\cline{1-7}
$0.05-0.20$ & 
$12679$ & 
$0.10$ & 
$10.40$ & 
$9.091\pm0.010$ & 
$8.963\pm0.010$ & 
$9.047\pm0.010$ \\
$0.05-0.15$ & 
$37$ & 
$0.12$ & 
$10.39$ & 
$9.082\pm0.021$ & 
$8.963\pm0.041$ & 
$9.020\pm0.043$ \\
$0.15-0.25$ & 
$120$ & 
$0.20$ & 
$10.39$ & 
$9.075\pm0.011$ & 
$8.937\pm0.020$ & 
$9.012\pm0.018$ \\
$0.25-0.35$ & 
$114$ & 
$0.30$ & 
$10.39$ & 
$9.049\pm0.010$ & 
$8.899\pm0.012$ & 
$8.977\pm0.010$ \\
$0.35-0.45$ & 
$93$ & 
$0.39$ & 
$10.39$ & 
$9.053\pm0.010$ & 
$8.899\pm0.011$ & 
$8.977\pm0.012$ \\
$0.45-0.55$ & 
$28$ & 
$0.50$ & 
$10.39$ & 
$9.020\pm0.018$ & 
$8.847\pm0.026$ & 
$8.931\pm0.030$
\\
\cline{1-7}
\multicolumn{7}{c}{$10.1<\log\,(\mass/\msun)<10.3$} \\
\cline{1-7}
$0.05-0.20$ & 
$14559$ & 
$0.08$ & 
$10.20$ & 
$9.076\pm0.010$ & 
$8.935\pm0.010$ & 
$9.014\pm0.010$ \\
$0.05-0.15$ & 
$44$ & 
$0.11$ & 
$10.20$ & 
$9.072\pm0.010$ & 
$8.922\pm0.012$ & 
$9.007\pm0.014$ \\
$0.15-0.25$ & 
$161$ & 
$0.21$ & 
$10.19$ & 
$9.053\pm0.010$ & 
$8.896\pm0.011$ & 
$8.974\pm0.012$ \\
$0.25-0.35$ & 
$100$ & 
$0.30$ & 
$10.19$ & 
$9.037\pm0.010$ & 
$8.872\pm0.012$ & 
$8.952\pm0.012$ \\
$0.35-0.45$ & 
$65$ & 
$0.39$ & 
$10.20$ & 
$9.025\pm0.010$ & 
$8.855\pm0.015$ & 
$8.925\pm0.016$ \\
$0.45-0.55$ & 
$12$ & 
$0.50$ & 
$10.22$ & 
$8.965\pm0.014$ & 
$8.784\pm0.016$ & 
$8.854\pm0.019$
\\
\cline{1-7}
\multicolumn{7}{c}{$ 9.9<\log\,(\mass/\msun)<10.1$} \\
\cline{1-7}
$0.05-0.20$ & 
$13523$ & 
$0.08$ & 
$10.00$ & 
$9.053\pm0.010$ & 
$8.891\pm0.010$ & 
$8.967\pm0.010$ \\
$0.05-0.15$ & 
$93$ & 
$0.12$ & 
$ 9.99$ & 
$9.032\pm0.011$ & 
$8.855\pm0.017$ & 
$8.924\pm0.016$ \\
$0.15-0.25$ & 
$151$ & 
$0.21$ & 
$10.00$ & 
$9.017\pm0.012$ & 
$8.837\pm0.018$ & 
$8.903\pm0.021$ \\
$0.25-0.35$ & 
$76$ & 
$0.30$ & 
$ 9.99$ & 
$9.012\pm0.010$ & 
$8.838\pm0.013$ & 
$8.919\pm0.013$ \\
$0.35-0.45$ & 
$37$ & 
$0.38$ & 
$ 9.99$ & 
$8.953\pm0.015$ & 
$8.764\pm0.017$ & 
$8.832\pm0.018$
\\
\cline{1-7}
\multicolumn{7}{c}{$ 9.7<\log\,(\mass/\msun)< 9.9$} \\
\cline{1-7}
$0.05-0.20$ & 
$10378$ & 
$0.07$ & 
$ 9.81$ & 
$9.016\pm0.010$ & 
$8.833\pm0.010$ & 
$8.904\pm0.010$ \\
$0.05-0.15$ & 
$105$ & 
$0.12$ & 
$ 9.80$ & 
$9.014\pm0.034$ & 
$8.827\pm0.035$ & 
$8.894\pm0.038$ \\
$0.15-0.25$ & 
$162$ & 
$0.20$ & 
$ 9.80$ & 
$8.986\pm0.010$ & 
$8.797\pm0.012$ & 
$8.866\pm0.013$ \\
$0.25-0.35$ & 
$63$ & 
$0.29$ & 
$ 9.81$ & 
$8.979\pm0.019$ & 
$8.780\pm0.024$ & 
$8.850\pm0.025$ \\
$0.35-0.45$ & 
$15$ & 
$0.37$ & 
$ 9.82$ & 
$8.955\pm0.022$ & 
$8.766\pm0.019$ & 
$8.833\pm0.018$
\\
\cline{1-7}
\multicolumn{7}{c}{$ 9.5<\log\,(\mass/\msun)< 9.7$} \\
\cline{1-7}
$0.05-0.20$ & 
$6474$ & 
$0.06$ & 
$ 9.61$ & 
$8.961\pm0.010$ & 
$8.763\pm0.010$ & 
$8.828\pm0.010$ \\
$0.05-0.15$ & 
$107$ & 
$0.12$ & 
$ 9.60$ & 
$8.958\pm0.020$ & 
$8.764\pm0.022$ & 
$8.829\pm0.026$ \\
$0.15-0.25$ & 
$118$ & 
$0.20$ & 
$ 9.59$ & 
$8.931\pm0.015$ & 
$8.736\pm0.014$ & 
$8.804\pm0.018$ \\
$0.25-0.35$ & 
$33$ & 
$0.28$ & 
$ 9.62$ & 
$8.880\pm0.020$ & 
$8.691\pm0.020$ & 
$8.747\pm0.023$
\enddata
\tablenotetext{a}{Mean metallicity of galaxies in multiple bins of stellar mass and redshift, based on the KK04, M91, and T04 calibrations.  Note that the metallicities in the first row of each stellar mass interval correspond to our SDSS sample, while the metallicities in the other rows are based on our AGES sample.}
\tablenotetext{b}{Mean number of galaxies in this redshift interval and stellar mass bin.  Note that the number of objects varies slightly depending on the abundance calibration (see, e.g., Table~\ref{table:samples}), so we show here the average number.  We only provide redshift and stellar mass bins with at least ten galaxies.}
\tablenotetext{c}{Mean redshift of the galaxies in this subsample.}
\tablenotetext{d}{Mean stellar mass of the galaxies in this subsample.}
\end{deluxetable*}

\noindent where $w_{i}$ is that object's statistical weight (see
\S\ref{sec:redux}), and $\sigma_{\log\,(\rm{O}/\rm{H})_{i}}$ is the
statistical uncertainty on the metallicity (see
\S\ref{sec:ohsummary}).  We list the measured mean metallicities in
Table~\ref{table:oh_bymass}, and plot the results in
Figure~\ref{fig:mzevol} using filled (open) red points for stellar
masses above (below) our completeness limit (see
\S\ref{sec:properties}).  Examining Figure~\ref{fig:mzevol}, we find a
well-defined \mz{} relation that changes remarkably little over this
redshift range.  Nevertheless, at fixed stellar mass the mean
metallicity of galaxies clearly decreases with increasing redshift,
although it is not yet clear whether the observed evolution varies
with mass.

\begin{figure}
\centering
\includegraphics[angle=0,scale=0.4]{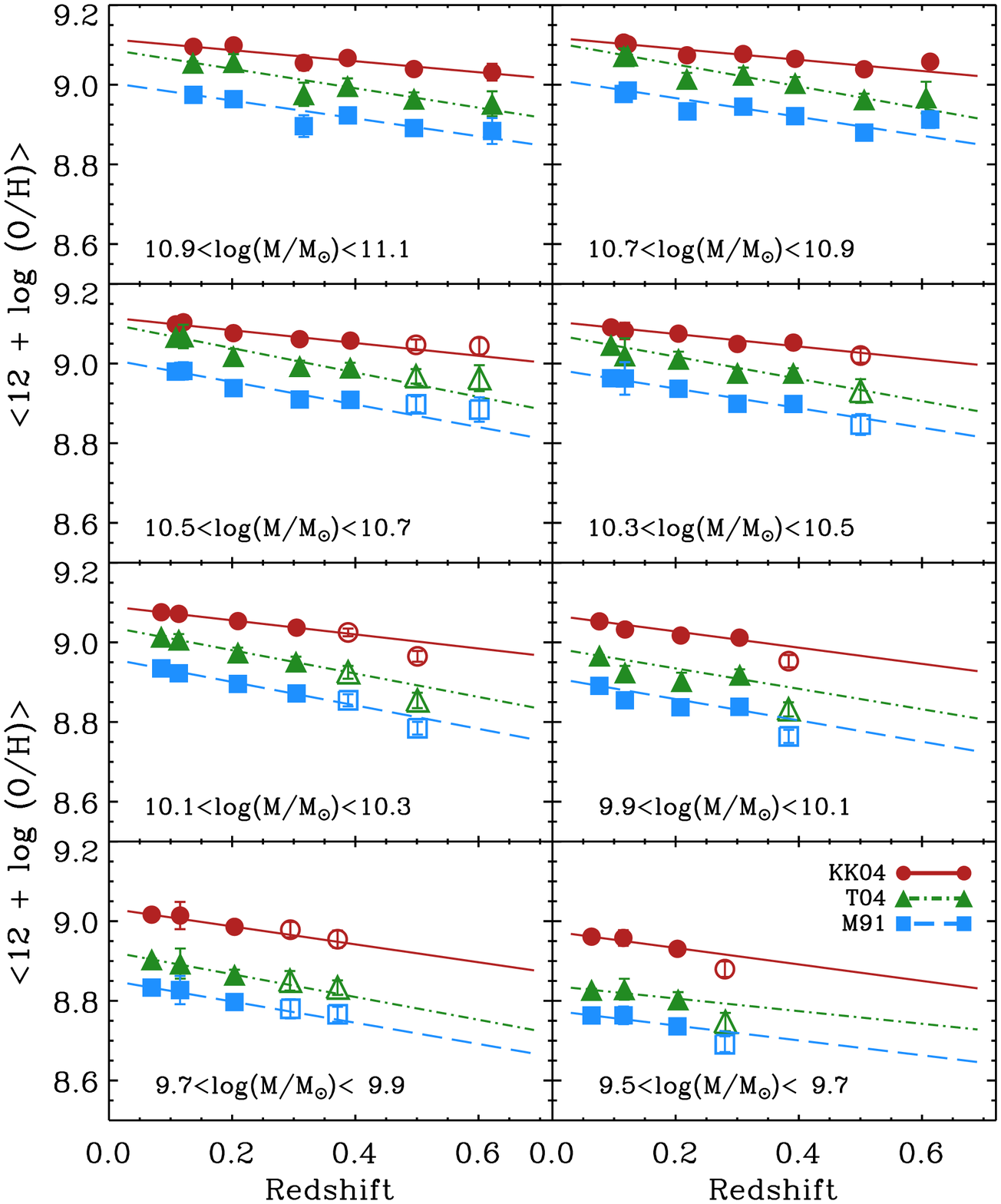}
\caption{Mean metallicity of galaxies at $z=0.05-0.75$ as a function
  of stellar mass and redshift.  The red points, green diamonds, and
  blue squares correspond to the KK04, T04, and M91 abundance
  calibration, respectively.  Open and filled symbols represent
  stellar masses above and below our completeness limit, respectively.
  We model the observed evolution as a linear function of redshift
  given by equation~(\ref{eq:z_vs_oh}), and show the best fitting
  lines using solid red, dot-dashed green, and dashed blue lines for
  each calibration.  We find that galaxies in all eight stellar mass
  bins become progressively more metal-poor with increasing redshift
  at comparable rates.
\label{fig:zvsoh_bymass}}
\end{figure}

We note in passing that the \emph{distribution} of oxygen abundance at
fixed stellar mass also appears to vary with redshift.  For example,
in the two highest redshift bins we see a tail of massive galaxies
with lower-than-average oxygen abundances, $\logoh\approx8.8-9$, that
are not present at lower redshift.  We find a comparable number of
objects based on both the T04 and M91 abundance calibrations.  These
objects may be similar to the population of ``metal-poor'' galaxies
reported by \citet{lilly03a} and \citet{maier05a}.  However, we defer
a more detailed analysis of this population of objects, and of the
full metallicity distributions to a future paper, and focus instead on
the evolution of the mean metallicity with redshift.

To investigate the trends seen in Figure~\ref{fig:mzevol} in more
detail, in Figure~\ref{fig:zvsoh_bymass} we show the mean metallicity
$\langle 12+\log\,(\textrm{O}/\textrm{H})\rangle$ versus redshift for
both our SDSS and AGES samples in eight bins of stellar mass spanning
$10^{9.5}-10^{11.1}$~\msun.  We show results based on the KK04, T04,
and M91 calibrations using red points, green triangles, and blue
squares, respectively, and render stellar mass bins above and below
our completeness limit using filled and open symbols.  We model the
observed evolution, separately for each calibration and stellar mass
interval, as a linear function of redshift given by

\begin{equation}
\langle 12+\log\,(\textrm{O}/\textrm{H})\rangle = \langle
12+\log\,(\textrm{O}/\textrm{H})\rangle_{z=0.1} +
\frac{{\mathrm d}[\log\,(\textrm{O}/\textrm{H})]}{{\mathrm
    d}z}\,(z-0.1),
\label{eq:z_vs_oh}
\end{equation}

\begin{figure}[b]
\centering
\includegraphics[angle=0,scale=0.4]{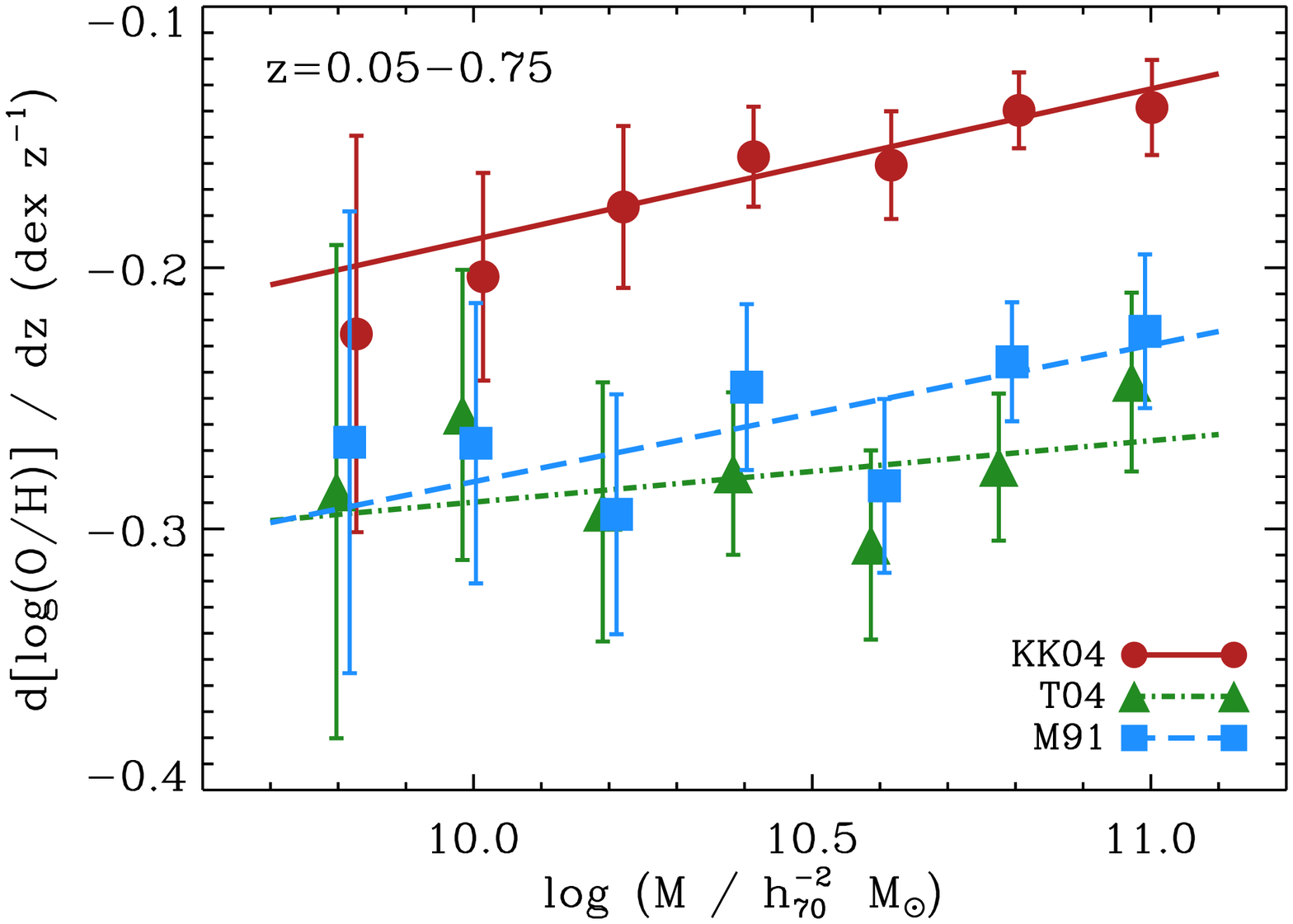}
\caption{Redshift-dependent rate of change in the mean oxygen
  abundance for star-forming galaxies at $z=0.05-0.75$ vs.~stellar
  mass.  The red points, green diamonds, and blue squares correspond
  to the KK04, T04, and M91 calibrations, respectively, while the
  corresponding solid red, dot-dashed green, and dashed blue lines
  show the best-fitting linear relation (see
  Table~\ref{table:oh_bymass_coeff}).  The KK04 calibration yields the
  smaller \emph{absolute} rate of metallicity evolution,
  $-0.160$~dex~$z^{-1}$ at $\mass=10^{10.5}~\msun$, but the steepest
  stellar mass dependence.  By contrast, the T04 and M91 calibrations
  indicate a large rate of metallicity evolution, $\sim-0.26$~dex per
  unit redshift, and no significant dependence of this rate on stellar
  mass. \label{fig:massslope}}
\end{figure}

\noindent where
$\langle12+\log\,(\textrm{O}/\textrm{H})\rangle_{z=0.1}$ is the mean
metallicity at $z=0.1$ and ${\mathrm
  d}[\log\,(\textrm{O}/\textrm{H})]/{\mathrm d}z$ is the logarithmic
rate of metallicity evolution.  We tabulate the best-fitting linear
coefficients in Table~\ref{table:oh_bymass_coeff}. 

Figure~\ref{fig:zvsoh_bymass} conveys several interesting results.
First, we find that the KK04, T04, and M91 abundance calibrations
yield mean metallicities that differ in an absolute sense by
significant amounts (no vertical offsets have been applied to the
measurements).  On average, the KK04-based metallicities are
$0.093$~dex ($24\%$) and $0.16$~dex ($45\%$) higher than the
metallicities derived using the T04 and M91 calibration, respectively.
These systematic differences likely originate from the different
assumptions and ingredients underlying the photoionization models used
to calibrate these strong-line methods \citep[e.g.,][]{kewley02a}.
Understanding the origin of these discrepancies is beyond the scope of
this paper, although these results highlight that even within the
class of theoretical abundance calibrations there are non-negligible
systematic differences (see, e.g., the discussion in
\citealt{moustakas10a}).

\begin{deluxetable*}{cccccc}[!h]
\tablecaption{Linear Evolution of the Mean Metallicity of Galaxies at $z=0.05-0.75$\tablenotemark{a}\label{table:oh_bymass_coeff}}
\tablewidth{0pt}
\tablehead{
\colhead{} & 
\colhead{$\langle12+\log\,(\textrm{O}/\textrm{H})\rangle_{z=0.1}$} & 
\colhead{${\mathrm d}[\log\,(\textrm{O}/\textrm{H})]/{\mathrm d}z$} & 
\colhead{} & 
\colhead{$\langle12+\log\,(\textrm{O}/\textrm{H})\rangle_{z=0.1}$} & 
\colhead{${\mathrm d}[\log\,(\textrm{O}/\textrm{H})]/{\mathrm d}z$} \\
\colhead{Calibration} & 
\colhead{(dex)} & 
\colhead{(dex $z^{-1}$)} & 
\colhead{} & 
\colhead{(dex)} & 
\colhead{(dex $z^{-1}$)} 
}
\startdata
\cline{1-6}
\colhead{} &  \multicolumn{2}{c}{$10.9<\log\,(\mass/\msun)<11.1$} & 
\colhead{} &  \multicolumn{2}{c}{$10.7<\log\,(\mass/\msun)<10.9$} \\
\cline{2-3}
\cline{5-6}
KK04 & 
$9.101\pm0.01$ & $ -0.139\pm  0.018$ & & $9.104\pm0.01$ & $ -0.140\pm  0.015$  \\
M91 & 
$8.983\pm0.01$ & $ -0.224\pm  0.029$ & & $8.990\pm0.01$ & $ -0.236\pm  0.023$  \\
T04 & 
$9.064\pm0.01$ & $ -0.244\pm  0.034$ & & $9.079\pm0.01$ & $ -0.276\pm  0.028$  \\
\cline{1-6}
\colhead{} &  \multicolumn{2}{c}{$10.5<\log\,(\mass/\msun)<10.7$} & 
\colhead{} &  \multicolumn{2}{c}{$10.3<\log\,(\mass/\msun)<10.5$} \\
\cline{2-3}
\cline{5-6}
KK04 & 
$9.100\pm0.01$ & $ -0.161\pm  0.021$ & & $9.090\pm0.01$ & $ -0.157\pm  0.019$  \\
M91 & 
$8.982\pm0.01$ & $ -0.284\pm  0.033$ & & $8.962\pm0.01$ & $ -0.246\pm  0.032$  \\
T04 & 
$9.069\pm0.01$ & $ -0.306\pm  0.036$ & & $9.045\pm0.01$ & $ -0.279\pm  0.031$  \\
\cline{1-6}
\colhead{} &  \multicolumn{2}{c}{$10.1<\log\,(\mass/\msun)<10.3$} & 
\colhead{} &  \multicolumn{2}{c}{$ 9.9<\log\,(\mass/\msun)<10.1$} \\
\cline{2-3}
\cline{5-6}
KK04 & 
$9.073\pm0.01$ & $ -0.177\pm  0.031$ & & $9.048\pm0.01$ & $ -0.203\pm  0.040$  \\
M91 & 
$8.930\pm0.01$ & $ -0.294\pm  0.046$ & & $8.884\pm0.01$ & $ -0.267\pm  0.054$  \\
T04 & 
$9.010\pm0.01$ & $ -0.294\pm  0.050$ & & $8.960\pm0.01$ & $ -0.256\pm  0.056$  \\
\cline{1-6}
\colhead{} &  \multicolumn{2}{c}{$ 9.7<\log\,(\mass/\msun)< 9.9$} & 
\colhead{} &  \multicolumn{2}{c}{$ 9.5<\log\,(\mass/\msun)< 9.7$} \\
\cline{2-3}
\cline{5-6}
KK04 & 
$9.010\pm0.01$ & $ -0.225\pm  0.076$ & & $8.954\pm0.01$ & $ -0.207\pm  0.107$  \\
M91 & 
$8.825\pm0.01$ & $ -0.267\pm  0.088$ & & $8.757\pm0.01$ & $ -0.187\pm  0.101$  \\
T04 & 
$8.895\pm0.01$ & $ -0.286\pm  0.094$ & & $8.822\pm0.01$ & $ -0.159\pm  0.128$ 
\enddata
\tablenotetext{a}{We model the measured change in the mean metallicity as a linear function of redshift given by: $\langle 12+\log\,(\textrm{O}/\textrm{H})\rangle = \langle 12+\log\,(\textrm{O}/\textrm{H})\rangle_{z=0.1} + {\mathrm d}[\log\,(\textrm{O}/\textrm{H})]/{\mathrm d}z\times(z-0.1)$, where $\langle12+\log\,(\textrm{O}/\textrm{H})\rangle_{z=0.1}$ is the mean metallicity at $z=0.1$ and ${\mathrm d}[\log\,(\textrm{O}/\textrm{H})]/{\mathrm d}z$ is the logarithmic rate of metallicity evolution.}
\end{deluxetable*}

Despite the significant differences in the \emph{absolute} abundances
of the galaxies in our sample, all three calibrations reveal that
star-forming galaxies of a given stellar mass become increasingly more
metal poor with increasing redshift.  The \emph{rate} of metallicity
evolution, however, clearly depends on the adopted abundance
calibration (see Table~\ref{table:oh_bymass_coeff}).  Specifically, we
find that ${\rm d}[\log\,(\textrm{O}/\textrm{H})]/{\rm d}z$ based on
the KK04 calibration is systematically shallower (by a factor of
$\sim1.7$ at $>3\sigma$ significance) than the slopes measured using
the other two calibrations.  We attribute this difference to the
relatively flat \mz{} relation for massive galaxies implied by the
KK04 calibration (see Fig.~\ref{fig:mzlocal}, top-left panel), which
has the effect of reducing the rate of metallicity evolution we
measure compared to the other two calibrations.  From these data alone
it is impossible to ascertain which of these three strong-line
calibrations is ``right.''  The inevitable conclusion, therefore, is
that the uncertainties in both the \emph{absolute} metallicites and
the \emph{relative} rates of metallicity evolution we measure are
dominated by systematic errors in the strong-line abundance
calibrations.

With the preceding discussion in mind, we now explore in more detail
whether the chemical enrichment rate depends on stellar mass.  In
Figure~\ref{fig:massslope} we plot ${\rm
  d}[\log\,(\textrm{O}/\textrm{H})]/{\rm d}z$ versus stellar mass for
galaxies with $\mass=10^{9.7}-10^{11.1}$~\msun{} using the KK04 ({\em
  red points}), T04 ({\em green triangles}), and M91 ({\em blue
  squares}) abundance calibration.  We exclude our lowest stellar mass
bin in the following because it is most affected by the limited
redshift coverage of AGES in this stellar mass range.  In each stellar
mass interval we use the average mass of all the galaxies in that bin
(i.e., across all redshifts), although we have applied small
horizontal offsets to the points in this figure for clarity.

To quantify the observed trends we fit a linear model of the form

\begin{equation}
\frac{{\mathrm d}[\log\,(\textrm{O}/\textrm{H})]}{{\mathrm d}z} =
a_{0} + a_{1} \log{\left(\frac{\mass}{10^{10.5}~\msun}\right)},
\label{eq:dlogohdz_bymass}
\end{equation}

\noindent where $a_{0}$ is the rate of evolution in dex per unit
redshift, and $a_{1}$ characterizes the power-law dependence of this
rate on stellar mass.  We list the best-fitting coefficients and
uncertainties for the individual calibrations in
Table~\ref{table:dlogohdz}.

Figure~\ref{fig:massslope} nicely encapsulates many of the principal
results of this section, and of this paper.  First, the factor of
$\sim1.7$ shallower rate of metallicity evolution implied by the KK04
calibration compared with the T04 and M91 calibrations is clearly
apparent in this figure.  Specifically, the KK04 calibration indicates
a metal enrichment rate of $-0.160\pm0.009$~dex~$z^{-1}$ for
star-forming galaxies with $\mass=10^{10.5}$~\msun, while the T04 and
M91 calibrations indicate more rapid evolution, $-0.278\pm0.015$ and
$-0.256\pm0.014$~dex~$z^{-1}$, respectively.  The KK04 calibration
also indicates a steeper dependence of the rate of metallicity
evolution on stellar mass, whereas the T04 and M91 calibrations show
no statistically significant ($\lesssim1\sigma$) dependence on mass
for galaxies with $\mass=10^{9.8}-10^{11}$~\msun.  We emphasize,
however, that even the stellar mass dependence implied by the KK04
calibration, $0.058\pm0.029$, is only marginally statistically
significant ($2\sigma$).  We conclude, therefore, that there is no
compelling evidence that the \mz{} relation evolves differentially,
that is, in a mass-dependent way, over this range of stellar masses
and redshifts.

We conclude this section by examining whether we can synthesize the
preceding results in terms of evolution in the physical parameters
\ohstar{} and \mstar{} given by equation~(\ref{eq:mzclosedbox}) (see
also Appendix~\ref{appendix:mzform}).  Recall that \ohstar{} is the
asymptotic metallicity of the \mz{} relation, and \mstar{} represents
the stellar mass at which the relation begins to turn over.  We begin
with the simplest possible model by fixing \mstar{} and $\gamma$ at
their local values and allowing \ohstar{} to vary linearly with
redshift as

\begin{equation}
\ohstar_{z} = \ohstar_{z=0.1} + \mathcal{P}(z-0.1),
\label{eq:ohstar}
\end{equation}

\noindent where $\ohstar_{z=0.1}$ is the characteristic metallicity at
$z=0.1$ (see Table~\ref{table:mzlzlocal}), and $\mathcal{P}$ is the
rate of chemical evolution in dex per unit redshift.  Solving for the
maximum likelihood value of $\mathcal{P}$ using {\sc mpfit}\footnote{A
  Levenberg-Marquardt least-squares minimization routine available at
  \url{http://cow.physics.wisc.edu/$\sim$craigm/idl}.}, we obtain
$\mathcal{P} = -0.137\pm0.017$, $-0.258\pm0.034$, and
$-0.250\pm0.044$~dex~z$^{-1}$ using the KK04, T04, and M91
calibration, respectively.  These results are statistically consistent
with the rate of chemical evolution given in
Table~\ref{table:dlogohdz} (ignoring the weak dependence on stellar
mass), and therefore provide a convenient way of evaluating the \mz{}
relation for star-forming galaxies at any redshift between $z=0.05$
and $z=0.75$.

\begin{figure}
\centering
\includegraphics[angle=0,scale=0.4]{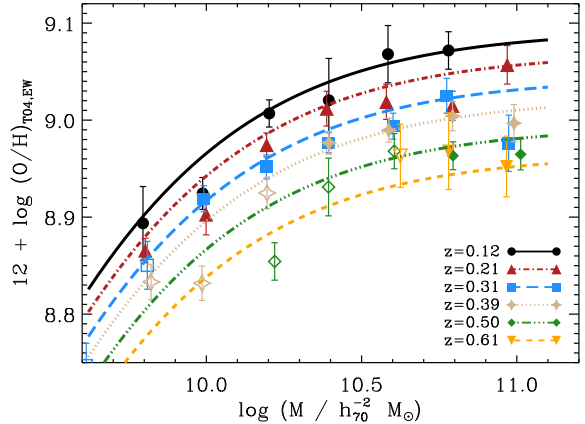}
\caption{Comparison of the T04-based mean metallicities of AGES
  galaxies in six redshift bins between $z=0.1$ and $z=0.6$ to the
  empirical model proposed in \S\ref{sec:mzevol}.  The filled (open)
  symbols represent stellar mass bins above (below) our completeness
  limit in each redshift bin, and to best match the plotted
  measurements we have evaluated the model at the median redshift of
  the galaxies in each sample, as indicated in the legend.  We find
  that this simple model, in which the \mz{} relation simply shifts
  toward lower metallicity without changing its shape, reproduces the
  observations well.  \label{fig:mzevol_model}}  
\end{figure}

In Figure~\ref{fig:mzevol_model} we show the evolution of the
T04-based \mz{} relation predicted by this empirical model, along with
the mean metallicities from AGES in our six redshift bins.  As before,
the filled and open symbols represent stellar mass bins above and
below our completeness limit, respectively.  To optimally match the
measurements plotted, we evaluate the model at the median redshift of
the galaxies in each redshift bin, as indicated in the legend.  We
find that this simple model, which consists of a linear decrease in
the oxygen abundance of star-forming galaxies as a function of
redshift and an invariant \mz{} relation \emph{shape}, reproduces the
observed data reasonably well.

We also attempt to fit the data with a more complex model, in which
the characteristic stellar mass \mstar{} also evolves linearly with
redshift as

\begin{equation}
\log\,(\mstar_{z}/\msun) = \log\,(\mstar_{z=0.1}/\msun) +
\mathcal{R}(z-0.1).  
\label{eq:massevol}
\end{equation}

\noindent Combining this relation with
equation~(\ref{eq:mzclosedbox}), the redshift-dependent \mz{} relation
becomes

\begin{eqnarray}
12 + \log\,(\textrm{O}/\textrm{H})_{z} & = &
12+\log\,(\textrm{O}/\textrm{H})^{\ast}_{z=0.1} + \mathcal{P}(z-0.1) \nonumber \\
- & \log\, & \left[1 + \left(\frac{\mstar\times10^{R(z-0.1)}}{10^{9}\,
    \msun}\right)^{\gamma}\right], 
\label{eq:mzevol}
\end{eqnarray}

\noindent where once again we fix $\gamma$ at its local value.
Finding the maximum likelihood values of the free parameters, we find
that the data are only marginally better fit, but not by a
statistically meaningful amount, compared to the simpler model in
which $\mathcal{R}=0$.  We conclude that our metallicity measurements
from the SDSS and AGES are statistically consistent with an \mz{}
relation that shifts toward lower metallicity with decreasing
redshift, but whose shape does not evolve.  

\begin{deluxetable}{ccc}
\tablecaption{Stellar Mass Dependence of the Rate of Metallicity Evolution\tablenotemark{a}\label{table:dlogohdz}}
\tablewidth{0pt}
\tablehead{
\colhead{Calibration} & 
\colhead{$a_{0}$} & 
\colhead{$a_{1}$} 
}
\startdata
KK04 &  $-0.160\pm 0.009$ &  $0.058\pm0.029$ \\ 
T04 &  $-0.278\pm 0.015$ &  $0.024\pm0.046$ \\ 
M91 &  $-0.256\pm 0.014$ &  $0.052\pm0.042$
\enddata
\tablenotetext{a}{Mass-dependent rate of metallicity evolution given by ${\mathrm d}[\log\,(\textrm{O}/\textrm{H})]/{\mathrm d}z = a_{0} + a_{1} \log\,(\mass/10^{10.5}~\msun)$, where the units of $a_{0}$ are dex per unit redshift.}
\end{deluxetable}

\subsubsection{Evolution of the $B$-Band \lz{}
  Relation}\label{sec:lzevol}

\begin{figure*}
\centering
\includegraphics[angle=0,scale=0.75]{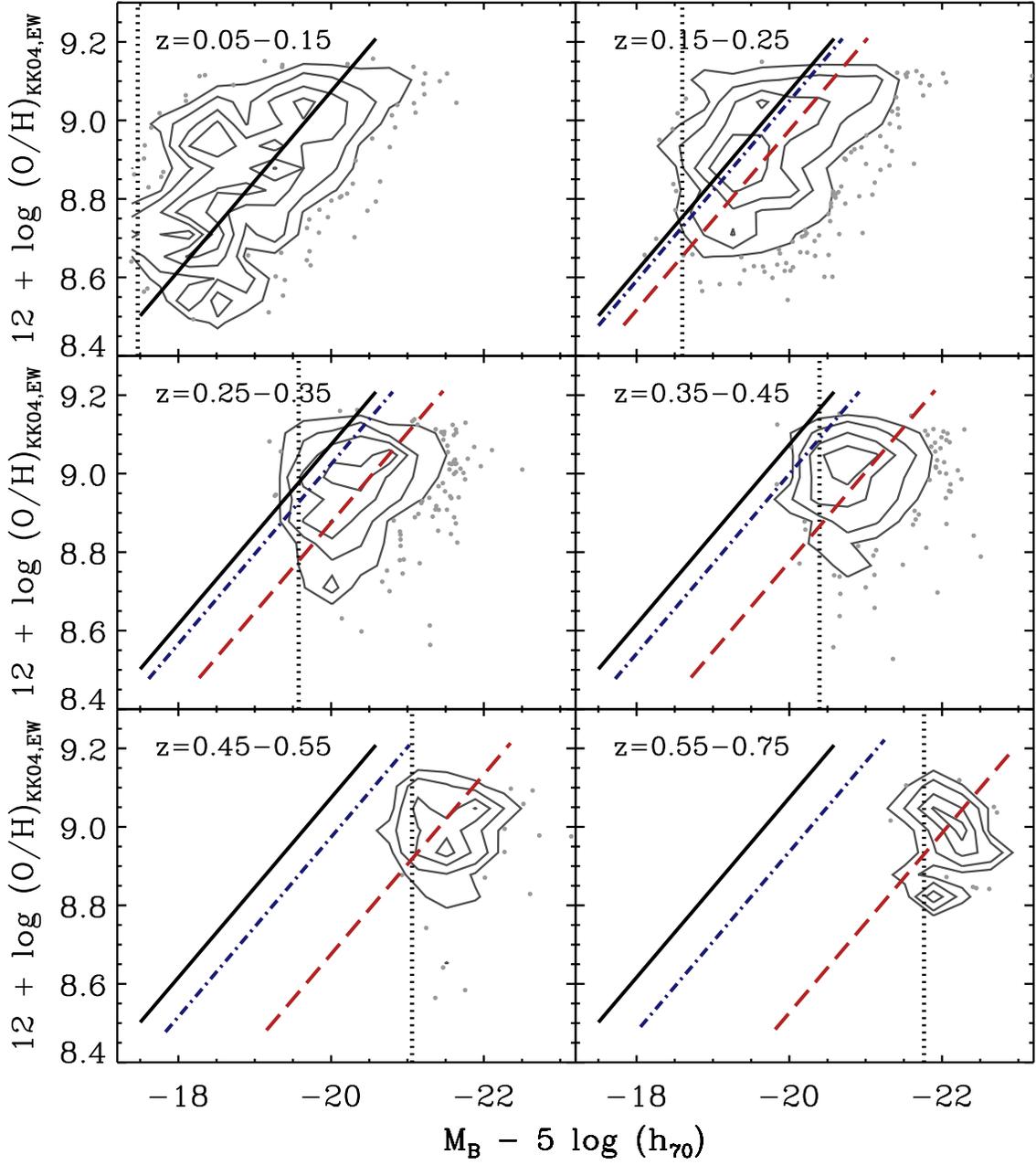}
\caption{Evolution of the $B$-band \lz{} relation at $z=0.05-0.75$
  based on the KK04 abundance calibration.  For reference, the
  contours enclose $25\%$, $50\%$, $75\%$, and $90\%$ of the galaxies
  in each redshift interval, and the small grey points represent
  individual galaxies that lie outside the $90\%$ contour level.  The
  solid black line reproduced in every panel shows the local \lz{}
  relation derived in \S\ref{sec:lzlocal}, and the vertical dashed
  lines show the completeness limits of AGES in each redshift interval
  (see \S\ref{sec:properties}).  Taken at face value, the observations
  suggest $-1.00$~dex~$z^{-1}$ of metallicity evolution at fixed
  luminosity ({\em dashed red line}), which significantly exceeds the
  much smaller rate of metallicity evolution inferred from the \mz{}
  relation ({\em dot-dashed blue line}).  We attribute the discrepancy
  to significant brightening in the $B$-band luminosities of
  star-forming galaxies over this redshift range, and conclude that
  the $B$-band \lz{} relation does not provide an effective means of
  constraining the chemical enrichment histories of
  galaxies.  \label{fig:lzevol}}
\end{figure*}

We turn next to the evolution of the $B$-band \lz{} relation at
intermediate redshift.  In Figure~\ref{fig:lzevol} we plot the
relationship between \mb{} and oxygen abundance in the same redshift
bins used in the previous section.  For clarity, we again only show
the \lz{} relation using the KK04 calibration, but we present results
from the other two calibrations below.  The solid black line
reproduced in every panel is the SDSS \lz{} relation derived in
\S\ref{sec:lzlocal}, and the vertical dotted lines show the absolute
magnitude limits of the survey in each redshift bin (see
Table~\ref{table:limits}).

Taken at face value, Figure~\ref{fig:lzevol} suggests that galaxies at
$z=0.05-0.75$ have experienced a significant amount of metallicity
evolution, luminosity evolution, or both.  To quantify the observed
trends we modify the local \lz{} relation given by
equation~(\ref{eq:lzfit}) to allow the characteristic metallicity at
$z=0.1$, $c_{0, z=0.1}$, to change linearly with redshift according to
$c_{0}(z) = c_{0, z=0.1} + \mathcal{S}(z-0.1)$, where the rate of
metallicity evolution, $\mathcal{S}$, is in units of dex~$z^{-1}$.  We
also allow the ensemble of star-forming galaxies in AGES to brighten
as $M_{B}(z) = M_{B,z=0.1} + Q_{B}(z-0.1)$, where $Q_{B}$ is the rate
of luminosity evolution in mag~$z^{-1}$.  Combining these relations
with equation~(\ref{eq:lzfit}), the $B$-band \lz{} relation at a given
redshift becomes

\begin{eqnarray}
12+\log\,(\textrm{O}/\textrm{H})_{z} & = &
c_{0, z=0.1} + c_{1}(\mb+20.5) + \nonumber \\
& & (\mathcal{S}+c_{1}\mathcal{Q_{B}})\,(z-0.1).
\label{eq:lzevol}
\end{eqnarray}

\noindent Note that we have implicitly assumed that the \emph{slope}
of the \lz{} relation, $c_{1}$, does not change with redshift, which
is likely an oversimplification; unfortunately, AGES does not extend
far enough down the luminosity function to test this assumption.  

Momentarily ignoring luminosity evolution (i.e., setting $Q_{B}=0$),
we find $\mathcal{S} = -1.00$, $-1.32$, and $-1.11$~dex~z$^{-1}$ using
the KK04, T04, and M91 calibration, respectively, with negligible
statistical uncertainties.  We show the results of this evolutionary
model as a dashed red line in Figure~\ref{fig:lzevol} for the KK04
calibration.  However, these results are clearly at odds with the
$\sim-0.2$~dex~$z^{-1}$ of evolution inferred from our analysis of the
\mz{} relation in \S\ref{sec:mzevol}, which we plot in
Figure~\ref{fig:lzevol} as a dot-dashed blue line for comparison.  We
conclude, therefore, that the evolution of the $B$-band \lz{} relation
\emph{requires} star-forming galaxies in AGES to have been brighter in
the past, that is $Q_{B}<0$.  Indeed, setting
$\mathcal{P}=\mathcal{S}$ and solving for $Q_{B}$, we obtain
$Q_{B}=-3.5\pm0.15$~mag~$z^{-1}$ averaged over all three calibrations.

Measurements of the $B$-band luminosity function for blue,
star-forming galaxies at intermediate redshift provide an independent
estimate of $Q_{B}$.  For example, \citet{blanton06a} report a
brightening of $\sim1.0$~mag for blue galaxies since $z=1$ using data
from the SDSS and DEEP2 surveys; \citet{faber07a} find a luminosity
evolution rate $\sim1.4$~mag~$z^{-1}$ over the same redshift interval
from an analysis of COMBO-17 and DEEP2 observations; and
\citet{cool12a} find $Q_{B}=-1.3\pm0.2$~mag~$z^{-1}$ based on data
from the SDSS and AGES.  Although we find a considerably larger rate
of luminosity evolution, recall that the luminosity function studies
cited above measure the brightening at the `knee' of the luminosity
function, $M_{B}^{\ast}$, whereas our sample is dominated galaxies
with $\mb\ll M_{B}^{\ast}$ at higher redshift.  Moreover, our
assumption that the \lz{} relation is linear is clearly not strictly
appropriate in light of our analysis of the \mz{} relation in
\S\ref{sec:mzevol}.  The principal conclusion from the preceding
analysis is that the $B$-band \lz{} relation does \emph{not} provide a
robust means of quantifying the evolution in the chemical abundances
of star-forming galaxies at higher redshift, as has been previously
emphasized by \citet{tremonti04a}, \citet{salzer05a},
\citet{zahid11a}, and others.

\subsection{Comparison with Previous Results}\label{sec:compare} 

A detailed comparison of our results with previous studies poses
numerous challenges.  Among the issues to consider are the wide
variations in sample selection, sample size, and survey area (i.e.,
cosmic variance), differences in the methods used to estimate oxygen
abundances (i.e., choice of abundance diagnostics), and differences in
the treatment of systematic effects like dust attenuation, stellar
absorption, and AGN contamination.  Moreover, most previous abundance
studies have measured the evolution of the $B$-band \lz{} relation,
which we have seen is not a robust means of quantifying the chemical
evolution histories of star-forming galaxies.  And finally, different
studies have relied on disparate measurements of the local \mz{} and
\lz{} relations, which can also influence the amount of chemical
evolution inferred.

With these issues in mind, we focus our comparison on three recent
measurements of the \mz{} relation at intermediate redshift by
\citet[hereafter CB08]{cowie08a}, \citet[hereafter L09]{lama09a}, and
\citet[hereafter ZKB11]{zahid11a}.  We select these studies because
they are based on a large number of homogenously selected galaxies
over a relatively wide area, and they make use of state-of-the-art
techniques for measuring stellar mass from broadband photometry.  CB08
derived gas-phase oxygen abundances for roughly $200$ galaxies at
$z=0.05-0.9$ in the $145$~arcmin$^{2}$ GOODS-N \citep{giavalisco04a}
field, L09 measured the \mz{} relation at $z=0.0-0.9$ based on optical
spectroscopy of $\sim3000$ galaxies in the $0.61$~deg$^{2}$ VVDS-DEEP
and $6.1$~deg$^{2}$ VVDS-WIDE survey fields \citep{lefevre05a,
  garilli08a}, and ZKB11 measured the metallicities of $1350$ galaxies
at $z=0.75-0.82$ selected from the $3.5$~deg$^{2}$ DEEP2 survey
\citep{davis03a}.  For simplicity, we concentrate here on a single
redshift interval, $z=0.6-0.8$, for which all three studies have
measured the \mz{} relation.  Although this redshift range is the
highest redshift bin probed by AGES, in which our sample spans a
relatively limited range of stellar mass, the evolutionary model of
the \mz{} relation we derive in \S\ref{sec:mzevol} is nevertheless
constrained by our larger sample of lower redshift galaxies.

\begin{figure}
\centering
\includegraphics[angle=0,scale=0.4]{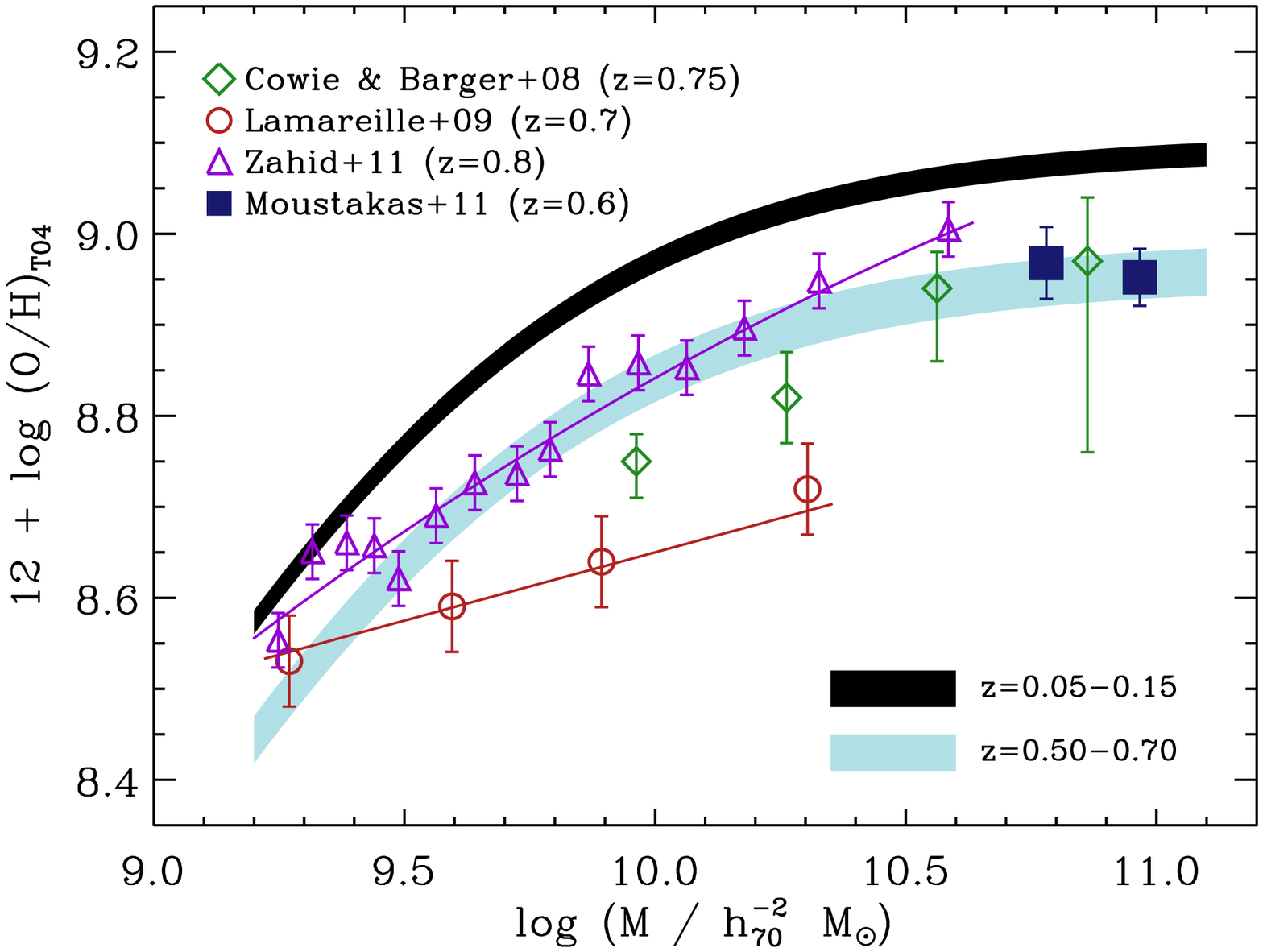}
\caption{Comparison of the \mz{} relation we infer at intermediate
  redshift from AGES ({\em filled blue squares}) to previous
  measurements by \citet{cowie08a} ({\em open green diamonds}),
  \citet{lama09a} ({\em open red circles}), and \citet{zahid11a} ({\em
    open purple triangles}), all converted to the T04 abundance scale
  and the \citet{chabrier03a} IMF.  For comparison, we show the \mz{}
  relation at $z\sim0.1$ ({\em shaded black region}) and at $z\sim0.6$
  ({\em shaded light blue region}) based on the empirical evolutionary
  model derived in \S\ref{sec:mzevol}.  We find very good agreement
  between our empirical model, which is constrained at low stellar
  mass by our lower-redshift observations, and the detailed
  measurements by \citet{zahid11a}, and with \citet{cowie08a} above
  $\sim10^{10.5}$~\msun.  \label{fig:mzlit}}
\end{figure}

In Figure~\ref{fig:mzlit} we plot stellar mass versus oxygen abundance
for galaxies at $z\sim0.7$.  We plot the mean metallicity measurements
from CB08, L09, ZKB11, and AGES using green diamonds, red circles,
purple triangles, and filled dark blue squares, respectively.  For
consistency across all the studies we show metallicities derived using
the T04 calibration of \pagel.  CB08 and L09 derive their abundances
using this calibration, and we have applied the appropriate conversion
formula from \citet{kewley08a} to adjust the measurements from ZKB11.
Finally, we subtract $0.25$~dex from the CB08 stellar masses to
convert them to the \citet{chabrier03a} IMF.  We also show in
Figure~\ref{fig:mzlit} the \mz{} relation at $z=0.05-0.15$ as a shaded
black region, and the \mz{} relation at $z=0.50-0.70$ from AGES as a
light blue shaded region.

Examining Figure~\ref{fig:mzlit}, we find generally very good
agreement between the intermediate-redshift \mz{} relation we infer
from AGES, and the \mz{} relations measured by CB08 and ZKB11.  The
agreement with ZKB11 between $\sim10^{9.5}-10^{10.5}$~\msun{} is
especially striking since it represents an extrapolation of the \mz{}
relation we measure at lower redshift (AGES does not probe these
stellar masses at this redshift).  In particular, our conclusion that
the evolution of the \mz{} relation does not depend significantly on
stellar mass (see \S\ref{sec:mzevol}) is strengthened by this
comparison.  The evolution of the \mz{} relation since $z\sim0.7$ is
consistent with a progressive increase in metallicity, with no
significant change in shape, contrary to the conclusions of several
previous studies \citep[e.g.,][]{savaglio05a, lama09a}.  Our
intermediate-redshift \mz{} relation is also reasonably consistent
with the measurements by CB08, particularly at the massive end.
Between $\sim10^{10}-10^{10.3}$~\msun, however, CB08 measure slightly
lower mean metallicities, but given the uncertainties in this
comparison the differences are probably not significant.  On the other
hand, L09 measure mean abundances that are $\sim50\%$ lower at
$\sim10^{10.3}$~\msun{} and a considerably shallower \mz{} relation.
The origin of this discrepancy is not clear, although the relatively
low spectral resolution ($R\approx230$) of the VVDS spectra analyzed
by L09 may introduce unintended selection effects related to a
preferential loss of massive, metal-rich galaxies with relatively low
equivalent-width emission lines \citep[see, e.g.,][]{liang04b}.  On
balance, we conclude that our measurement of the \mz{} relation at
intermediate redshift is consistent with previous measurements.

\section{Systematic Uncertainties}\label{sec:syseffects}  

Before discussing our results in the next section, we first
investigate how emission-line selection effects, aperture bias, AGN
contamination, and other potential sources of systematic bias might
impact our conclusions.  Readers that are only interested in the
interpretation of our results can safely skip ahead to
\S\ref{sec:discussion}.

\subsection{Emission-Line Selection Effects}\label{sec:effects} 

We selected our SDSS and AGES emission-line galaxy samples using an
\hb{} flux cut, and by requiring well-measured \oiilam{} and
\oiiilam{} emission lines (see \S\ref{sec:selection}).  Here, we
examine how these emission-line selection criteria might affect our
conclusions.

At a basic level, gas-phase abundances can only be measured for
galaxies with ongoing star formation, otherwise \hb{} will be absent
from their integrated spectra.  In principal, the metallicity of the
gas during the last major epoch of star formation is encoded in the
distribution of \emph{stellar} metallicities of the stars responsible
for the continuum light; however, disentangling this relationship is
complicated \citep[although not intractable; see,
  e.g.,][]{gallazzi05a, tojeiro07a, cid07a, panter08a}, and outside
the scope of the present analysis.  At face value, therefore, our
\hb{} flux cut translates into an \hb{} luminosity cut, or, modulo
aperture effects and dust attenuation, a cut in absolute SFR
\citep{kenn98a}.

In detail, however, our AGES and SDSS samples are also $I$- and
$r$-band limited, respectively; therefore, \hb{} in a galaxy at a
given redshift must be sufficiently bright relative to the underlying
continuum to be detected and measured given the design of the
spectroscopic survey (instrumental setup, flux limit, exposure time,
etc.).  In other words, in a broadband flux-limited survey, an \hb{}
flux cut translates into a EW-limited sample as a function of redshift
(see Fig.~\ref{fig:zvhb}).  This EW(\hb) limit is related to the
minimum specific SFR \citep{kenn94a, brinchmann04a} a galaxy must
exceed to enter our sample, although, the minimum (redshift-dependent)
specific SFR depends on many details, including aperture effects, dust
attenuation, the frequency of starbursts, and the amount of luminosity
and SFR evolution in galaxies.  Nevertheless, comparison of our
results against theoretical models should be aware of the systematic
loss of galaxies with low specific SFRs with increasing redshift.

In addition to excluding galaxies with low specific and absolute SFRs,
our \hb{} flux cut may also exclude extremely dusty galaxies.  In an
integrated spectrophotometric survey of star-forming galaxies in the
nearby Universe that included a significant number of
infrared-luminous galaxies, \citet{moustakas06a} found that galaxies
with significant \ha{} emission but with \hb{} too buried in the
stellar continuum to be measurable constituted $\lesssim10\%$ of the
sample with $L_{\rm IR}>10^{11}~\lsun$ \citep[so-called
  LIRGs;][]{sanders96a}.  Given the rapid increase in the incidence of
LIRGs to $z=1$ \citep{lefloch05a, rodighiero10a}, our sample may be
missing the dustiest galaxies at each redshift.  On the other hand, we
do not expect the metallicity properties of the small fraction of very
dusty galaxies to be significantly different with respect to the
galaxies that satisfy our \hb{} selection.  Nevertheless, future
studies of distant emission-line galaxies that are \ha- or even
Pa$\alpha$-selected would mitigate the selection effects due to dust
attenuation.

\begin{figure*}
\centering
\includegraphics[angle=90,scale=0.65]{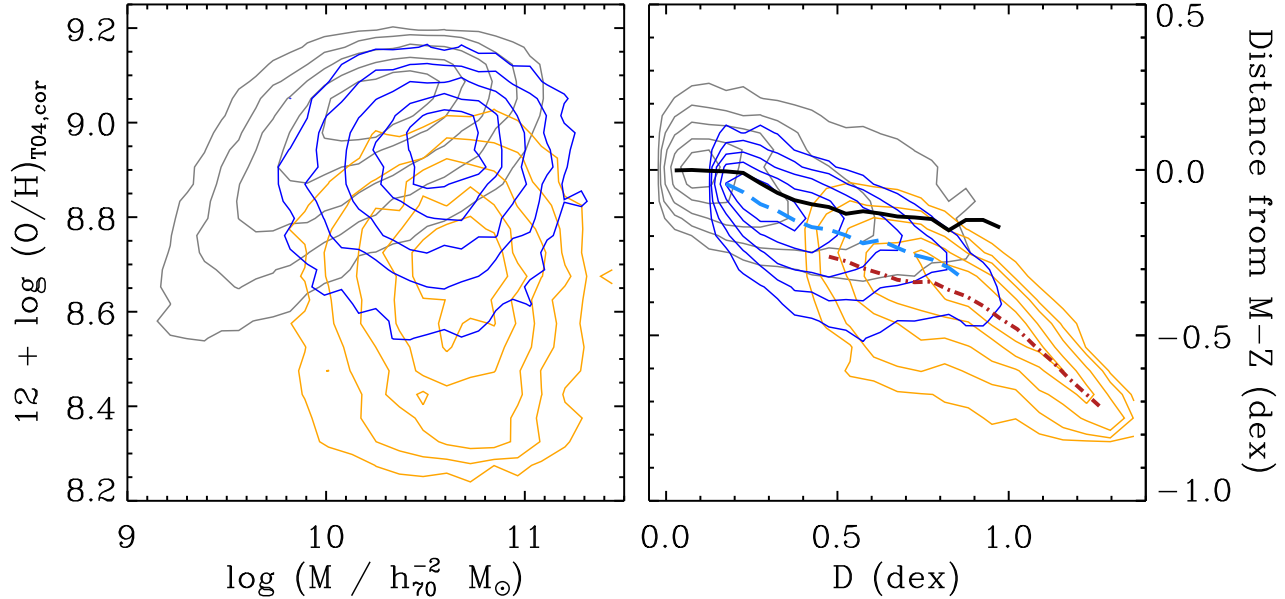}
\caption{Effect of AGN contamination on metallicities derived using
  the \pagel{} parameter.  The grey, blue, and orange contours,
  enclosing $25\%$, $50\%$, $75\%$, $90\%$, and $97.5\%$ of the
  galaxies in each sample, represent galaxies classified as SF,
  SF/AGN, and AGN, respectively (see \S\ref{sec:agncontam} for the
  definition of these classes).  (\emph{Left}) Inspection of the
  T04-based \mz{} relation clearly demonstrates that applying an
  \pagel-based abundance calibration to galaxies hosting AGN results
  in metallicities that are spuriously low.  (\emph{Right}) Residuals
  from the \mz{} relation vs.~distance, $D$, from
  $(\nii/\ha,\oiii/\hb) = (-0.45, -0.5)$~dex in the BPT diagram.  The
  solid black, dashed blue, and dot-dashed red curves show the median
  abundance residuals as a function of $D$ for the SF, SF/AGN, and AGN
  class, respectively.  We find that the metallicity bias for AGN is
  always $\gtrsim0.2$~dex, and can be as high as $\sim0.6$~dex;
  meanwhile, the metallicity bias for SF/AGN is always
  $\lesssim0.1$~dex. \label{fig:agncontam}}
\end{figure*}

Next, we also required our sample to have a well-measured \oii{}
emission line.  In \S\ref{sec:selection} we found that galaxies that
failed this cut constituted $\sim6\%$ of the \hb-selected sample and
were frequently dusty starbursts.  In a volume-limited sample of SDSS
galaxies, \citet{yan06a} found that galaxies with significant \ha{}
emission but without \oii{} emission constituted $\sim15\%$ of their
sample, the overwhelming majority of which had AGN-like forbidden
emission-line ratios, and the remainder likely dusty star-forming
galaxies.  Therefore, we conclude that requiring a well-measured
\oii{} emission line has a negligible effect on our results.  Indeed,
the high detection rate of \oii, due to its intrinsic strength and
relative insensitivity to variations in metallicity and excitation
across a wide range of physical conditions, is why this line remains
an appealing SFR diagnostic at both low and high redshift
\citep{moustakas06b, gilbank10a, mostek11a}.

Finally, we also required our sample of emission-line galaxies to have
a significant \oiiilam{} emission line.  Unlike \oii, the strength of
\oiii{} decreases rapidly with increasing metallicity and decreasing
excitation \citep{kewley02a}; therefore, an \oiii-limited sample could
be biased against massive, metal-rich galaxies.  To estimate the
magnitude of this potential systematic bias, we appeal to the
T04-based \mz{} relation plotted in the lower-left panel of
Figure~\ref{fig:mzlocal}.  Unlike the T04 calibration of \pagel{} (see
Appendix~\ref{appendix:calib}), the MPA-JHU metallicities do not
explicitly require significant \oiii{} emission.  We find that the T04
and MPA-JHU \mz{} relations agree very well, and only begin to deviate
from one another around $\mass\approx10^{10.6}$~\msun.  Furthermore,
the magnitude of the deviation, in the sense that the T04-based \mz{}
relation at large masses might be underestimated, is always
$\lesssim0.05$~dex to $10^{11.1}$~\msun.  We conclude, therefore, that
requiring well-measured \oiii{} emission line in our AGES and SDSS
samples does not significantly bias our conclusions.

\subsection{Strong-Line Abundance Calibration}\label{sec:calib}

We have shown that the choice of strong-line abundance calibration has
a significant effect on the absolute normalization of the \mz{}
relation and, to second order, its shape (see Fig.~\ref{fig:mzlocal}).
The KK04 abundance calibration, for example, yields oxygen abundances
that are, on average, $15\%-40\%$ higher than abundances derived using
the T04 and M91 calibrations, and results in a markedly shallower
\mz{} relation.  At fixed stellar mass, these differences translate
into a factor of $1.5-2$ systematic variation in the \emph{rate} at
which the mean metallicity of star-forming galaxies changes with
redshift (see \S\ref{sec:mzevol} and Fig.~\ref{fig:zvsoh_bymass}).

These results highlight two important points.  First, great care must
exercised when examining chemical evolutionary trends with redshift
based on oxygen abundances derived using different strong-line
calibrations.  For example, the abundances of distant galaxies derived
using the KK04 calibration should not, in general, be compared blindly
to the \mz{} relation published by \citet{tremonti04a}.  And second,
even differential trends in metallicity can depend on the adopted
abundance calibration.  An analogous result was found by
\citet{bresolin09a} and \citet{moustakas10a}, who showed that the
slope of the radial abundance gradient in nearby disk galaxies depends
on the adopted strong-line calibration.  Obviously both these effects
have wide-ranging implications for using abundance measurements to
constrain theoretical models \citep[e.g.,][]{dutton10a, dave11b}.

Various techniques have been devised to deal with these issues.  For
example, \citet{kewley08a} provide polynomial expressions to transform
abundances derived using different strong-line calibrations onto a
common abundance scale.  Other authors have used a combination of
abundances derived using electron temperature measurements and
photoionization models to attempt to generate a self-consistent suite
of abundance diagnostics based on different strong-line ratios
\citep{nagao06a, maiolino08a}, although these methods are susceptible
to systematic errors and inconsistencies in the photoionization models
and abundance measurements based on electron temperature estimates.
However, these calibrations should not applied beyond the range of
parameter space for which they were developed \citep[see especially
  the discussion in][]{stasinska10a}.  Moreover, these techniques
implicitly ignore the \emph{origin} of the systematic differences
between abundance derived using theoretical and empirical strong-line
methods \citep[see][and references therein]{moustakas10a}, which may
introduce metallicity-dependent biases.  These issues are particularly
relevant when applying strong-line calibrations based on the strength
of the \emph{nitrogen} emission line \niilam{} to infer the
\emph{oxygen} abundances of galaxies, such as \nii/\oii, \nii/\ha, and
(\oiii/\hb)/(\nii/\ha) \citep[also known as the N2O2, N2, and O3N2
  diagnostics, respectively;][and references therein]{kewley08a}.
Because nitrogen is both a primary and secondary nucleosynthetic
product, the N/O abundance ratio in galaxies has a well-known
second-order dependence on the recent star formation history, which
can introduce a non-negligible systematic bias in the inferred
metallicities \citep{contini02a, pilyugin03a, yin07a,
  perez-montero09b, thuan10a}.  Exploring these issues in more
quantitative detail is beyond the scope of this paper, suffice it to
say that we have avoided most of these problems by using three
independent abundance calibrations and applied them to our full sample
of SDSS and AGES galaxies consistently.

\subsection{Residual AGN Contamination}\label{sec:agncontam} 

In \S\ref{sec:agn} we identified AGN among the AGES emission-line
galaxies using a combination of the BPT diagram, the \citet{yan11a}
empirical diagnostic diagram, and a variety of complementary
multiwavelength diagnostics from the X-ray to the radio.  We
conservatively estimated, however, that up to $\sim50\%$ of the
emission-line galaxies in our $z>0.4$ sample may host composite (weak)
AGN.  Of course, the true fraction may be higher or lower depending on
a variety of different factors, such as evolution in the luminosity
and optical color distribution of galaxies hosting AGN, dilution of
the AGN spectrum due to aperture effects, and so forth.  Nevertheless,
oxygen abundances derived from the \pagel{} parameter are particularly
susceptible to AGN contamination because line-ratios involving
high-ionization lines such as \oiii/\hb{} are preferentially elevated
in the narrow-line regions of AGN \citep[e.g.,][]{osterbrock06a}.  For
galaxies on the upper \pagel{} branch, an unidentified contribution
from an AGN to the integrated emission-line spectrum will cause
\pagel{} to be \emph{overestimated}, and therefore the metallicity to
be systematically \emph{underestimated} (see \S\ref{sec:ohpagel} and
Appendix~\ref{appendix:calib}).

We use our SDSS sample to quantify the effect of unidentified AGN on
our \pagel-based oxygen abundances.  In Figure~\ref{fig:agncontam}
(\emph{left}) we plot the T04-based \mz{} relation for three sets of
objects: (1) galaxies classified as star-forming using the BPT diagram
({\em grey contours}, hereafter SF galaxies); (2) objects classified
as AGN using the BPT diagram, but as star-forming using the
\citet{yan11a} diagnostic diagram ({\em blue contours}, hereafter
SF/AGN); and (3) objects classified as AGN using both the BPT and the
\citet{yan11a} diagnostic diagram ({\em orange contours}, hereafter
AGN).  In this example we adopt the T04 calibration, but the same
conclusions would hold using any other \pagel-based abundance
calibration.  Figure~\ref{fig:agncontam} (\emph{right}) plots the
residuals from the \mz{} relation against the logarithmic distance,
$D$, from $\log\,(\nii/\ha)=-0.45$, $\log\,(\oiii/\hb) = -0.5$ in the
BPT diagram, at the base of the sequence of star-forming galaxies (see
Fig.~\ref{fig:bpt} and \citealt{kauffmann03c}).  For example, a galaxy
with $\log\,(\nii/\ha)=+0.2$ and $\log\,(\oiii/\hb)=+0.6$ would have
$D=\sqrt{(0.2+0.45)^2+(0.6+0.5)^2}=0.66$~dex.  In essence, powerful
AGN have large values of $D$ because they lie far from the tight
sequence of star-forming emission-line galaxies, while normal
star-forming galaxies have small $D$ values.  The solid black, dashed
blue, and dot-dashed red curves show the median abundance residuals as
a function of $D$ for the SF, SF/AGN, and AGN, respectively.

Figure~\ref{fig:agncontam} clearly reveals why AGN and, to a lesser
extent SF/AGN, must be removed from any \pagel-based abundance study:
the inferred metallicities are spuriously low because the emission
lines are not due to photoionization from massive stars.
Quantitatively, we find a very strong correlation between $D$ and the
metallicity bias.  Among AGN, the metallicity can be (mistakenly)
underestimated by as much as $\sim0.6$~dex for objects with
$D\sim1.2$, with a typical bias of $\sim0.2$~dex for most objects.
Meanwhile, the metallicity bias for SF/AGN is considerably more
modest, $\lesssim0.1$~dex (the separation betwen the solid black and
dashed blue curves in the right panel) over the full range of $D$
values.  A $50-50$ mixture of SF and SF/AGN, therefore, would exhibit
a mean metallicity bias of $\lesssim0.05$~dex.  We conclude,
therefore, that unidentified SF/AGN in our sample of AGES galaxies,
even those at $z>0.4$, should not significantly bias our measurement
of the evolution of the \mz{} relation.

\subsection{Aperture Bias}\label{sec:apbias}

\begin{figure}
\centering
\includegraphics[angle=0,scale=0.4]{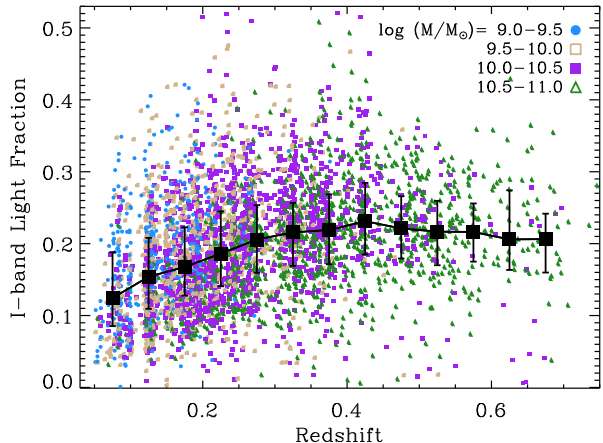}
\caption{Fraction of $I$-band light subtended by the AGES $1\farcs5$
  diameter fiber-aperture for galaxies in four bins of stellar mass
  vs.~redshift.  The large solid black squares correspond to the
  median light fraction (for all objects) in fixed $0.05$-wide bins of
  redshift, and the error bars indicate the interquartile range of the
  light fraction in each bin.  The median light-fraction is a weak
  function of redshift, increasing from $\sim13\%$ at $z\sim0.1$, to
  $\sim20\%$ above $z\gtrsim0.25$.  \label{fig:z_vs_lfrac}}
\end{figure}

Most star-forming disk galaxies in the local Universe exhibit radial
abundance gradients, in the sense that the metallicity in their
central regions is higher than in their outskirts \citep{zaritsky94a,
  vanzee98a, moustakas10a}.  Therefore, oxygen abundances derived from
fiber-optic spectra, which target the inner, metal-rich regions of the
galaxy, may be biased relative to abundances derived from integrated
spectroscopy \citep{zaritsky95a, kochanek01a, kewley05a,
  moustakas06b}.  If the amount of light subtended by the
spectroscopic aperture is a strong function of redshift then so-called
aperture bias (or aperture effects) could mimic genuine chemical
evolution.

In Figure~\ref{fig:z_vs_lfrac} we explore this question by plotting
the $I$-band light fraction versus redshift for galaxies in our
abundance sample divided into four bins of stellar mass, \logmmsun:
$9-9.5$ ({\em blue points}), $9.5-10$ ({\em open tan squares}),
$10-10.5$ ({\em filled purple squares}), and $10.5-11$ ({\em green
  diamonds}).  We define the light-fraction as the flux measured from
our unconvolved $I$-band mosaics inside a $1\farcs5$ diameter aperture
centered on each galaxy, divided by the total galaxy flux.  The median
light-fraction for the whole sample is $0.19$, with an interquartile
range of $0.14-0.24$; however, Figure~\ref{fig:z_vs_lfrac} reveals
that the light-fraction is indeed a (weak) function of redshift.  The
large solid black squares correspond to the median light fraction in
$0.05$-wide bins of redshift, while the error bars indicate the
interquartile range in each bin.  The median light-fraction increases
from $\sim13\%$ at $z\sim0.1$, and then remains approximately constant
at $\sim20\%$ above $z\gtrsim0.25$.  For reference, the physical
diameter subtended by the AGES fiber-aperture is $5.9$~kpc at
$z=0.25$, ranging from $2.8-10.7$~kpc between $z=0.1-0.7$.

Taking these results at face value, the \emph{relative} metallicities
we measure in AGES should be insensitive to aperture bias between
$z=0.25-0.75$.  Below $z\sim0.25$ we expect aperture effects to be
more severe; however, at these redshifts we rely on the abundances
inferred from the SDSS spectra, which were obtained through a
$3\arcsec$ diameter fiber-optic aperture.  The median $i$-band
light-fraction for our SDSS sample is $0.24$, with an interquartile
range of $0.18-0.31$.  The physical diameter subtended by the SDSS
fiber is $5.5$~kpc at $z=0.1$, ranging from $2.9-9.9$~kpc between
$z=0.05-0.2$.  In other words, by combining our SDSS sample at low
redshift with AGES at higher redshift the relative metallicity
evolution we measure should be insensitive to aperture effects.  Of
course, because of the existence of abundance gradients, the
\emph{absolute} metallicities we measure may be susceptible to
aperture bias at all redshifts.

We can place these results on a more quantitative footing by
performing some simple simulations.  For convenience, in the following
we adopt the notation $Z\equiv\logoh$.  The inputs to our simulation
are the (assumed linear) abundance gradient $Z(r)=Z_{0}+(dZ/dr)r$,
where $r$ is the galactocentric radius, $Z_{0}$ is the metallicity at
$r=0$, and $dZ/dr$ is the slope; and the radial surface-brightness
profile, $I(r)\propto\exp[-(r/r_{0})^{1/n}]$, where $n$ is the
\citet{sersic68a} index, and $r_{0}$ is a scale factor that is related
to the half-light radius, $r_{50}$ (see \citealt{graham05a} for the
relationship).  The \sersic{} index indicates the \emph{concentration}
of the galaxy, with increasing $n$ corresponding to a more centrally
concentrated surface-brightness profile \citep[e.g.,][]{graham01a,
  peng02a}.  For example, $n=1$ implies a pure exponential disk, while
$n=4$ corresponds to a classical \citet{devac48a} $r^{1/4}$
light-profile.

At a given radius the enclosed light-fraction, $F(r)$, is given by

\begin{equation}
F(r) = \frac{\int_{0}^{r} I(r)rdr}{\int_{0}^{\infty} I(r)rdr}, 
\label{eq:apbias1}
\end{equation}

\noindent and the surface-brightness weighted metallicity is given by 

\begin{equation}
\langle Z(r) \rangle \propto \frac{\int_{0}^{r} Z(r)\,I(r)rdr}
        {\int_{0}^{r} I(r)rdr}.
\label{eq:apbias2}
\end{equation}

\noindent The \emph{integrated} metallicity, $\langle Z_{\rm int}
\rangle$, that would be measured from a spectroscopic aperture that
captures all the light of the galaxy \citep[e.g.,][]{jansen00a,
  moustakas06a} follows from equation~(\ref{eq:apbias2}) by letting
$r\rightarrow\infty$.

\begin{figure}
\centering
\includegraphics[scale=0.4]{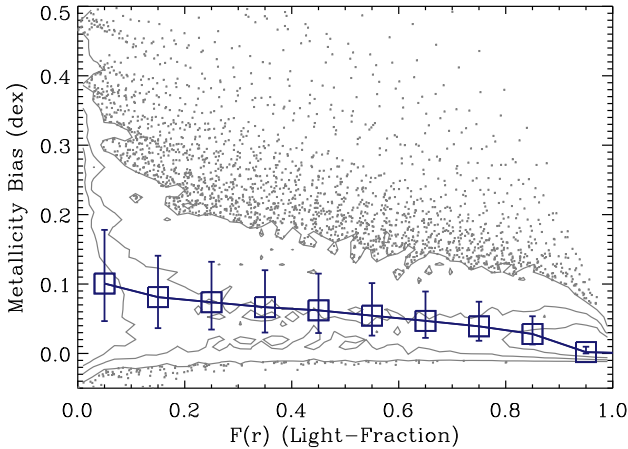}
\caption{Simulated effect of aperture bias on the metallicity inferred
  from spectra that enclose a fraction of the integrated light of the
  galaxy (see \S\ref{sec:apbias} for a complete description of the
  parameters of the simulation).  For a given light-fraction,
  $0<F(r)<1$, the abscissa gives the difference between the
  metallicity one would infer from a spectrum that encloses $F(r)$ of
  the total light of the galaxy and the integrated metallicity.  The
  metallicity bias is positive because disk galaxies exhibit radial
  abundance gradients; therefore, their centers are generally more
  metal-rich than their outskirts.  The contours and small grey points
  correspond to simulated galaxies with a range of concentrations
  (\sersic{} indices), sizes, and abundance gradient slopes.  For
  reference, the contours enclose $50\%$, $75\%$, and $95\%$ of the
  simulated galaxies and the points are objects that lie outside the
  $95\%$ contour level.  The large blue squares show the median
  metallicity bias in fixed bins of $F(r)$, and the error bars
  indicate the interquartile range.  We find that the metallicity bias
  is typically $\lesssim0.1$~dex, and a relatively weak function of
  enclosed light fraction. \label{fig:apbias}}
\end{figure}

With these simple ingredients in-hand we can now estimate the relative
effects of aperture bias on the inferred metallicities for a wide
range of surface-brightness profiles, sizes, and abundance gradients.
To constrain the range of $n$ and $r_{50}$ we use the distribution of
physical properties for late-type galaxies recently compiled by
\citet[see especially \S3.1 and Fig.~8 in that review]{blanton09a} for
SDSS galaxies at $z<0.05$.  Specifically, we adopt a normally
distributed prior on $n$ with mean $\langle n\rangle=1.8$ and standard
deviation $\sigma_{n}=0.6$, and a log-normal prior on $r_{50}$ with
$\langle \log\,(r_{50})\rangle=0.35$~kpc and
$\sigma_{\log\,(r_{50})}=0.25$~kpc.  For the range of abundance
gradient slopes we use the recent measurements by \citet{moustakas10a}
based on a sample of $21$ nearby disk galaxies; they find typical
values $\langle dZ/dr\rangle=-0.03$~dex~kpc$^{-1}$ and
$\sigma_{dZ/dr}=0.02$~dex~kpc$^{-1}$.  For each mock galaxy, we
simulate the effects of metallicity scatter at fixed galactocentric
radius (e.g., due to azimuthal metallicity variations) by perturbing
the metallicity at each radius by a Gaussian distribution with mean
$0.06$~dex and standard deviation $0.02$~dex \citep{moustakas10a}.
Finally, for each simulated galaxy we compute, as a function of $r$,
the enclosed light fraction, $F(r)$, and the {\em metallicity bias},
$\Delta \langle Z(r) \rangle\equiv \langle Z(r) \rangle-\langle Z_{\rm
  int} \rangle$, or the difference between the surface-brightness
weighted metallicity at that radius and the integrated metallicity
using equations~(\ref{eq:apbias1}) and (\ref{eq:apbias2}).  We perform
$1000$ Monte Carlo simulations and plot the results in
Figure~\ref{fig:apbias}.

Figure~\ref{fig:apbias} reveals that, for our fiducial set of
simulations, the metallicity bias is typically $\lesssim0.1$~dex;
moreover, it is a relatively weak function of enclosed light fraction.
The principal reason for this result is that, although disk galaxies
do indeed possess radial abundance gradients, they are not especially
steep in absolute terms (i.e., in dex~kpc$^{-1}$).  For example,
making the mean abundance gradient slope a factor of three steeper,
$\langle dZ/dr\rangle=-0.09$~dex~kpc$^{-1}$ \citep[e.g., if abundance
  gradients were steeper at higher redshift;][]{boissier01a},
increases the absolute metallicity bias to $0.2-0.3$~dex for
$F(r)\lesssim0.5$, and makes it a much stronger of light-fraction;
however, the \emph{relative} metallicity bias over the relevant range
of light-fractions (see Fig.~\ref{fig:z_vs_lfrac}) is still
$\lesssim0.1$~dex.  Finally, we tested making our simulated galaxies a
factor of $2-3$ larger and a factor of $2-3$ more concentrated and the
basic results do not change.  Based on these tests, we conclude that
aperture bias should not significantly affect our measurement of the
relative change in the abundances of star-forming galaxies from
$z=0.05-0.75$.

\section{Discussion}\label{sec:discussion} 

We have measured the evolution of the \mz{} relation from
$z=0.05-0.75$ using statistically complete samples of star-forming
galaxies selected from AGES and the SDSS, carefully accounting for
both random and systematic sources of uncertainty.  Our analysis has
yielded three key results.  First, we find that star-forming galaxies
obey a well-defined \mz{} relation over the entire redshift range
studied, $z=0.05-0.75$.  This result is generally consistent with
previous measurements of the \mz{} relation at intermediate redshift,
although the size and statistical completeness of our sample places it
on a firmer quantitative footing.  Second, we find that at fixed
stellar mass, $\mass=10^{10.5}$~\msun, the mean metallicity of
star-forming galaxies changes by just $30\%-60\%$ since $z\sim0.7$.
The factor of two uncertainty on the amount of metallicity evolution
is due to the systematic differences in the abundances inferred using
the three strong-line calibrations investigated in this paper (KK04,
T04, and M91).  Finally, we find no statistically significant evidence
that the \mz{} relation evolves in a mass-dependent way over this
redshift range for galaxies with $\mass\simeq10^{9.8}-10^{11}$~\msun.

Despite considerable effort in recent years, a thorough theoretical
explanation of the physical origin and evolution of the \mz{} relation
remains elusive.  However, models now provide quantitative estimates
not only of the \mz{} relation, but also auxiliary relations between
stellar mass, SFR, and gas fraction, that can be utilized in concert
to test the underlying model assumptions.  Although we postpone a
detailed comparison of our results on the evolution of the \mz{}
relation with various theoretical models to a forthcoming paper, a
brief discussion of one class models is suggestive of what can be
done.

\citet{dave11c} have proposed a simple analytic framework for
interpreting the \mz{} relation based on insights gained analyzing a
suite of cosmological hydrodynamic simulations with different
prescriptions for galactic winds \citep{dave06a, dave11b,
  oppenheimer08a, finlator08a, oppenheimer10a}.  Their central
hypothesis is that most galaxies are in a slowly evolving equilibrium
with the intergalactic medium (IGM), balanced between the competing
effects of accretion, outflows, and star formation.  They derive an
expression for the \mz{} relation given by

\begin{equation}
Z_{\rm g} = \frac{y}{(1+\eta)(1-\alpha_{Z})},
\label{eq:mzdave}
\end{equation}

\noindent where $\eta$ (the {\em mass loading factor}) is a
proportionality constant relating the outflow-driven mass-loss rate to
the instantaneous SFR, and $\alpha_{Z}$ is the ratio of the
metallicity of the gas being accreted from the IGM relative to the
gas-phase metallicity of the galaxy \citep{finlator08a}.  Their
simulations show that a significant fraction of the metals previously
ejected from galaxies in star-formation driven outflows are reaccreted
in a mass-dependent way \citep[{\em wind
    recycling};][]{oppenheimer08a, oppenheimer10a}, or $\alpha_{Z}>0$.
The result is an \mz{} relation whose shape is predominantly
determined by the fraction of accreted gas that forms stars (modulated
by $\eta$), while its evolution is govered by $\alpha_{Z}$, the
relative enrichment level of the infalling gas.  Their favored model
based on momentum-driven galactic winds \citep{murray_norman05a}
grossly reproduces both the shape and evolution of the \mz{} relation
from $z=0-3$ (with some notable discrepancies; see \citealt{dave11b}
for details).

An interesting consequence of this model is that deviations from
equilibrium at fixed stellar mass should correlate with metallicity,
SFR, and gas fraction.  For example, galaxies that have recently
accreted significant amounts of cold gas from the IGM should have
slightly lower metallicities (via dilution, assuming
$\alpha_{Z}\sim0$), and should be forming stars at slightly higher
rates (due to the renewed gas supply) compared to galaxies of the same
stellar mass still at equilibrium.  These results provide a compelling
physical explanation for the so-called fundamental metallicity
relation \citep{mannucci10a}, which posits that galaxies occupy a
three-dimensional plane linking stellar mass, gas-phase metallicity,
and SFR.  Specifically, the fundamental metallicity relation reveals
that at fixed stellar mass, galaxies with high SFRs tend to be
metal-poor, while galaxies that lie above the median \mz{} relation
generally have lower SFRs.

This type of comparison can, and will, be done with our SDSS and AGES
observations.  By incorporating accurate SFRs and approximate gas
fractions, the latter of which we intend to estimate by inverting the
Kennicutt-Schmidt \citep{kenn98b} law \citep[see, e.g., Appendix~A
  in][]{peeples11a}, our analysis will provide powerful new insights
into the physical origin and evolution of the \mz{} relation, and,
more generally, on our theoretical understanding of galaxy formation.

We conclude this section with a brief discussion of previous and
ongoing efforts to measure the \mz{} relation at even higher redshift,
$z>1$.  Abundance studies of star-forming galaxies at these redshifts
are especially challenging because they require near-infrared
spectroscopy, which remains difficult and time-consuming to obtain
from the ground.  Nevertheless, building on a handful of pioneering
abundance studies of $z\sim2$ galaxies \citep{pettini98a, pettini01a,
  kobulnicky00b, shapley04a, swinbank04a}, measurements of the \mz{}
relation for star-forming galaxies now exist for galaxies at $1<z<1.6$
\citep{shapley05a, maier06a, liu08a, perez-montero09a, queyrel09a},
Lyman-break and near-infrared selected galaxies at $z\sim2$
\citep{erb06a, hayashi09a}, and star-forming galaxies at redshifts as
high as $z\sim3.5$ \citep{maiolino08a, mannucci09a, lemoine10a,
  richard11a}.

It is instructive to highlight two important, outstanding issues
confronting observational abundance studies of star-forming galaxies
at $z>1$.  First, the availability of different emission lines
depending on the redshift range of the sample under investigation and
the spectral range of the observations means that metallicities in
general cannot be derived using the same strong-line calibration at
all redshifts.  For example, the abundances of star-forming galaxies
at $z=1-2$ have been based, in most cases, on the N2 and O3N2
diagnostics \citep[see \S\ref{sec:calib}; e.g.,][but see
  \citealt{perez-montero09a} for application of the so-called O2Ne3
  diagnostic]{shapley05a, erb06a, queyrel09a}, while the \pagel{}
diagnostic has been used for galaxies at $z=3-3.5$ \citep{mannucci09a,
  richard11a}.  Consequently, efforts to compare the \mz{} relation
for star-forming galaxies at different epochs is complicated by the
systematic differences among strong-line abundance calibrations (see
the discussion in \S\ref{sec:calib}).  Second, observations suggest
that the physical conditions (ionization parameter, electron density
or interstellar pressure, prevalence of shocks, etc.) in star-forming
galaxies at $z\gtrsim1$ may be dramatically different compared to
local galaxies \citep{shapley04a, liu08a, brinchmann08a, lehnert09a,
  hainline09a}.  It remains to be seen, therefore, whether using
strong-line abundance diagnostics that have been calibrated against
\hii{} regions in nearby galaxies introduces systematic errors in the
metallicity measurements of high-redshift galaxies.

Addressing these issues requires concerted observational effort on two
fronts.  First, high-quality observations of \hii{} regions in nearby
galaxies, coupled with more detailed photoionization models, are
needed to understand the physical origin of the factor of $\sim5$
systematic uncertainty in the nebular abundance scale (see the
discussion in \citealt{moustakas10a}, and references therein).  In
addition, the effect of secondary nitrogen enrichment on oxygen
abundances derived using the N2 and O3N2 abundance diagnostics must be
better understood \citep{yin07a, perez-montero09b, thuan10a}.  And
second, dedicated optical and near-infrared spectroscopic observing
campaigns of large samples of higher redshift galaxies are needed to
measure the full suite of rest-frame optical emission-line
diagnostics.  In particular, the advent of highly multiplexed
near-infrared spectrographs on $8-10$-m class telescopes like LBT/{\sc
  lucifer} \citep{mandel00a}, Keck/{\sc mosfire} \citep{mclean08a},
GTC/{\sc emir} \citep{garzon03a} will facilitate the kinds of detailed
studies that until recently have only been possible for lower-redshift
galaxies.

\section{Summary}\label{sec:summary}

We have measured the gas-phase oxygen abundances, stellar masses, and
$B$-band luminosities of $\sim3000$ star-forming galaxies at
$z=0.05-0.75$ observed as part of AGES, a redshift survey of $I_{\rm
  AB}<20.45$ galaxies over $7.9$~deg$^{2}$ in the NDWFS \bootes{}
field.  This sample is among the largest statistically complete,
wide-area samples of intermediate-redshift galaxies with measured
nebular abundances assembled.  We use state-of-the-art techniques
coupled to high-resolution population synthesis models to measure the
nebular emission lines free from the systematic effects of underlying
stellar absorption.  In addition, we model the observed deep,
aperture-matched broadband optical and near-infrared photometry of the
galaxies in our sample using {\tt iSEDfit}, a new Bayesian SED-fitting
code, to infer their stellar masses.  Using multiple complementary
multiwavelength diagnostics based on optical line-ratios and ancillary
X-ray, mid-infrared, and radio observations, we identify and remove
AGN from our sample.  We combine volume-limited observations from AGES
with a complementary, statistically complete sample of $\sim75,000$
star-forming galaxies at $z=0.05-0.2$ selected from the SDSS, which we
analyze in the identical manner.  We use the joint SDSS and AGES
sample to measure the evolution of the \mz{} and $B$-band \lz{}
relations between $z=0.05$ and $z=0.75$ using three independent
strong-abundance calibrations (KK04, T04, and M91) of the
metallicity-sensitive \pagel{} parameter.

We divide our principal findings into two categories, lessons learned
and quantitative results.  Among the lessons we have learned:

\begin{itemize}
\item[1.]{EWs provide a powerful, reliable means of inferring oxygen
  abundances from nebular emission lines free from the systematic
  effects of dust attenuation.  They also have the added advantage of
  being measureable in spectra that have not been flux-calibrated,
  although it is critical that the emission lines, especially \hb, be
  corrected for underlying stellar absorption.}

\item[2.]{A thorough culling of AGN is necessary to ensure that
  metallicities derived using the \pagel{} parameter are not
  spuriously underestimated; the metallicity bias for galaxies with a
  subdominant ($\lesssim30\%$) contribution from an AGN is typically
  $\lesssim0.1$~dex, but can be $\gg0.2$~dex for powerful AGN.}

\item[3.]{We find that aperture effects are not a significant source
  of systematic error in the gas-phase metallicities derived from
  fiber-optic spectroscopic surveys like the SDSS and AGES, owing to
  the relatively shallow radial abundance gradients in disk galaxies.}

\item[4.]{The choice of strong-line calibration significantly affects
  not only the absolute normalization and shape of the \mz{} relation,
  but also the \emph{rate} of chemical evolution inferred, and its
  stellar mass dependence.  Consequently, we caution strongly against
  comparing the metallicity measurements across redshifts based on
  different emission-line ratios (see especially the discussion in
  \S\ref{sec:calib}).}

\item[5.]{We propose a new, physically motivated parametric model for
  the \mz{} relation given by equation~(\ref{eq:mzclosedbox}).  This
  parameterization has the same number of free parameters as the more
  commonly used polynomial model but has whose asymptotic behavior is
  much more stable.}

\item[6.]{We confirm that stellar mass provides a much less ambiguous
  measurement of metallicity evolution than optical luminosity, which
  itself evolves strongly.  The $B$-band \lz{} relation in particular
  does not provide a reliable means of constraining the chemical
  evolution of star-forming galaxies.}
\end{itemize}

\noindent And our principal quantitative conclusions are:

\begin{itemize}
\item[1.]{We find no statistically significant evidence for evolution
  in the shape of the \mz{} relation to $z=0.75$ for
  $\mass\simeq10^{9.8}-10^{11}$~\msun{} star-forming galaxies,
  contrary to previous findings based on smaller, incomplete samples.}

\item[2.]{We find only modest evolution in the overall normalization
  of the \mz{} relation, suggesting a metallicity increase of between
  $30\%-60\%$ for $\mass\simeq10^{9.8}-10^{11}$~\msun{} galaxies since
  $z=0.75$.  The uncertainty in our measurement is entirely due to the
  choice of strong-line calibration.}

\end{itemize}

Our results indicate that massive star-forming galaxies at
intermediate redshift are chemically evolved, implying that they
synthesized the bulk of their metals at higher redshift, $z>1$.  Given
their stellar masses, this result is perhaps not too surprising.  On
the other hand, a challenge for theoretical galaxy formation models
that incorporate chemical evolution will be to explain not only the
relatively slow chemical enrichment rate for star-forming galaxies
over this redshift range, but also the lack of evolution in the
\emph{shape} of the \mz{} relation.  In a forthcoming paper we will
combine our metallicity and stellar mass estimates with SFRs and
approximate gas fractions to directly test the predictions of several
of these theoretical models \citep[e.g.,][]{brooks07a, dave11c,
  peeples11a}.

On the observational side, joint optical and near-IR spectroscopy of
intermediate-redshift star-forming galaxies would be especially
useful.  These observations would provide the full suite of rest-frame
optical emission-line diagnostics, from \oiilam{} to \ha{} and
\niidoublet, thereby enabling optical AGN to be identified more
securely, and allowing gas-phase metallicities to be estimated using
multiple independent strong-line abundance calibrations.  Finally,
abundance studies of statistically complete samples of
intermediate-redshift galaxies spanning a wider range of stellar
masses, $10^{8}-10^{11}$~\msun, should be a top priority
\citep[e.g.,][]{zahid11a}.  Follow-up high-resolution optical and
near-IR spectroscopy of lower-mass galaxies preselected from the
latest generation of wide-area galaxy redshift surveys like GAMA
\citep{driver11a}, zCOSMOS \citep{lilly09a}, and PRIMUS
\citep{coil11a} would be one path forward.  These abundance
measurements, coupled with detailed stellar masses and SFRs, would
provide valuable insight into the star formation and chemical
enrichment histories of star-forming galaxies over a significant
fraction of the age of the Universe.

\acknowledgements

This paper has benefited from insightful suggestions, discussions, and
encouragement from James Aird, Roberto Assef, Michael Blanton, Alison
Coil, Aleks Diamond-Stanic, Ryan Hickox, David Hogg, Robert Kennicutt,
Alexander Mendez, Leonidas Moustakas, Casey Papovich, and Christy
Tremonti.  We gratefully acknowledge the MMT staff and Hectospec
instrument team for their help in carrying out the AGES survey, and
J.~M. acknowledges financial support through the National Science
Foundation grant AST-0908246.  We kindly thank Len Cowie for providing
an electronic table of the metallicity measurements from
\citet{cowie08a} shown in Figure~\ref{fig:mzlit}, and Fuyan Bian and
Xiaohui Fan for providing access to their LBT/LBC $U$-band imaging of
the \bootes{} field.

Observations reported here were obtained at the MMT Observatory, a
joint facility of the Smithsonian Institution and the University of
Arizona.  This work made use of images and/or data products provided
by the NOAO Deep Wide-Field Survey \citep{jannuzi99a}, which is
supported by the National Optical Astronomy Observatory (NOAO). NOAO
is operated by AURA, Inc., under a cooperative agreement with the
National Science Foundation.  This work is based in part on
observations made with the {\em Spitzer Space Telescope}, which is
operated by the Jet Propulsion Laboratory, California Institute of
Technology under a contract with NASA. Support for this work was
provided by NASA through an award issued by JPL/Caltech.  The X-ray
data were obtained using the Chandra X-ray Observatory operated by the
Chandra X-ray Center at the Harvard-Smithsonian Center for
Astrophysics, funded by NASA COntract NAS8-03060.

Funding for the Sloan Digital Sky Survey (SDSS) and SDSS-II has been
provided by the Alfred P. Sloan Foundation, the Participating
Institutions, the National Science Foundation, the U.S. Department of
Energy, the National Aeronautics and Space Administration, the
Japanese Monbukagakusho, and the Max Planck Society, and the Higher
Education Funding Council for England. The SDSS Web site is
\url{http://www.sdss.org}.  

The SDSS is managed by the Astrophysical Research Consortium (ARC) for
the Participating Institutions. The Participating Institutions are the
American Museum of Natural History, Astrophysical Institute Potsdam,
University of Basel, University of Cambridge, Case Western Reserve
University, The University of Chicago, Drexel University, Fermilab,
the Institute for Advanced Study, the Japan Participation Group, The
Johns Hopkins University, the Joint Institute for Nuclear
Astrophysics, the Kavli Institute for Particle Astrophysics and
Cosmology, the Korean Scientist Group, the Chinese Academy of Sciences
(LAMOST), Los Alamos National Laboratory, the Max-Planck-Institute for
Astronomy (MPIA), the Max-Planck-Institute for Astrophysics (MPA), New
Mexico State University, Ohio State University, University of
Pittsburgh, University of Portsmouth, Princeton University, the United
States Naval Observatory, and the University of Washington.

{\it Facilities:} 
\facility{MMT, Spitzer, Chandra, Mayall, LBT}

\setcounter{figure}{0}
\renewcommand\thefigure{A\arabic{figure}}

\begin{figure*}
\centering
\includegraphics[angle=0,scale=0.75]{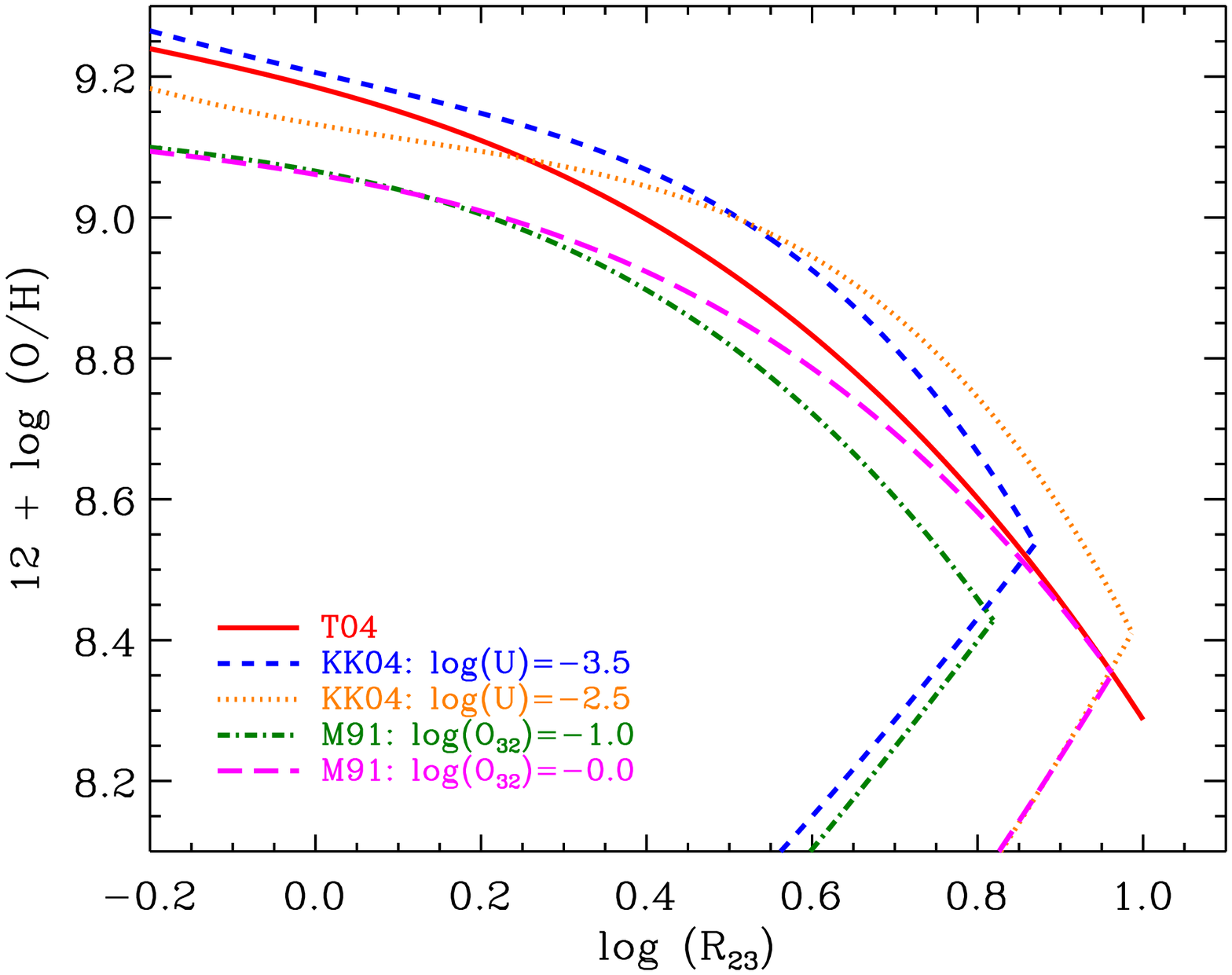}
\caption{Relationship between \pagel{} and oxygen abundance for the
  three theoretical strong-line abundance calibrations adopted in this
  paper: T04 ({\em solid red curve}); KK04 ({\em short-dashed blue and
    dotted orange curve, corresponding to $\logu=-3.5$ and $-2.5$,
    respectively}); and M91 ({\em dot-dashed green and long-dashed
    purple curve, corresponding to $\logioniz=-1.0$ and $0.0$,
    respectively}). \label{fig:r23}}
\end{figure*}

\begin{appendix}
\section{Abundance Calibrations}\label{appendix:calib}

For completeness and the convenience of the reader, we describe and
compare the three strong-line calibrations adopted in this paper.  For
the KK04 calibration, we have

\begin{equation}
12 + \log\,({\rm O/H})_{\rm KK04,lower} = 9.4 + 4.65x - 3.17x^2 -
\logq\,(0.272 + 0.547x -0.513x^2),  
\label{eq:kk04_lower}
\end{equation}

\noindent and 

\begin{eqnarray}
12 & + & \log\,({\rm O/H})_{\rm KK04,upper} = 9.72 - 0.777x - 0.951x^2
-0.072x^3 - 0.811x^4 \nonumber \\ &-& \logq\,(0.0737 -0.0713x -
0.141x^2 + 0.0373x^3 - 0.058x^4),
\label{eq:kk04_upper}
\end{eqnarray}

\noindent for galaxies on the lower and upper branch, respectively,
where $x\equiv\log\,(\pagel)$.  The ionization parameter $q$ in
${\rm cm~s}^{-1}$ is given by

\begin{eqnarray}
\logq & = & 32.81 - 1.153y^2 + z(-3.396 - 0.025y + 0.1444y^2) \nonumber \\ 
      & \times & [4.603 - 0.3119y - 0.163y^2 + z(-0.48+0.0271y+0.02037y^2)]^{-1},
\label{eq:kk04_logu}
\end{eqnarray}

\noindent where $z\equiv\logoh$ and $y\equiv\log\,(\ioniz)$
characterizes the hardness of the ionizing radiation field.  The
ionization parameter is also frequently written as a dimensionless
quantity $U\equiv q/c$, where $c=2.99\times10^{10}$~\cms{} is the
speed-of-light \citep{shields90a, kewley02a}.  Note that equations
(\ref{eq:kk04_lower})-(\ref{eq:kk04_logu}) must be solved iteratively
for both the ionization parameter and the oxygen abundance;
convergence is typically achieved in a handful of iterations.

For the M91 calibration, we adopt the parameterizations proposed by
\citet{kobulnicky99a}, but see \citet{kuzio04a} for alternative
parameterizations based on trigonometric functions.  For the lower
branch, we have

\begin{equation}
12 + \log\,({\rm O/H})_{\rm M91,lower} = 12 - 4.944 + 0.767x +
0.602x^2 - y(0.29 + 0.332x - 0.331x^2),
\label{eq:m91_lower}
\end{equation}

\noindent and for the upper branch we have

\begin{eqnarray}
12 + \log\,({\rm O/H})_{\rm M91,upper} & = & 12 - 2.939 - 0.2x -
0.237x^2 - 0.305x^3 - 0.0283x^4 \nonumber \\ &-& y(0.0047 -
0.0221x - 0.102x^2 - 0.0817x^3 - 0.00717x^4).
\label{eq:m91_upper}
\end{eqnarray}

And finally the T04 calibration is only defined for galaxies on the
upper \pagel{} branch:

\begin{eqnarray}
12 & + & \log\,({\rm O/H})_{\rm T04,upper} =
9.185 - 0.313x - 0.264x^2 - 0.321x^3
\label{eq:t04_upper}
\end{eqnarray}

In Figure~\ref{fig:r23} we compare the three separate strong-line
calibrations for two values of the ionization parameters $U$ and
\ioniz.  

\section{Proposed New Functional Form for the \mz{}
  Relation}\label{appendix:mzform}

Previous studies have parameterized the \mz{} relation using a
quadratic or third-order polynomial \citep[e.g.,][]{tremonti04a,
  kewley08a}.  A polynomial model is unsatisfactory, however, because
it can lead to an \mz{} relation that \emph{decreases} with increasing
mass (see, e.g., Fig.~2 of \citealt{kewley08a}).  This artificial
turn-over in the polynomial \mz{} relation is especially problematic
when the model is extrapolated beyond the range of stellar masses for
which it was calibrated.  Therefore, in \S\ref{sec:mzlocal} we fit the
local \mz{} relation with a new functional form, which we reproduce
here for the convenience of the reader:

\begin{equation}
12+\log\,(\textrm{O}/\textrm{H}) = 12+\log\,(\textrm{O}/\textrm{H})^{\ast} -
\log\, \left[1+\left(\frac{\mstar}{10^{9}\, \msun}\right)^{\gamma}\right].
\label{eq:appendix:mzclosedbox}
\end{equation}

\noindent Unlike the polynomial model, this parameterization
encapsulates our physical intuition that the \mz{} relation should
vary monotonically with stellar mass.  In our proposed model, \mstar{}
refers to the stellar mass where the \mz{} relation begins to bend or
flatten at a rate that is controlled by the power-law slope, $\gamma$.
For example, an \mz{} relation with $\gamma\ll1$ remains steep well
above \mstar, while an \mz{} relation with $\gamma\gg1$ is essentially
flat for $\mass>\mstar$.  The parameter \ohstar{} is the asymptotic
metallicity of the \mz{} relation, i.e., the metallicity for galaxies
with $\mass\gg\mstar$.  

The form and asymptotic limit of our model is motivated by the
closed-box equations of chemical evolution \citep{talbot71a,
  binney98a}.  If the SFR declines exponentially with a characteristic
timescale $\tau$, $\sfr(t)\propto\exp(-t/\tau)$, then it is
straightforward to show that the gas-phase metallicity at time $t$,
$Z(t)$, is given by

\begin{equation}
Z(t) = -y\,\ln\, [1-(1-R)\, (1-\exp^{-t/\tau})],
\label{eq:zoft}
\end{equation}

\noindent where $y$ is the nucleosynthetic yield, and $R$ is the {\em
  return fraction}, the fraction of mass returned by stars to the
interstellar medium by supernovae and stellar winds
\citep{pagel97a}.\footnote{In detail, the return fraction is also a
  function of time \citep{leitner11b}, although we ignore its
  time-dependence here.}  Note that in the absence of inflow and
outflow, the return fraction simply equals the baryonic gas fraction.
In the limit $t\gg\tau$ equation~(\ref{eq:zoft}) becomes
$Z=-y\ln\,(R)$, i.e., the gas-phase metallicity becomes a constant
equal to a fraction of the yield, which is analogous to our
characteristic metallicity, \ohstar.

In addition to equation~(\ref{eq:appendix:mzclosedbox}) and the
traditional polynomial parameterization of the \mz{} relation, we also
investigated a smooth double power-law model, which is commonly used
to fit the X-ray and infrared luminosity functions
\citep[e.g,][]{aird08a, rujopakarn10a}, and a broken double power-law.
Unfortunately, the double power-law model exhibits the same general
behavior as the polynomial model (i.e., it turns over rapidly at large
stellar mass), while the broken power-law has a sharp, artificial
break with no physical basis.

\end{appendix}

\end{document}